\documentclass[aps,pra,amssymb,twocolumn,showpacs,a4paper]{revtex4-1}

\usepackage{amsmath,amssymb,epsfig,graphicx,color,framed}
\usepackage{hyperref}
\usepackage{amsfonts} 
\usepackage{amsmath}
\usepackage{amssymb}
\usepackage{graphicx}
\usepackage{hyperref}
\usepackage{color}
\usepackage{natbib}  
\usepackage{bm} 
\usepackage{isomath} 
\usepackage{braket}
\usepackage{booktabs}

\newcommand{\vect}[1]{\vectorsym{#1}} 

\definecolor{dblue}{RGB}{0,0,0.8}

\newcounter{lecture}
\begin{document}

\title{Gaussian phase sensitivity of boson-sampling-inspired strategies}

\author{Antonio A. Valido}
\email{a.valido@iff.csic.es}
\author{Juan Jos\'e Garc\'ia-Ripoll}

\affiliation{Instituto de F\'isica Fundamental IFF-CSIC, Calle Serrano 113b, 28006 Madrid, Spain}

\date{\today}


\begin{abstract}
In this work we study the phase sensitivity of generic linear interferometric schemes using Gaussian resources and measurements. Our formalism is based on the Fisher information. This allows us to separate the contributions of the measurement scheme, the experimental imperfections, and auxiliary systems. We demonstrate the strength of this formalism using a broad class of multimode Gaussian states that includes well-known results from single- and two-mode metrology scenarios. Using this, we prove that input coherent states or squeezing beat the non-classical states proposed in preceding boson-sampling-inspired phase-estimation schemes. We also develop a novel polychromatic interferometric protocol, demonstrating an enhanced sensitivity with respect to two-mode squeezed-vacuum states, for which the ideal homodyne detection is formally shown to be optimal.
\end{abstract}

\maketitle

\section{Introduction}

During the last decade a considerable attention has been devoted to figure out the optimal phase-estimation scheme for a (linear) photonic interferometer using Gaussian states and ideal quadrature measurements\ \cite{serafini20171, jiang20141,safranek20151,gao20141,monras20131} by means of the celebrated parameter estimation theory\ \cite{giovannetti20041,giovannetti20111,paris20091,toth20141,dowling20081,braun20181,pirandola20181,lee20021,demkowicz20151,sidhu20191,ataman20191,ataman20201,gessner20201}. In this context, most theoretical and experimental treatments have paid attention to the so-called quantum Fisher information (QFI), which dictates the ultimate phase sensitivity under generic measurements\ \cite{paris20091,toth20141,giovannetti20111,braunstein19941}. Interestingly, the optimal phase scheme able to attain the QFI could be determined via the Symmetric Logarithmic Derivative (SLD) \ \cite{safranek20151,jiang20141,monras20131,gao20141,safranek20181}, though it displays an intricate dependence on the desired parameter, which represents a major obstacle at the experimental level. For instance, the QFI has been intensively studied for noisy and lossy two-mode Mach-Zenhder interferometers (MZI) pumped by either a cross-product of coherent and squeezed-vacuum state\ \cite{gard20171,oh20171}, or a two-mode squeezed-vacuum state\ \cite{oh20171,steuernagel20041,bondurant19841,li20141}.
The phase sensitivity of the multimode scenario is less understood\ \cite{pinel20121,gagatsos20161,nichols20181, matsubara20191,oh20191,safranek20151,monras20131,safranek20161,oh20201}. Recent work suggests that, in the case of decoherence-free Gaussian resources with fixed average number of photons, the optimal Heisenberg limit is reached with a trivial squeezed-vacuum state\ \cite{matsubara20191}.

Most optimal Gaussian protocols relying on the QFI involve non-trivial technical challenges \cite{yonezawa20121,slussarenko20171,aasi20131}, such as engineering the passive transformation and generating high-intensity \cite{pinel20121}, or highly-squeezed light beams. In view of these problems, when working with experimental constraints we must focus on the Fisher Information (FI)\ \cite{takeoka20171,jarzyna20121} for the resources at hand---families of states, transformations and measurements---. This task has been completed in the single-mode MZI scenario\ \cite{oh20181}, and in some cases also for the multimode setup\ \cite{safranek20181,gessner20201,chaboyer20151}. More recently, the FI has permitted to show the Heisenberg scaling in multimode interferometric schemes endowed with single-mode squeezing resources and some preliminary classical knowledge about the parameter \cite{gramegna20201,gramegna20211}. The FI approach is a versatile treatment to study the phase resolution of general circuits, such as reconfigurable photonic circuits\ \cite{polino20191,aasi20131,yonezawa20121,paesani20171} with homodyne measurements. It is complementary to earlier and more difficult studies based on the quantum fidelity\ \cite{takeoka20171,pinel20131,pinel20121,oh20191}.

In this work we compute the Fisher information of arbitrary multimode interferometers working with Gaussian input states and Gaussian measurements (without necessarily assuming non-passive resources and pre- or post- processing treatments). This allows to envisage strategies retrieving a reasonable compromise between the phase sensitivity and the technical constraints upon the experimental resources. For instance, we tackle the question whether multimode setups can provide a metrological advantage, or beat the shot-noise limit (SNL) with less demanding components than single photon sources \cite{you20171, olson20171, motes20151,su20171}. Our formalism also quantifies deviations from the ideal limits provided by the QFI and gives insight on the interplay between the experimental resources and imperfections, such as losses and non-ideal detectors. On top of that, we analyze various interferometric schemes in terms of the resolution-energy trade-off, and introduce a new polychromatic protocol providing a multiplicative enhancement of the phase sensitivity with respect to the conventional strategy.

This work is divided in two parts. In Sect.\ \ref{sec:gaussian-sensitivity} we study the phase sensitivity of a linear, passive interferometer with Gaussian resources and measurements. In Sect.\ \ref{SPre} we introduce the phase-space formalism\ \cite{ferraro20051,serafini20171,braunstein20071,wedbrook20111}. In Sect.\ \ref{SPSTh} we review the connection between phase estimation and the Fisher information. Sect.\ \ref{SPSTh} uses the phase-space formalism to compute the Fisher information of an arbitrary linear and passive interferometer with Gaussian input states and measurements. Our results connects the FI to the QFI, identifying contributions from the ancillas, the interferometer and the measurement setup. The second part of this work illustrates how the FI formalism can be applied to various setups. Sect.\ \ref{SQM} discusses an input state formed by a single-mode squeezed vacuum and coherent states on $N-1$ auxiliary modes. Sect.\ \ref{PSNDR} introduces a new interferometric scheme with polychromatic light. Finally, in Sect.\ \ref{SDEFED} we show how to introduce losses and non-ideal detectors.

\section{Gaussian phase sensitivity}
\label{sec:gaussian-sensitivity}

\begin{figure}
\includegraphics[width=0.99\columnwidth]{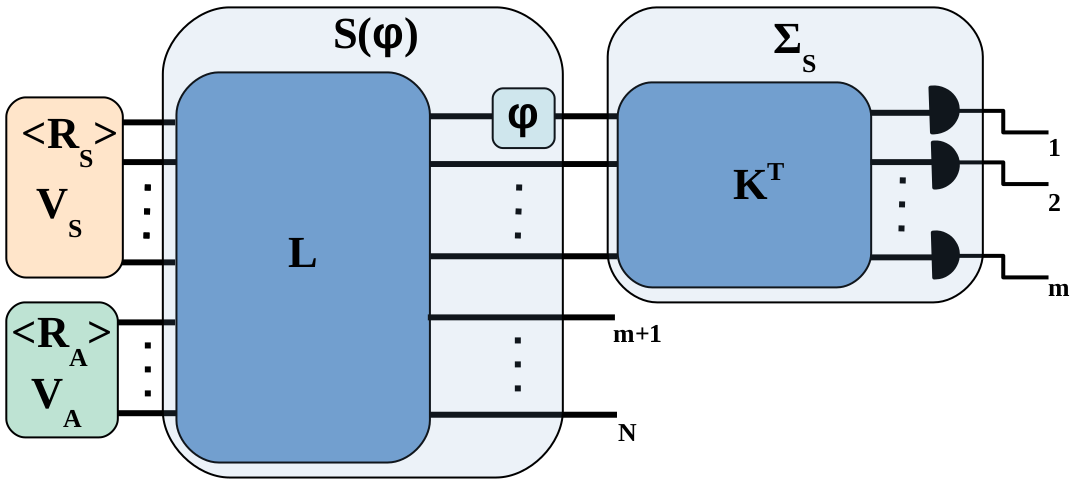}
\caption{(color online). Sketch of a generic $N$-mode Gaussian phase-estimation strategy consisting of a probe $m$-mode state, characterized by $\left\langle \vect R_{S}\right\rangle$ and $\vect V_{S}$ (orange), and an ancilla $(N-m)$-mode state, characterized by $\left\langle \vect R_{A}\right\rangle$ and $\vect V_{A}$ (green). Both probe and ancilary systems interact via the interferometer modeled by $\vect L$, whereafter the first probe mode undergoes the (single) phase rotation $\varphi$, such that the whole propagation is described by $\vect S(\varphi)$. The output modes of the probe system are finally assessed by a generic quadrature measurement determined by $\vect \Sigma_{S}$. \label{Fig1}} 
\end{figure}

Our work is devoted to studying an interferometric setup\ \cite{braun20181,lee20021,sidhu20191,demkowicz20151} such as the one in Fig.\ \ref{Fig1}. This phase-estimation scheme consists of: (i) an $N$-mode input state of light $\hat\rho$ with $m$ principal models and $N-m$ auxiliary degrees of freedom that will be eventually discarded\ \cite{boixo20071,fraisse20171,ataman20191,ataman20201}, (ii) an interferometer $L$ that prepares the state of light prior to interaction, (iii) the actual phase transformation $\varphi$ that we wish to detect, (iv) a final measurement stage that combines a linear transformation with local homodyne measurement on $m$ modes. This general scheme contains the MZI, and the vast majority of Gaussian (single) phase-estimation previously treated as particular instances\ \cite{monras20061,aspachs20091,matsubara20191,sparaciari20161,gard20171,oh20171,oh20181,pinel20121}. Our work focuses on a family of Gaussian input states\ \cite{safranek20151,monras20131} which we denote $\mathcal{G}(m,\bar{n}_{t})$ and which is a product $\hat{\rho}=\hat\rho_S \otimes \hat\rho_A,$ of a general Gaussian state  $\hat \rho_{A}$ for the ancilla, and an isothermal state $\hat \rho_{S}$ for the system---i.e. an $m$-mode Gaussian state with a uniform number of thermal photons $\bar{n}_{t}$ on each mode. On the output of the interferometer, we consider a general $m$-mode homodyne detection scheme, engineered by an interferometer $K$ and local homodyne measurements. Finally, without loss of generality, we assume that the measured phase $\varphi$ acts as a local rotation $\hat U=\exp(-i\varphi \hat{H})$ on one of the modes.

We will now proceed in three steps. The following section will introduce the phase space formalism, explaining how to express states $\hat\rho,$ interferometers, local phase rotations and measurements. Later in Sect.\ \ref{SPSTh}, we will introduce the Cramer-Rao bound and how the Fisher Information determines the maximum achievable sensitivity of our interferometer. Finally, Sect.\ \ref{SGLI} connects both formalisms, providing an explicit formula for the Fisher information and the phase sensitivity of our setup, expressed in terms of the first and second moments of the input state, the covariance matrix of the measurement and the passive transformations $L$ and $K.$

\subsection{Phase-space formalism}\label{SPre}

We model the light using two quadratures per mode, $\hat{q}_i$ and $\hat{p}_i,$ which satisfy the canonical commutation relations $\left[ \hat q_{i},\hat p_{j} \right] = i\left[\vect J_{N} \right]_{ij}.$ Here we have introduced the symplectic form \citep{wedbrook20111,ferraro20051} $\vect J_{N} = \bigoplus_{i=1}^{N}\vect J$, expressed in terms of $\left[\vect J\right]_{\alpha\beta}=\varepsilon_{\alpha\beta}$, the Levi-Civita symbol in two dimensions $\varepsilon_{\alpha\beta}.$ Any operator $\hat O$ is described in terms of the Weyl Symbol $W_{O}(\vect R)$ spanned by the phase-space basis $\vect R=(q_{1},p_{1},\dots,q_{N},p_{N})^{T}\in \mathbb{R}^{2N}$ with support in the real symplectic space $(\mathbb{R}^{2N}, \vect J_{N})$ \cite{wedbrook20111,braunstein20071}. Gaussian states are those whose density matrix has a Weyl symbol $W(\vect V,\left\langle \vect R\right\rangle)$ that is fully determined by the first moments $\left\langle \vect R\right\rangle \in \mathbb{R}^{2N}$ and the covariance (CV) matrix
\begin{equation}
\vect V=\frac{1}{2}\left\langle \left\lbrace \vect R,\vect R^{T} \right\rbrace \right\rangle\in\mathbb{R}^{2N\times 2N}.
\end{equation}
In particular, our input state $\hat \rho$ is a tensor-product Gaussian state with first-moment vector $\left\langle \vect R\right\rangle=(\left\langle \vect R_{S}\right\rangle,\left\langle \vect R_{A}\right\rangle)^{T}$ and CV matrix $\vect V=\vect V_{S}\oplus \vect V_{A}$. Moreover, for our iso-thermal states $\vect V_{S}=(2\bar{n}_{t}+1)\vect S' \vect I_{m} \vect S'^{T}$, where $\vect S'$ is an arbitrary $m$-mode (active or passive) symplectic transformation and $\vect I_{m}$ is the $2m\times2m$ identity matrix. This set of states satisfy a symplectic-like identity $\vect V_{S}\vect J_{m}\vect V_{S}= (2\bar{n}_{t}+1)^2\vect J_{m},$ implying a relation 
\begin{equation}
\vect V_{S}=(2\bar{n}_{t}+1)^2\vect J_{m}\vect V_{S}^{-1} \vect J_{m}^{T}.
\label{SPCVII}
\end{equation}
that will be extensively used throughout this work.

The initial state undergoes a multimode interferometer transformation, given by an $2N\times 2N$ orthogonal, symplectic matrix\ \cite{ferraro20051}. For convenience, we split this matrix into system and ancilla
\begin{eqnarray}
\vect L&=&\left( \begin{array}{cc}
\vect L_{S} &  \vect L_{SA}  \\
\vect L_{AS}& \vect L_{A}
\end{array}\right) , \label{GLIT}
\end{eqnarray}
where $\vect L_{SA}$ is a $2m\times2(N-m)$ isometry, while $\vect L_{S}$ is a non-orthogonal $2m\times 2m$ matrix which satisfies a symplectic-like relation $\vect L_{S}=\vect J_{m}^{T}\vect L_{S}\vect J_{m}$ as well.

After this preparation, the bosonic system suffers an unknown phase shift $\hat U(\varphi),$ generated by the operator
\begin{equation}
\hat H=\frac{1}{4}\big( \hat q_{1}^{2}+\hat p_{1}^{2}\big)  -\frac{1}{2}.
\label{PSRT}
\end{equation}
The phase shift $\hat{U}$ induces a rotation in phase space $\vect U_{N}(\varphi)=\vect U(\varphi)\oplus \vect I_{N-1},$ with $\vect U(\varphi)$ given by Eq.~\eqref{URR}.

The combined $N$-mode transformation $\vect S(\varphi)$ is composed of an $2m\times 2m$ non-orthogonal (non-singular) matrix $\vect S_{S}$ acting solely upon the probe system, and an isometry $\vect S_{SA}(\varphi)$ describing the interference between the system and the ancillas
\begin{eqnarray}
\vect S_{S}(\varphi) &=&\vect U_{m}(\varphi)\vect L_{S},\label{SMS1} \\
\vect S_{SA}(\varphi) &=&\vect U_{m}(\varphi)\vect L_{SA}.\nonumber
\end{eqnarray}
Since $\vect S(\varphi)$ describes a \textit{passive} interferometric evolution, the following relations must hold \cite{serafini20171}
\begin{eqnarray}
\vect S_{S}(\varphi)\vect S_{S}^{T}(\varphi)&=&\vect I_{m}-\vect S_{SA}(\varphi)\vect S_{SA}^{T}(\varphi), \label{SPST0} \\
\vect S_{S}(\varphi)&=&\vect J_{m}^{T}\vect S_{S}(\varphi)\vect J_{m}, \ \ \text{for} \ \varphi\in \mathbb{R},
\label{SPST}
\end{eqnarray}
which also shall be used in the subsequent derivation.  

The phase-estimation task is finally accomplished by performing a $m$-mode Gaussian measurement with outcome $\vect \lambda\in\mathbb{R}^{2m}$. Any $m$-mode general-dyne measurement acting as a Gaussian POVM $\hat \Pi_{\vect \lambda}$ is characterized by a $2m\times2m$ real, symmetric, and positive-definite CV matrix
\begin{eqnarray}
\vect \Sigma_{S}=\vect K\bigoplus_{j=1}^{m}\left( \begin{array}{cc}
r_{j} & 0\\
0 & \frac{1}{r_{j}}\\
\end{array}\right)\vect K^{T}.
\label{CVMGS}
\end{eqnarray}
$\vect K$ is an orthogonal symplectic transformation\ \cite{olivares20121,giedke20021,wedbrook20111} that may be implemented by the same or a different interferometer. The squeezing parameter $r_{j}=(1-\tau_{j})/\tau_{j}$ is a function by the transmissivity $\tau_{j}$ of the measurement setup\ \cite{genoni20141,kim19961}. It includes the limit of an ideal homodyne measurement in the $q$- or $p$- quadratures as $r_{j}\rightarrow 0$ and $r_{j}^{-1}\rightarrow 0,$ respectively. In the applications of Sect.\ \ref{SANAGI} we will consider the \textit{ideal $N$-mode homodyne detection scheme} consisting of identical local quadrature measurements, with $\vect K=\vect I_{N}$ and $r_{i}=r.$

We can compute the probability $p(\vect \lambda|\varphi)$ of obtaining a measurement outcome $\vect \lambda$ conditioned to a phase shift $\varphi.$ This is a Gaussian function characterized by the first-moment vector and the CV matrix \cite{olivares20121}
\begin{align}
\left\langle \vect \lambda(\varphi)\right\rangle &= \underbrace{\vect S_{S}(\varphi)\left\langle \vect R_{S}\right\rangle}_{\left\langle \vect \lambda_{S}(\varphi)\right\rangle} +\vect S_{SA}(\varphi)\left\langle \vect R_{A}\right\rangle,  \label{GGMS1} \\
\vect \sigma (\varphi) &=\underbrace{\vect \Sigma_{S}+\vect S_{S}(\varphi)\vect V_{S}\vect S_{S}^{T}(\varphi)}_{\vect \sigma_{S} (\varphi) }+\vect S_{SA}(\varphi)\vect V_{A}\vect S_{SA}^{T}(\varphi),
\label{GGMS2}
\end{align}
Note how the probe system statistics $\hat\rho_S$ only appears in $\left\langle \vect \lambda_{S}(\varphi)\right\rangle$ and $\vect \sigma_{S}.$
 
\subsection{Basics of phase estimation theory}\label{SPSTh}

Using the so-called maximum likelihood and Bayesian estimators\ \cite{pezze20081,braun20181,monras20061}, we can approximate an unknown phase shift $\varphi$ from a set of measurement outcomes $\vect \lambda.$ The precision of this method will be determined by the conditional probability $p(\vect \lambda|\varphi),$ as well as the estimator strategy $p_{est}(\tilde\varphi|\vect\lambda).$ The statistical inference process is described by the probability distribution\ \cite{braunstein19961,giovannetti20111,sidhu20191},
\begin{equation}
P(\tilde{\varphi}|\varphi)=\int\!\!\mathrm{d}^{2m}\vect \lambda\ p_{est}(\tilde\varphi|\vect \lambda)p(\vect \lambda|\varphi).
\end{equation}
The quality of the estimator, or its precision, is given by the mean square error\ \cite{giovannetti20111,braun20181,demkowicz20151}
\begin{equation}
\text{var}(\varphi)=\langle\!\langle(\tilde{\varphi}-\varphi)^2\rangle\!\rangle =\int \!\mathrm{d}\tilde{\varphi}\ (\tilde \varphi-\varphi)^2 P(\tilde{\varphi}|\varphi).
\end{equation}
In particular, for any unbiased estimator function with $\varphi=\langle\!\langle \tilde{\varphi}\rangle\!\rangle,$ the ultimate precision satisfies the Cram\'er-Rao bound (CRB) \cite{toth20141,paris20091,giovannetti20111,demkowicz20151},
\begin{equation}
\text{var}(\varphi)\geq \frac{1}{F(\varphi)},
\label{CCRB}
\end{equation}
where $F(\varphi)$ is the Fisher information of the probability distribution $p(\vect \lambda|\varphi)$ \cite{toth20141,paris20091} (see the Eq.~\eqref{CFI} in the App.\ \ref{app1}). In our Gaussian scenario, the FI can be explicitly computed [See App.\ \ref{app1}] as
\begin{equation}
F(\varphi)=\partial_{\varphi}\left\langle \vect \lambda^{T}\right\rangle\vect \sigma^{-1}\partial_{\varphi}\left\langle \vect \lambda\right\rangle-\frac{1}{2}\text{Tr}\Big(\partial_{\varphi}\vect \sigma^{-1}\partial_{\varphi}\vect \sigma \Big),\label{CFGEX}
\end{equation}
This includes earlier results for single-\ \cite{oh20181,pinel20121}, two\ \cite{sparaciari20161}, and multimode Gaussian metrology scenarios\ \cite{monras20131,sidhu20191}. {Notice that $F(\varphi)$ in the multiphase scenario is replaced by the Fisher information matrix \cite{safranek20181,pezze20171}, which could be expressed as Eq.(\ref{CFGEX}) up to minor changes: $\partial_{\varphi}$ should be substituted by a gradient in the vector parameter, while $\vect \sigma$ would be replaced by a larger matrix containing the parameter correlations due to the interferometic transformation. Since our framework relies on Eq.(\ref{CFGEX}), it could be equivalently adapted to the multiphase situation as well by following the procedure illustrated in Sec.\ref{SGLI}.}

The Fisher information is particularized for a measurement strategy. The Quantum Fisher Information (QFI) is an upper bound over all POVM strategies, Gaussian or not\ \cite{paris20091,toth20141,demkowicz20151,giovannetti20111},
\begin{equation}
\mathcal{F}=\text{max}_{\hat \Pi_{\vect \lambda}}[F(\varphi)].
\end{equation}
Since by definition $F(\varphi)\leq \mathcal{F},$ it follows that the ultimate sensitivity \cite{braunstein19941,braunstein19961} for any quantum or classical measurement strategy is dictated by the quantum Cram\'er-Rao bound (QCRB) \cite{paris20091,toth20141,giovannetti20111},
\begin{equation}
\mathrm{var}(\varphi) \geq \frac{1}{\mathcal{F}}.
\label{QCRB}
\end{equation}
As shown in App.\ \ref{app1}, there is a closed-form formula for the QFI when working with isothermal Gaussian input states and passive linear transformations\ \cite{pinel20131,pinel20121,jiang20141}
\begin{eqnarray}
\mathcal{F}&=&\frac{1}{(2\bar{n}_{t}+1)^2}\Bigg[\left\langle \vect R_{1}'\right\rangle^{T} \vect V_{1}'\left\langle \vect R_{1}'\right\rangle \label{QFIGGS} \\
& +&\frac{1}{1+(2\bar{n}_{t}+1)^{-2}}\Big(\text{Tr}\big(\vect V_{1}'\vect V_{1}'\big)-2(2\bar{n}_{t}+1)^2\Big)\Bigg].
\nonumber
\end{eqnarray}
The first-moment $\left\langle \vect R_{1}'\right\rangle$ and CV $\vect V_{1}'$ belong to the probe mode immediately before undergoing the phase-shift rotation. This expression is independent of $\varphi$ because of the  phase-shift generator $\hat{H}(\varphi)=\hat{H}$\ \cite{toth20141,braunstein19961}. 

While $\mathcal{F}$ dictates the ultimate sensitivity limit, this limit requires implementing a measurement strategy that can depend on the estimator $\varphi.$ This can involve elaborate transformations $\vect L$ and $\vect K$ and measurements of second or higher order moments of the quadrature. For this reason, unlike the vast majority of the previous works\ \cite{braun20181,paris20091,toth20141,giovannetti20111}, we will center on discussing the FI and the attainable limits of phase sensitivity under given experimental setups and constraints. As we will show below, this is not a severe restriction. We can compute the sensitivity of protocols that are experimentally feasible [cf. Fig.\ \ref{Fig1}]. We can also show that it saturates the QCRB around certain strategies, and we can manipulate\ \eqref{CFGEX} to separate the contributions of the probe, the ancillary Gaussian state, the interferometer and the Gaussian measurement scheme. 

\subsection{FI analysis}\label{SGLI}

We now present the main result which is the basis of the future analysis. Starting from the identity\ \eqref{CFGEX}, in App.\ref{app2} we decompose the FI as follows,
\begin{eqnarray}
&&F(\varphi)=\mathcal{F}_{S}+F_{\text{Anc}}(\varphi)+F_{\text{Int}}(\varphi)-F_{\text{Meas}}(\varphi) \label{IFSST} \\
&+&\frac{(2\bar{n}_{t}+1)^2}{1+(2\bar{n}_{t}+1)^{2}}\Bigg(\frac{\text{Tr}\big(\vect V_{1}'\vect V_{1}'\big)}{(2\bar{n}_{t}+1)^4}+2\Bigg) -\text{Tr}\Big(\vect P_{0}\vect L_{S}\vect L_{S}^{T}\Big),\nonumber
\end{eqnarray}
where $\mathcal{F}_{S}$ is the QFI associated to the $m$-mode probe system alone (which is obtained from \eqref{QFIGGS} in the absence of the ancilla), and $\vect P_{\varphi}=\vect U(\varphi) \oplus \vect 0_{m-1}$ is a $2m\times 2m$ projection matrix. The new functions $F_{\text{Anc}}(\varphi)$, $F_{\text{Meas}}(\varphi)$, and $F_{\text{Int}}(\varphi)$  respectively encode the influence of the input ancilla state, the $m$-mode quadrature measurement, and the interference between the ancilla and system. The measurement contribution reads 
\begin{eqnarray}
F_{\text{Meas}}(\varphi)&=& \left\langle \vect R_{S}^{T}\right\rangle\partial_{\varphi}\vect S_{S}^{T}\tilde{\vect \Sigma}_{S}\partial_{\varphi}\vect S_{S}\left\langle \vect R_{S}\right\rangle  \nonumber \\
&-&\frac{1}{2}\text{Tr}\Big(\partial_{\varphi}\tilde{\vect \Sigma}_{S}\partial_{\varphi}\Big(\vect S_{S} \vect V_{S}\vect S_{S}^{T} \Big)\Big).
\label{CFIMM}
\end{eqnarray}
The symmetric and symplectic  $2m\times 2m$ matrix
\begin{eqnarray}
\tilde{\vect \Sigma}_{S}&=&\big(\vect S_{S}^{T}\big)^{-1}\vect V_{S}^{-1}\vect S_{S}^{-1}\Big(\vect \Sigma^{-1}_{S} \nonumber \\
&+& \big(\vect S_{S}^{T}\big)^{-1}\vect V_{S}^{-1}\vect S_{S}^{-1}\Big)^{-1}\big(\vect S_{S}^{T}\big)^{-1}\vect V_{S}^{-1}\vect S_{S}^{-1},
\end{eqnarray}
is manifestly independent of the input ancilla state. The influence of the ancilla is fully contained in
\begin{eqnarray}
F_{\text{Anc}}(\varphi)&=&  2\left\langle \vect R_{S}^{T}\right\rangle\partial_{\varphi}\vect S_{S}^{T}\vect \sigma_{S}^{-1}\partial_{\varphi}\vect S_{SA}\left\langle \vect R_{A}\right\rangle  \label{IFINT1}\\
&+&\left\langle \vect R_{A}^{T}\right\rangle\partial_{\varphi}\vect S_{SA}^{T}\vect \sigma_{S}^{-1}\partial_{\varphi}\vect S_{SA}\left\langle \vect R_{A}\right\rangle  \nonumber\\
&-&\partial_{\varphi}\left\langle \vect \lambda^{T}\right\rangle\tilde{\vect V}_{A}\partial_{\varphi}\left\langle \vect \lambda\right\rangle   +\frac{1}{2}\text{Tr}\Big(\partial_{\varphi}\tilde{\vect V}_{A} \partial_{\varphi}\vect \sigma\Big) \nonumber \\
&-&\frac{1}{2} \text{Tr}\Big(\partial_{\varphi}\vect \sigma_{S}^{-1} \partial_{\varphi}\Big(\vect S_{SA}\vect V_{A}\vect S_{SA}^{T}\Big) \Big).\nonumber 
\end{eqnarray}
with
\begin{equation}
\tilde{\vect V}_{A}=\vect \sigma_{S}^{-1}\vect S_{SA}(\vect V_{A}^{-1}+\vect S_{SA}^{T}\vect \sigma_{S}^{-1}\vect S_{SA})^{-1}\vect S_{SA}^{T}\vect \sigma_{S}^{-1}.
\end{equation}
Similarly, the function $F_{\text{Int}}(\varphi)$ only depends on the input system state and system-ancilla interference $\vect S_{SA}$ (see Eq.~\eqref{IFINT2} in App.\ \ref{app2}). Note that both $F_{\text{Anc}}(\varphi)$ and $F_{\text{Int}}(\varphi)$ vanish when the system-ancilla interference cancel (which corresponds to the non-assisted scenario without ancilla system).

Let us give a brief overview about the derivation of the expression \eqref{IFSST}. From Eqs.\ \eqref{GGMS1} and \eqref{GGMS2} we may separate the contribution of the ancilla state. Indeed, using the so-called Woodbury identity (cf. Eq.~\eqref{WOODI} in App.\ \ref{app2}) \cite{petersen20121,bernstein20051},
\begin{eqnarray}
F(\varphi)&=&F_{S}(\varphi)+F_{\text{Anc}}(\varphi),
\label{CFI2}
\end{eqnarray}
we can separate the contribution $F_{S}(\varphi)$ from the first-moment $\left\langle \vect \lambda_{S}(\varphi)\right\rangle$ and CV $\vect \sigma_{S} (\varphi)$. Collecting all remaining terms that depend on the auxiliary system, $F_\text{Anc}$ adopts the form in Eq.~\eqref{IFINT1}. This procedure may be repeated, using the symplectic-like identities\ \eqref{SPCVII} and\ \eqref{SPST}, to separate from $F_{S}(\varphi)$ the interference $F_{\text{Int}}(\varphi)$ and measurement terms $F_{\text{Meas}}(\varphi),$ as shown in Eq.\ \eqref{CFIMS}. Finally, using property\ \eqref{SMS1D}, one may group the remaining terms into the QFI $\mathcal{F}_{S}$ (see Eq.\ (\ref{IFSS1})) plus additional corrections, as show in Eq.\ \eqref{IFSST}.

The closed-form expression \eqref{IFSST} is valid for any probe isothermal Gaussian state $W(\vect V_{S},\left\langle \vect R_{S}\right\rangle) \in \mathcal{G}(m,\bar{n}_{t})$, and for {single-phase interferometric schemes, Gaussian ancilla states as well as measurements}. For the sake of clarity, we pay special attention to input coherent resources and a particular subset of Quantum Uniform Multimode Interferometers (QUMI) recently studied in the context of boson-sampling inspired phase-estimation strategies \cite{you20171,olson20171,motes20151}. These are further discussed in the following section.

\subsubsection{Coherent ancilla state and QUMI}\label{SCAQUMI}

In the simple scenario in which the ancillary system are coherent states $\vect V_{A}=\vect I_{N-m},$ that interfere with the system through a simple QUMI device---cf. the linear transformation $\vect L$ from Eq.~\eqref{DLQUMI}---, the FI simplifies to
\begin{eqnarray}
F(\varphi)&=&\tilde{F}_{S}(\varphi)+ \left\langle \vect R_{A}^{T}\right\rangle\partial_{\varphi}\vect S_{SA}^{T}\vect \sigma^{-1}\partial_{\varphi}\vect S_{SA}\left\langle \vect R_{A}\right\rangle \nonumber \\
&+&2\left\langle \vect R_{S}^{T}\right\rangle\partial_{\varphi}\vect S_{S}^{T}\vect \sigma^{-1}\partial_{\varphi}\vect S_{SA}\left\langle \vect R_{A}\right\rangle,
\label{ASCS}
\end{eqnarray}
with 
\begin{equation}
\vect \sigma=\underbrace{\big(\vect \Sigma_{S}+\vect I_{m}\big)}_{\tilde{\vect \Sigma}_{S}}+\vect S_{S}\underbrace{\big(\vect V_{S}-\vect I_{m}\big)}_{\tilde{\vect V}_{S}}\vect S_{S}^{T},
\end{equation}
The term $\tilde{F}_{S}(\varphi)$ is the FI of a phase estimation scheme that uses a Gaussian input state with first-moment $\left\langle \vect R_{S}\right\rangle$ and CV  $\tilde{\vect V}_{S}$, along with a Gaussian measurement with a white background noise $\tilde{\vect \Sigma}_{S}$.

 For a state with homogeneous input intensity $\bar{n}_{c}$---i.e. $\left\langle \vect R_{i}\right\rangle=(\sqrt{2\bar{n}_{c}},\sqrt{2\bar{n}_{c}})$ for $i\in\left[ 1,N\right] $---, it turns out that $\vect S_{SA}\left\langle \vect R_{A}\right\rangle=(N-m)/m\vect S_{S}\left\langle \vect R_{S}\right\rangle$ (which follows from the transformation \eqref{DLQUMI}). This means that the ancillary terms in Eq.\ \eqref{ASCS} are positive and increase the FI---provided $\tilde{\vect V}_{S}$ is a positive semi-definite matrix---. The auxiliary coherent state improves the phase sensitivity, although it introduces some background noise in the measurement outcome.

This result simplifies in the ideal homodyne detection in which the system, not only the ancilla, is in a coherent state $\vect V_{S}=\vect I_{m}.$ Furthermore, the studied subset of QUMI schemes has the property that it maps a superposition of all input modes to the single mode that experiences the phase transformation\ \cite{olson20171}. In our notation, this mode has label 1 [cf. Fig.~\ref{Fig1}], so that
\begin{equation}
    \bar{n}_{1} =\frac{1}{4} \Bigg(\left\langle\sum_{i=1}^N\frac{1}{\sqrt{N}} \  \vect R_{i}\right\rangle\Bigg)^2=\frac{N}{2}\big(\sqrt{2\bar{n}_{c}}\big)^2=N\bar{n}_{c}.\label{PSACQ1}
\end{equation}
The ancilla proves beneficial still increases phase sensitivity, since $\tilde{F}_{S}(\varphi)$ becomes $4m\bar{n}_{c}$ in the optimal operating points $\varphi_{\text{opt}}=\mp\pi/4$ (see the discussion around Eq.\ (\ref{FIAS1}) in Sect.\ \ref{SQM}). Using Eq.\ \eqref{ASCS} we obtain the phase sensitivity for the QUMI assisted coherent setup 
\begin{equation}
(\delta \varphi)^2=\frac{1}{4\bar{n}_{c}N},
\label{PSACQ}
\end{equation}
in agreement with previous results for single-parameter schemes with an external phase reference \cite{ataman20191,ataman20201}. This coincides with the phase sensitivity of a single-mode coherent state with input intensity $\bar{n}_{c}N$. Moreover, Eq.\ \eqref{PSACQ} shows that input coherent resources outperform earlier  QUMI-based phase-estimation using single-photon states\ \cite{you20171, olson20171, motes20151,su20171}, for any size of the interferometer. For more general assisted phase-estimation schemes, it is less clear to see the influence owing to the interferometer $F_{\text{Int}}(\varphi)$ and ancilla $F_{\text{Anc}}(\varphi)$ contributions at first sight, instead they deserve a more profound analysis that is beyond the scope of the present treatment \cite{boixo20071,fraisse20171,ataman20191,ataman20201}. 

\section{Application: $N$-mode homodyne detection without ancilla system}\label{SANAGI}

We will now compare the strength of our treatment with earlier Gaussian phase-estimation analysis\ \citep{gard20171,pinel20121,matsubara20191,oh20171,oh20181}, using no auxiliary modes $(N=m)$, Gaussian pure input states $(\bar{n}_{t}=0)$ and an ideal $N$-mode homodyne measurement. Since there are no ancillas, we can eliminate the subscript $S$, $\vect S_{S}(\varphi)\rightarrow\vect S(\varphi)$, $\left\langle \vect R_{S}^{T}\right\rangle\rightarrow\left\langle \vect R^{T}\right\rangle$,$\vect V_{S}\rightarrow\vect V$, and $\mathcal{F}_{S}\rightarrow\mathcal{F}$. Both the FI\ \eqref{IFSST}
\begin{equation}
F(\varphi)=\mathcal{F}-F_{\text{Meas}}(\varphi) +\frac{1}{2}\Big(\text{Tr}\Big(\vect V_{1}'\vect V_{1}'\Big)-2\Big), \label{IFSSTPW}
\end{equation}
and the contribution from the measurement radically simplify [cf. Eq.~\eqref{FMEAS} in App.\ref{app2}], 
\begin{eqnarray}
F_{\text{Meas}}(\varphi)&=&\left\langle \vect R^{T}\right\rangle\vect L^{T}\vect P_{\varphi}^{T}\vect S\vect V\vect S^{T}\tilde{\vect \Sigma}\vect S\vect V\vect S^{T}\vect P_{\varphi}\vect L\left\langle \vect R\right\rangle \nonumber \\
&-&\frac{1}{2}\text{Tr}\Bigg(\Big(\tilde{\vect \Sigma}\partial_{\varphi}\Big(\vect S\vect V\vect S^{T} \Big)\Big)^2\Bigg)  \nonumber \\
&+&\text{Tr}\Bigg(\tilde{\vect \Sigma}\vect S\vect V\vect S^{T}\Big(\partial_{\varphi}\Big(\vect J_{N}\vect S\vect V\vect S^{T} \Big)\Big)^2\Bigg).
\label{FICHS}
\end{eqnarray}
The matrix $\vect \Sigma$ that characterizes the Gaussian measurement appears in the new matrix $\tilde{\vect \Sigma}= \big(\vect \Sigma+\vect S\vect V\vect S^{T}\big)^{-1}$. For an ideal homodyne detection in either position or momentum quadrature, $\vect \Sigma$ effectively becomes a projection matrix, $\vect \pi^{(x)}=\text{diag}(1,0,1,0,\cdots,1,0)$ or $\vect \pi^{(p)}=\text{diag}(0,1,0,1,\cdots,0,1)$ respectively. In this case $\tilde{\vect \Sigma}$ must be understood as a Moore-Penrose (MP) inverse\ \cite{giedke20021,wedbrook20111,eisert20021}, computed as follows\ \cite{petersen20121}
\begin{equation}
\tilde{\vect \Sigma}^{(x/p)}=\Big(\vect \pi^{(x/p)}\vect S\vect V\vect S^{T}\vect \pi^{(x/p)}\Big)^{\text{MP}}.
\label{SMDAM}
\end{equation}
Note also that the CV matrix of the chosen measurement remains invariant $\vect \Sigma=\vect K\vect \Sigma\vect K^{T}$ under any interferometric transformation $\vect K,$ rendering this choice irrelevant \footnote{Physically, the tensorial product of identical Gaussian states (up to an arbitrary displacement) with diagonal CV matrix remains invariant under a beam splitter transformation (which induces no phase shift between transmitted and reflected modes) for any transmission coefficient \cite{springer20091,kim20021}.}. 

\subsection{Coherent and one-mode squeezed resources}\label{SQM}

Let us analyze a collection of independent single-mode squeezed states, characterized by an arbitrary displacement $\left\langle \vect R\right\rangle\in \mathbb{R}^{2N}$ and the CV matrix,
\begin{equation}
\vect V=\vect V_1(s_1) \bigoplus\vect V_{N-1}(s_{2}),
\label{SMSSM}
\end{equation}
with
\begin{equation}
\vect V_{l}(s)=\bigoplus_{i=1}^{l}\left(\begin{array}{cc}
s & 0 \\
0 & \frac{1}{s}
\end{array}\right),
\nonumber
\end{equation}
The squeezing of the first and of the remaining $N-1$ modes are given by the parameters $s_{1},s_{2}\in \mathbb{R}^{+}.$ When $N=2,$ this state reduces to the vast majority of non-entangled Gaussian states previously studied: when $s_1=s_2=s,$ it maps to studies of single-mode squeezed states\ \cite{gard20171, oh20181,monras20061,olivares20091,safranek20151,pasquale20151,gaiba20091,lang20141,sparaciari20151,sparaciari20161}, when $s_2=1$ we have the squeezed mode combined with a coherent state from Refs.\ \cite{ataman20191,ataman20201,matsubara20191,oh20171,sparaciari20151,sparaciari20161,caves19811,pezze20081,pinel20121}, and for $s_1=s_2=1$ we recover the coherent phase-estimation scenario and the SNL scaling.

The QFI depends of the average number of photons on the mode that undergoes the phase rotation  (concretely, the QFI for pure probe states is proportional to the variance of the photon number  \cite{toth20141,olson20171,monras20061}). We can therefore concentrate on the previously introduced QUMI setup, which maximizes this intensity. For this we find
\begin{equation}
\vect S_{\text{QUMI}}\vect V\vect S_{\text{QUMI}}^{T}=\left(\begin{array}{cc}
\vect \Omega_{N} (\varphi,s_{1},s_{2}) & \vect 0 \\
\vect 0 & \vect V_{N-2}(s_{2})
\end{array}\right),
\label{CVMSMI}
\end{equation}
where $\vect \Omega_{N}$ is a $4\times 4$ real, symmetric matrix whose representation does not affect the discussion [cf. App.\ \ref{app3} and Eq.~\eqref{CVMSMII}]. Note how the size of \eqref{CVMSMI} grows as $2(N_{s_{1}}+1)\times 2(N_{s_{1}}+1)$ for a large number $N_{s_{1}}$ of states with squeezing $s_{1}$.  Replacing \eqref{CVMSMI} in \eqref{SMDAM}, we further obtain
\begin{equation}
\tilde{\vect \Sigma}^{(x/p)}= \vect A^{(x/p)}\oplus \text{diag}(1,0,\cdots,1,0).
\label{SXSQCS}
\end{equation}
Here, $\vect A^{(x/p)}$ is a $4\times 4$ real, matrix given by Eqs.\ \eqref{MSAX} and \eqref{MSAP}. By paying attention to \eqref{CVMSMI}, it is clear to see that the matrices within the trace in the expression \eqref{FICHS} effectively play the role of a projection operator in the phase space supporting the mode undergoing the rotation, i.e. $\partial_{\varphi}(\vect S_{\text{QUMI}}\vect V\vect S_{\text{QUMI}}^{T})=\partial_{\varphi}\vect \Omega_{N}\oplus \vect 0_{N-1}$. Having evaluated the quantities \eqref{CVMSMI} and \eqref{SXSQCS}, after substitution in \eqref{FICHS} one obtains the FI 
\begin{eqnarray}
F^{(x/p)}(\varphi)&=&\mathcal{F}-\frac{1}{2N^2}\left(a_{N}^2(s_{2},s_{1})+\frac{a_{N}^2(s_{1},s_{2})}{(s_{1}s_{2})^2}-2 N^2  \right) \nonumber \\
&+&f_{N}^{(x/p)}(\sin^2\varphi,s_{1},s_{2})\nonumber \\
&-&\left\langle \vect R_{1}'\right\rangle \vect W_{N}^{(x/p)}(\varphi, s_{1},s_{2})\left\langle \vect R_{1}'\right\rangle,
\label{FICHSXS}
\end{eqnarray}
where we have introduced $a_{N}(s_{1},s_{2})=(N-1)s_{1}+s_{2}$, two auxiliary functions $f_N^{(x)}$, $f_{N}^{(p)}$  [cf. Eqs.\ \eqref{FFXCS} and \eqref{FFPCS}] and a real symmetric matrix $\vect W_{N}^{(x/p)}\in\mathbb{R}^{2\times2}$ [cf. Eqs.\ \eqref{MWCS1}-\eqref{MWCS3}].

\begin{figure*}
\includegraphics[scale=0.34]{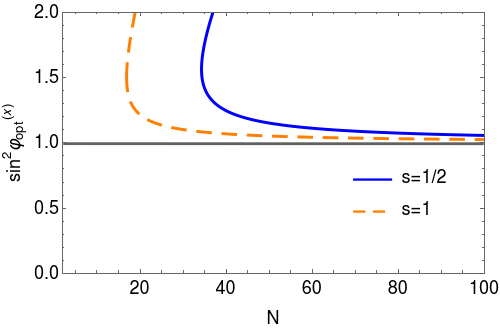}
\includegraphics[scale=0.33]{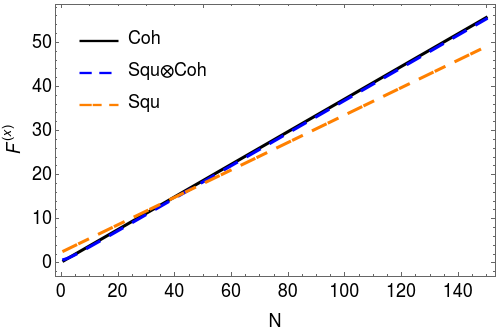} 
\includegraphics[scale=0.33]{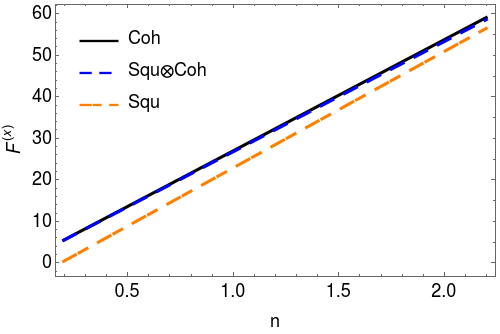} 
\caption{(color online). (Left) The roots of the polynomial \eqref{QCRBC} as functions of the interferometer size $N$, where the horizontal black line represents the unit value $\sin^2\varphi_{0}^{(x)}=1$. Notice that the solid blue and dashed orange lines do not regard injective functions because the polynomial has two distinct real roots. (Central) The FI associated to the position quadrature measurement as a function of the interferometer size and for distinct input probe states: the black-solid, blue-dashed, and orange-dot-dashed lines correspond to the tensor product of coherent (i.e. $s_{1}=s_{2}=1$), one-mode squeezed$\otimes$coherent (i.e. $s_{1}=e^{2s'}$ and $s_{2}=1$), and single-mode squeezed (i.e. $s_{1}=s_{2}=e^{2s'}$) states, respectively. For a fair comparison, we have fixed the input mean photon number per mode to an identical value for all input states, i.e. $\bar{n}\approx 1.38 $, as well as we have chosen the unknown phase shift $\varphi=\pi/3$ and the squeezing parameter $s'=1/2$. (Right) Similarly, the FI as a function of the mean photon number per mode for a fixed squeezing parameter. We have taken the same values for the rest of parameters.\label{Fig2}} 
\end{figure*}

The optimal phase-estimation strategy for a given $\varphi$ must saturate the QCRB \eqref{QCRB}. In that case the last three terms in Eq.\ \eqref{FICHSXS} cancel each other and $F=\mathcal{F}.$ To illustrate, let us consider an input beam with the same coherent low-intensity $\bar{n}_{c}\ll N$ in each mode (i.e. $\left\langle \vect R_{i}\right\rangle=(\sqrt{2\bar{n}_{c}},\sqrt{2\bar{n}_{c}})$). For this choice, the QFI takes the form 
\begin{eqnarray}
\mathcal{F}&=&2\bar{n}_{c}N\frac{1+s_{2}^2}{s_{2}}  \nonumber \\
&+&\frac{1}{2} \left(4 \bar{n}_{c} \left(s_{1}+\frac{1}{s_{1}}-s_{2}-\frac{1}{s_{2}}\right)+s_{2}^2+\frac{1}{s_{2}^2}-2\right) \nonumber \\
&+&\mathcal{O}\big(N^{-1}\big).
\label{QFIAS1}
\end{eqnarray}
The first term, proportional to $N,$ reproduces the QFI of a coherent input state, and the second and third term cancel precisely for that type of input $s_{1}=s_{2}=1$. Moreover, we may expand Eq.\ \eqref{FICHSXS} in the limit of large interferometers $1\ll N$ with finite energy $1/N\ll s_{1/2} \ll N$
\begin{align}
&F^{(x/p)}(\varphi)= \frac{4\bar{n}_{c} N s_{2}(1\mp\sin (2 \varphi ))}{1+s_{2}^2\mp(1-s_{2})\cos(2\varphi)} \label{FIAS1}\\
&+ \frac{2s_{1}^{-1}(s_{1}(1\mp  1)+s_{2}(1\pm 1))}{(1+s_{2}^2\mp(1-s_{2}^2)\cos(2\varphi))^2}\Bigg(\frac{s_{1}(1-s_{2}^2)^2\sin^2(2\varphi)}{s_{1}(1\mp  1)+s_{2}(1\pm 1)} \nonumber \\
&+\bar{n}_{c}(s_{1}-s_{2})\big(1\mp\sin(2\varphi))(1-s_{2}^{2}\mp (1+s_{2}^2)\cos(2\varphi)\big)\Bigg)\nonumber \\
&+\mathcal{O}\big(N^{-1}\big).
\nonumber
\end{align}
Here the signs $\mp$ correspond to the use of position and momentum quadratures, respectively.

Inspecting Eq.\ \eqref{FIAS1} reveals that the leading sensitivity in $F^{(x/p)}$ resembles the QFI of coherent states\ \eqref{QFIAS1} around the optimal working points $\varphi_{\text{opt}}^{(x/p)}=\mp\pi/4$, i.e.
\begin{equation}
    F^{(x/p)}(\mp\pi/4)= 4\bar{n}_{c} N+\mathcal{O}\big((s_1-1)N^{-1},(s_2-1)N^{-1}\big).\label{FIAS1a}
\end{equation} 
In other words, the combination of ideal homodyne detection and squeezed input resources with $s_{1}\neq s_2$ can approach the QCRB for large interferometers, though it never saturates the QFI except in the strict coherent limit ($s_1\rightarrow 1$, $s_2\rightarrow 1$) in agreement with previous results for two-mode \cite{gard20171,oh20171} and multimode interferometric schemes \cite{pinel20121}. On the other hand, if we use displaced single-mode squeezed states $s_{1}=s_{2}=s$, the ideal homodyne detection is never an optimal measurement scheme: the three last terms in the right-hand side of Eq.~\eqref{FICHSXS} never cancel each other if $0<|\!\left\langle \vect R\right\rangle\!|$ and $0< s $. 

For input resources with vanishing displacement, the optimal working point $\varphi_{\text{opt}}^{(x/p)}$ is found by solving second order equations in the variable $y=\sin^2\varphi\rightarrow$ [cf. Eqs.\ \eqref{FFXCS} and \eqref{FFPCS}]. For instance, the condition to saturate the QCRB for a position quadrature measurement is
\begin{equation}
y^2 \alpha_{N}^{(x)}(s_{1},s_{2}) + y \beta_{N}^{(x)}(s_{1},s_{2})+\delta_{N}^{(x)}(s_{1},s_{2})=0,
\label{QCRBC}
\end{equation}
with coefficients $\alpha_{N}^{(x)}$, $\beta_{N}^{(x)}$, and $\delta_{N}^{(x)}$ given by Eqs.\ \eqref{PSCCS1}-\eqref{PSCCS3}.In the particular situation of an homogeneous squeezing $s_{1}=s_{2}= e^{-2s'}$ with $s'\in \mathbb{R}$, the QCRB is saturated for
\begin{equation}
\cos\Big(2\varphi^{(x/p)}_{\text{opt}}\Big)=\pm\tanh (2s'),
\label{QCRBSSND}
\end{equation}
 and Eq.(\ref{FICHSXS}) returns
\begin{equation}
   F^{(x/p)}\Big(\varphi^{(x/p)}_{\text{opt}}\Big)=8\bar{n}_{s'}(\bar{n}_{s'}+1),
   \label{QCRBSSNDF}
\end{equation}
with $\bar{n}_{s'}$ denoting the input average photon number per mode (i.e. $\bar{n}_{s'}=\sinh^2 s'$). Notice that this result holds for any choice of the interferometric transformation [66]. This coincides with the single-mode Gaussian state results, found with alternative methods based on the fidelity\ \cite{monras20061,olivares20091,aspachs20091,safranek20151} or the SLD\ \cite{oh20181}.

The subsidiary condition\ \eqref{QCRBC} proves that a quadrature detection  in position (or equivalently, in momentum) is no longer optimal for a tensor product of zero-displacement states with $s_{1}=s$ and $s_{2}=1$. We see this in the left panel of Fig.\ \ref{Fig2}, which shows the real roots of \eqref{QCRBC} as a function of $N$ for two fixed squeezing values $s$. Note how these roots are always above or at most equal to $1$ for all problem sizes $N$. Consequently, there is no value $\varphi_{\text{opt}}^{(x)}$ for which the QCRB is saturated except for the single-mode Gaussian metrology setup $N=1$. This observation is also confirmed by computing the roots in the limits of extreme squeezing in either position or momentum, i.e. $\lim_{s\rightarrow \infty}\sin^2\varphi_{\text{opt}}^{(x)}=N^2/(2N-1)$. 
All these findings are consistent with results obtained in the single- and two- mode phase-estimation analysis based on the SLD\ \cite{aspachs20091,gard20171,oh20181,oh20191, pinel20121}: for displaced
squeezed states the SLD is a quadratic operator in terms of the quadrature operators \citep{serafini20171,monras20131,jiang20141} (which means that the optimal measurement scheme is non-Gaussian), however it becomes linear when dealing with either coherent or squeezed-vacuum resources \cite{oh20181}.

The FI is also plotted in Fig.\ \ref{Fig2} for the purpose of comparison. The central panel depicts this in terms of the interferometer size $N$ for a given homogeneous intensity $\bar{n}$, while the right panel illustrates it as a function of $\bar{n}$ at a fixed interferometer size $N=100$. In summary, these figures outline the main conclusion from Eq.~\eqref{FIAS1}: that is, none of the non-entangled Gaussian states along with the QUMI architecture provide a better scaling than the SNL (see, the black solid line) in the finite energetic regime and for large interferometer sizes. In other words, our analysis indicates that QUMI-based phase-estimation strategies provide no real advantage w.r.t. the resolution-energy trade-off\ \cite{olson20171,motes20151,su20171,you20171}.

\subsection{Two-mode squeezed resources and polychromatic phase generator}\label{PSNDR}

\begin{figure*}
\includegraphics[scale=0.29]{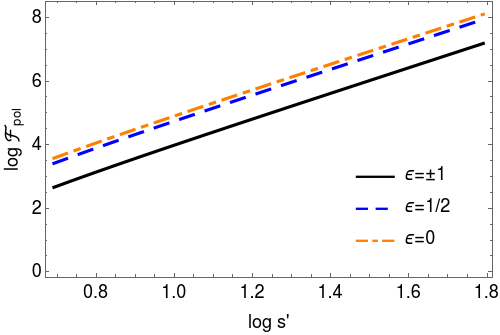}
\includegraphics[scale=0.29]{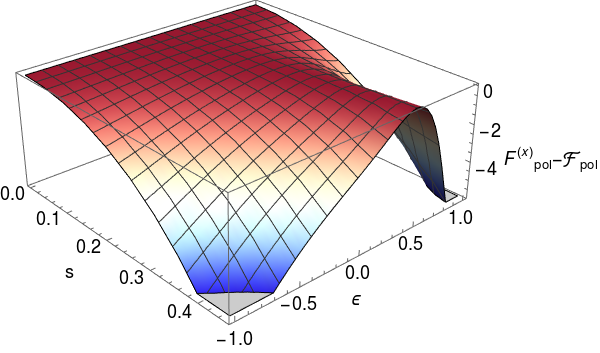}
\includegraphics[scale=0.27]{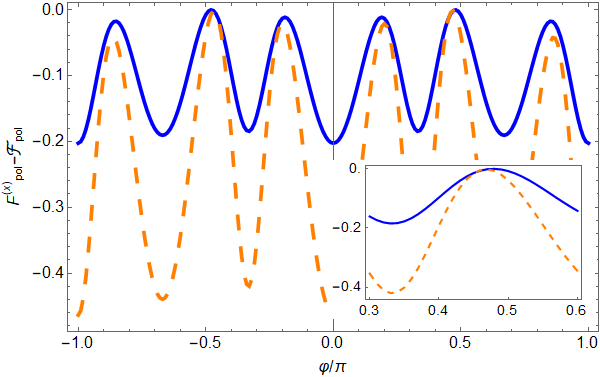}
\caption{(color online). (Left) Log plot of the polychromatic QFI as a function of $s'$ and for distinct values of the modulation parameter $\epsilon$, shown in the large squeezing regime. We have fixed the transmissivity $\tau=0$. (Central) Three-dimensional plot of the deviation associated to the position quadrature measurement for input two-mode squeezed-vacuum states and fixed value of the phase shift $\varphi=\pi/4$. (Right) Similarly, the deviation of the FI as a function of $\varphi$ for a fixed modulation frequency $\epsilon=1/2$ and two given values of the squeezing parameter: the blue and dashed orange lines correspond to $s=0.1$ and $s=0.15$, respectively. In the central and left panels, the transmissivity was chosen $\tau=1/2$. \label{Fig3}} 
\end{figure*}

Let us now study a metrology setup using two-mode non-degenerate squeezed states as resources. These states have been shown to overcome the SNL in estimation errors or phase sensitivities when using homodyne\ \cite{bondurant19841,hang20181}, intensity\ \cite{yurke19861}, or parity measurements\ \cite{anisimov20101,plick20101,birrittella20151}. The input state will be described by the first-moment vector $\left\langle \vect R\right\rangle\in \mathbb{R}^4$ and a CV matrix\ \cite{demkowicz20151,steuernagel20041,kim20021},
\begin{eqnarray}
\vect V=\left( \begin{array}{cccc}
\cosh 2s' & 0 & \sinh 2s' & 0 \\
0 & \cosh 2s' & 0 & -\sinh 2s' \\
\sinh 2s' & 0 & \cosh 2s' & 0 \\
0& -\sinh 2s' & 0 & \cosh 2s'  
\end{array}\right),
\label{CVMTMS}
\end{eqnarray}
that depends on the squeezing parameter $s'\in\mathbb{R}^{+}.$ Notice that one obtains the result related to the coherent resources discussed previously for the choice $s'=0$.

These states can be generated using the well established procedure of pumping a \textit{non-degenerate optimal parametric amplifier} (OPA) with a strong coherent beam, say at frequency $2\omega_{0}.$ These input photons are split into highly correlated pairs that conserve the total energy $\omega_{1,2}=\omega_{0}\pm \Omega.$ Here $\Omega< \omega_{0}$ is a small modulation frequency that renders the photons distinguishable\ \cite{braunstein20071,anisimov20101,plick20101}. To the best of our knowledge, there is no previous treatment that studied the influence of such modulation from the metrological point of view (for instance, see Refs.\ \cite{pasquale20151,you20191,nichols20181}). 

We will now go beyond previous phase-estimation analysis, addressing a \textit{polychromatic metrology} scenario in which each port of the two-mode interferometric setup is fed with beams at two different frequencies. We label those modes with the annihilation operators $\hat a_{\omega_{1}}$ and $\hat a_{\omega_{2}}$, and consider that different frequencies may experience a different single-mode phase-shift, generated by 
\begin{equation}
\hat{H}_{\text{pol}}(\epsilon)=(1+\epsilon)\hat{n}_{\omega_{1}}+(1-\epsilon)\hat{n}_{\omega_{2}}.
\label{HPC}
\end{equation}
The parameter $-1\leq \epsilon\leq 1$ can be regarded as a frequency-dependent index of refraction or optical path, and $\hat{n}_{\omega_{i}}=\hat{a}^{\dagger}_{\omega_i}\hat{a}_{\omega_i}.$ As the total average energy $\left\langle \hat{H}_{\text{pol}}(\epsilon)\right\rangle $ remains constant for distinct $\epsilon$, we can compare the resolution-energy trade-off retrieved by polychromatic Gaussian phase-estimation scenarios. The choice \eqref{HPC} returns an extension of the phase-shift generator that is 
\begin{equation}
\vect U_{\text{pol}}(\varphi,\epsilon)=\vect U((1+\epsilon)\varphi)\oplus\vect U((1-\epsilon)\varphi),
\label{PIGP}
\end{equation}
which reduces to the conventional generator \eqref{URR} for the choices $\epsilon=\pm 1$ \cite{hang20181}. Further, we shall consider that the transformations $\vect L$ represents a beam splitter with transmissivity $\tau$.

Returning to the phase space formalism, the polychromatic QFI can be expressed as follows,
\begin{eqnarray}
\mathcal{F}_{\text{pol}}(\epsilon)&=&(1+\epsilon)^2\mathcal{F}_{1}+(1-\epsilon)^2\mathcal{F}_{2}  \label{QFISPSII} \\
&+&4\big(1-\epsilon^2\big)\Big(\text{Tr}\Big(\vect V_{12}'\vect V_{12}'\Big)+2\left\langle \vect R_{1}'^T\right\rangle\vect V_{12}'\left\langle \vect R_{2}'\right\rangle\Big),
\nonumber
\end{eqnarray}
where $\mathcal{F}_{1}=\mathcal{F}_{2}=(1+4\tau(1-\tau))\bar{n}_{s'}(\bar{n}_{s'}+2)$ with $\bar{n}_{s'}$ denoting the total average number of photons, whereas $\left\langle \vect R_{i}'\right\rangle=(\vect L \left\langle \vect R\right\rangle)_{i}$ and $\vect V_{12}'=\text{diag}((1-2\tau)\sinh 2s',-(1-2\tau) \sinh 2s')$.

The left panel of Fig.\ \ref{Fig3} shows a log-log plot of the QFI in terms of the squeezing parameter, for distinct choices of the frequency modulation and vanishing input displacement. The polychromatic QFI is larger than the monochromatic counterpart for sufficient high squeezing $(1\ll s'),$ and the highest sensitivity is obtained for $\epsilon=0$. Interestingly, the sensitivity grows with the squeezing with an identical power for all values of $\epsilon,$ so that the polychromatic QFI may be approximately expressed as $\mathcal{F}_{\text{pol}}(\epsilon)\approx c(\epsilon,\tau)\mathcal{F}_{1}$ with $c$ being a multiplicative enhancement independent of $s'$. This factor is found to take values $2\lessapprox c\lessapprox 10$ for the available modulation frequencies and transmissivity, implying that a polychromatic setup can provide a significant improvement of the resolution-energy trade-off compared to the monochromatic MZI, e.g. $\mathcal{F}_{\text{pol}}\thicksim 10\bar{n}_{s'}^2$ for $1\ll s'$, $\epsilon=0$ and $\tau=0$.

The treatment about the FI presented in Sect.\ \ref{SGLI} holds for very general phase generators beyond \eqref{PSRT} and can be adapted to the polychromatic scenario. Going back to the general expression\ \eqref{CFIMS} and replacing the phase-shift generator\ \eqref{PIGP}, we obtain a closed-form expression of the FI associated to the polychromatic strategy by following a similar procedure as to compute the expression \eqref{IFSSTPW} discussed in Sect.\ \ref{SGLI}. The result is
\begin{eqnarray}
F_{\text{pol}}(\varphi,\epsilon)&=&\mathcal{F}_{\text{pol}}(\epsilon)-F_{\text{Meas}}(\varphi,\epsilon)-2(1+\epsilon^2)  \label{FICHSPpol}\\
&+&\frac{1}{2}\text{Tr}\Big((1+\epsilon)^2\vect V_{1}'\vect V_{1}'+(1-\epsilon)^2\vect V_{2}'\vect V_{2}'\Big) \nonumber \\
&-&2\big(1-\epsilon^2\big)\Big(\text{Tr}\big(\vect V_{12}'\vect V_{12}'\big)+3\left\langle \vect R_{1}'^T\right\rangle\vect V_{12}'\left\langle \vect R_{2}'\right\rangle\Big),
\nonumber
\end{eqnarray}
where $F_{\text{Meas}}(\varphi, \epsilon)$ is obtained from \eqref{FICHS} after substituting the CV matrix \eqref{CVMTMS}.

The central panel of Fig.\ \ref{Fig3} displays the deviation of the Fisher information from the Quantum limit $F_{\text{pol}}^{(x)}(\varphi,\epsilon)-\mathcal{F}_{\text{pol}}(\epsilon),$ in the case of position quadrature measurements, for a fixed unknown phase shift and vanishing input displacement. As expected, the deviation is always negative or zero. However, it also remains close to zero for a growing squeezing around $\epsilon\approx\pm 1/2$. This indicates that an ideal $N$-mode quadrature detection may constitute an optimal measurement scheme. We can verify this for two-mode squeezed vacuum states and $\epsilon=\pm1$. After a 50:50 beam splitter transformation (i.e.  $\tau=1/2$), the probe system is in the tensor product of single-mode squeezed vacuum states. In agreement with the discussion in the previous section, we may expect to recover an identical relation for the operating point as Eq.\ \eqref{QCRBSSND}. Indeed, after some manipulation Eq.\ \eqref{FICHSPpol} boils down to a simple algebraic expression in the argument $y=\cos(4\phi)$ (see Eqs.\ \eqref{FPOLNX} and \eqref{FPOLNP}), from which follows the subsidiary condition:  $\cos\Big(4\phi_{\text{opt}}^{(x/p)}\Big)=\mp\epsilon/|\epsilon|\tanh(2s')$. This is complementary to earlier findings for homodyne or intensity detection schemes combined with active interferometry\ \cite{sparaciari20161, yurke19861,li20141}.

The right panel in Fig.\ \ref{Fig3} also illustrates the saturation of the QCRB, as zeros of the difference $F_{\text{pol}}^{(x)}(\varphi,\epsilon)-\mathcal{F}_{\text{pol}}(\epsilon),$ for a strategy based on position measurements and vanishing input displacement. Note how this deviation is an oscillating function of the phase, with an amplitude that grows with the squeezing [cf. dashed vs. solid lines in Fig.\ \ref{Fig3}]. Upon a closer look we appreciate an optimal operating point around $\varphi_{\text{opt}}^{(x)}\approx\pi/2$ [see Fig.\ \ref{Fig3} inset], where the FI reaches the QFI, and thus, it takes the value
\begin{equation}
    F_{\text{pol}}\big(\varphi_{\text{opt}}^{(x)},1/2\big)=5\sinh^2(2s')\approx 5 \bar{n}_{s'}^{2}.
    \label{TMSFIR}
\end{equation}
This is an optimal measurement strategy for the polychromatic scenario (with $\tau=1/2$), a result which is also recovered in a setup with momentum-based measurements. 

Unfortunately, the quadrature measurement is no longer optimal in the case of a vanishing modulation frequency $\epsilon=0$, which is when the polychromatic scheme obtains the largest improvement over the conventional strategy. In this case, the optimal operating point is  determined by an algebraic equation $f_{0}^{(x)}(\cos(4\varphi),s')=0$ [cf. discussion around Eq.\ (\ref{QCRBTSI}) in App.\ \ref{app3}], as in preceding sections. The closed-form expressions for these roots given in Eq.~\eqref{FPOLROOTS}, shows that no value of squeezing $0<s'$ can saturate the QCRB for a given phase shift $\varphi$. 

\begin{table*}[t!]
\begin{center}
\begin{tabular}{c|c|c|c|c}
\hline
\hline
Input resources & Interferometric  &    Scaling  & Scaling with & QCRB  \\
& transformation & per mode energy & interferometer size  & \\
\hline
Coherent           & QUMI    & SNL    & SNL & Yes \\ 
  ($s_{1}=s_{2}=1$)         &     &   &  & See text around Eq.(\ref{PSACQ}) \\\hline
 single-mode  squeezed vacuum       & Any      &  HL    & Constant & Yes \\ 
   ($s_{1}=s_{2}=e^{-2s'}$)       &     &   &  & See Eq.(\ref{QCRBSSNDF})\\ \hline
one-mode squeezed$\otimes$coherent           & QUMI      & sub-SNL     &  sub-SNL & No (nearly optimal for $s'\ll N$) \\ 
     ($s_{1}=e^{-2s'}$ and $s_{2}=1$)      &     &   &  &  See Eq.(\ref{FIAS1a})\\\hline
two-mode squeezed vacuum ($N=2$)          & 50:50 beam splitter     &  HL  & -   & Yes (for $\epsilon\neq 0$) \\            &     &   &  & See Eq.(\ref{TMSFIR}) \\ \hline
\hline
\end{tabular}
\end{center}
\caption{Summary of the phase sensitivity retrieved by the distinct Gaussian interferometric phase-estimation strategies involving an ideal homodyne detection, expressed in terms of the interferometer size $N$ and the input average photon number per mode $\bar{n}$. The particular choice of the probe Gaussian state and the interferometric scheme are specified in the first two columns. The third column illustrate the scaling of the FI with respect to $\bar{n}$ at a fixed value of the interferometer size $N$. The fourth column represents instead the scaling in terms of $N$ and for a given input intensity $\bar{n}$, whilst the fifth column determines which strategies are enable to attain the QCRB. Accordingly, the SNL-type scaling in terms of the input intensity must be understood as $F\sim 4\bar{n} $, whereas the SNL-type scaling is similarly defined as $F\sim 4N $ in terms of the interferometer size.}
\label{tab:table1}
\end{table*}

\subsection{Photon-loss effects and non-unit efficiency detection}\label{SDEFED}

Finally, we address the degrading effects owing to the experimental imperfections, extending our treatment to include these in the analysis of the FI.
In most interesting cases, the photon-loss process, determined by a given strength $\eta_{\text{loss}}$, and the non-unit efficiency detection, designated by $\eta_{\text{eff}}$, can be regarded as the major limits to interferometric precision \cite{nichols20181,pinel20121, sparaciari20161,gard20171,oh20171,oh20181,jarzyna20171,gao20161,li20141,gagatsos20171,aspachs20091,escher20111}. Furthermore, it is customary to assume that the environmental noise and photon-loss mechanism act identically and independently upon each probe mode \cite{koodynski20131}, as well as the environment is in a thermal state at a temperature determined by the mean photon number $n_{\text{th}}$. Under these considerations, the light interferometric propagation is modified in the presence of decoherence as $\vect S(\varphi)\vect V\vect S^{T}(\varphi)\rightarrow \eta_{\text{loss}} \vect S(\varphi)\vect V\vect S^{T}(\varphi)+(1- \eta_{\text{loss}})(1+n_{\text{th}})\vect I_{N}$ \cite{valido20141,jarzyna20171,serafini20051,oh20171}. Combining this result with Eqs.\ \eqref{GGMS1} and \eqref{GGMS2}, we directly obtain
\begin{eqnarray}
\left\langle \vect \lambda(\varphi)\right\rangle &=& \sqrt{\eta_{\text{loss}}}\vect S(\varphi)\left\langle \vect R\right\rangle,  \label{GGMSPLID1} \\
\vect \sigma (\varphi) &=&\big(1-\eta_{\text{eff}}+(1-\eta_{\text{loss}})(1+n_{\text{th}})\big)\vect I_{N} \nonumber\\
&+&\underbrace{\eta_{\text{eff}}\vect \Sigma+\eta_{\text{\text{loss}}}\vect S(\varphi)\vect V\vect S^{T}(\varphi)}_{\vect \sigma_{\text{deco}}(\eta_{\text{loss}},\eta_{\text{eff}})} \label{GGMSPLID2}, 
\end{eqnarray}
where the CV matrix $\vect \sigma_{\text{deco}}$ solely regards photon-loss effects. By replacing \eqref{GGMSPLID1} and \eqref{GGMSPLID2} into the general equation \eqref{CFGEX} and doing some manipulation as illustrated in Sect.\ \ref{SGLI}, we obtain a closed-form expression of the FI in presence of these decoherence effects, say $F_{\text{deco}}$, similar in structure to \eqref{IFSST} (see Eq.~\eqref{TFIDISS} in App.\ \ref{app4}). In the particular case we assume the propagation photon losses and non-unit efficiency  contribute equally, i.e. $\eta_{\text{loss}}=\eta_{\text{eff}}=\eta$, the FI can be cast as follows, 
\begin{eqnarray}
&&F_{\text{deco}}(\varphi,\eta,n_{\text{th}})=\eta^2 F(\varphi) \label{TFIDISS2} \\
&-&\eta\left\langle \vect R^{T}\right\rangle\vect L^{T} \vect P_{\varphi}^{T}\vect \Sigma_{\text{deco}}^{-1}(\eta,n_{\text{th}})\vect P_{\varphi}\vect L\left\langle \vect R\right\rangle \nonumber \\
&+&\frac{1}{2}\text{Tr}\big(\partial_{\varphi}\vect \Sigma_{\text{deco}}^{-1}(\eta,n_{\text{th}})\partial_{\varphi}\vect \sigma_{\text{deco}}(\eta)\big) \nonumber \\
&-&(1-\eta^2)\Bigg(2+\frac{1}{2}\text{Tr}\Bigg(\Big(\tilde{\vect \Sigma}\partial_{\varphi}\Big(\vect S\vect V\vect S^{T} \Big)\Big)^2\Bigg)\Bigg),
\nonumber
\end{eqnarray}
where $\vect \Sigma_{\text{deco}}(\eta,n_{\text{th}})$ is a $2N\times 2N$ real, symmetric matrix (given by Eq.~\eqref{SDISS}) that fully contains the influence owing to the environmental thermal noise. Recall $F(\varphi)$ denotes the FI in the ideal scenario.

Eq.\ \eqref{TFIDISS2} manifests that the decoherence effects influence the phase resolution beyond a limiting constant factor of the phase sensitivity achievable in the ideal case \cite{escher20111}. Furthermore, this expression shows that the decoherence effects impact differently 
the phase sensitivity provided by distinct probe resources \cite{sparaciari20161,oh20171,demkowicz20151}: while the third and last terms in the right-hand side vanishes for input coherent states,  the second term cancel for probe resources without an initial displacement. For instance, Eq.~\eqref{TFIDISS2} indicates that thermal noise is specially detrimental for input displaced states, whilst the phase sensitivity due to coherent resources is apparently more tolerant to photon losses\ \cite{oh20171} (since the last term in\ \eqref{TFIDISS2} vanishes).

As a final remark, form Eq.~\eqref{TFIDISS2} it is clear that the Gaussian interferometric schemes in presence of experimental imperfections cannot reach the HL, instead they could be able to beat the SNL for moderate values of $\eta$, as well as saturate the QCRB for quadrature measurements \cite{jarzyna20171,sparaciari20161,gard20171,li20141,sidhu20191}. Rather than figuring out the strict homodyne measurement attaining the ultimate sensitivity given by the QFI in presence of photon loss and noise, from Eq.~\eqref{TFIDISS2} one may be tempted to look for an alternative "optimal" Gaussian measurement scheme where optimal is understood in the sense that $F_{\text{deco}}$ eventually converges to $\eta^{2} \mathcal{F}$ instead (notice that $\mathcal{F}$ denotes the QFI in the ideal scenario). The latter yields an algebraic subsidiary condition as well, from which we may determine the corresponding "optimal" operating point. For instance, for the probe coherent scheme we find out that this is given by the formula  $\sin(2\varphi^{(x/p)})=\mp(2(\eta/\tilde{\eta})^2-1)$ (see Eq.~\eqref{FIDISS}), with
\begin{equation}
 \tilde{\eta}^2= \eta^2+\frac{(2+n_{\text{th}})(1-\eta)}{\eta+(1-\eta)n_{\text{th}}-2}, 
\end{equation}
which significantly differs from the ideal scenario (i.e. $\varphi_{\text{opt}}^{(x/p)}=\mp\pi/4$). This manifests that the experimental imperfections substantially influence the optimal working point besides the ultimate sensitivity.

\section{Outlook and concluding remarks}\label{OCR}

In this work we have presented a theoretical framework to explore the metrological potential of generic Gaussian interferometric schemes accessible with current photonic technology. Our treatment 
proves convenient to address the optimal phase-estimation scheme and operating point: in particular, we recover the vast majority of previous well-known results in the single- and two- mode Gaussian metrology scenarios. In Table\ \ref{tab:table1}, we summarize the phase sensitivity provided by the choice of different input states and interferometric schemes in the finite energetic regime. To a large extent this table contains most of previous results related to Gaussian phase resolution in the absence of photon loss and for perfect detection schemes \cite{pinel20121,oh20181,gard20171, oh20171}.

Interestingly, input coherent resources were shown to outperform the probe non-classical states used in previous QUMI-based phase-estimation proposals. Moreover, our analysis revealed that in the low-intensity regime (e.g. when squeezing parameter is small compared to the interferometer size $N$) the QUMI architecture along with probe single-mode squeezed states is unable to provide a real metrological advantage with respect to the best classical strategy for a large $N$.

Additionally, we also developed a polychromatic version of the well-established MZI setup endowed with probe two-mode non-degenerate squeezed-vacuum states. We show that this setup can significantly improve the resolution-energy trade-off with optimal (ideal) quadrature measurements. Besides our treatment is a versatile approach to address the impact of experimental imperfections on the phase sensitivity unlike the analysis based on the complex SLD: e.g., we show that the optimal working point associated to coherent resources is significantly shifted by both the photon losses and the nonunit-efficiency detection.

Remarkably, the recent developments on the fabrication and manipulation of integrated photonic circuits \cite{harris20171,polino20191,paesani20171} makes them more resilient to phase stability, or photon losses and noise effects, which opens new avenues to implement higher sophisticated phase-estimation experiments with relatively little effort \cite{chaboyer20151} (e.g. endowed with current photon sources and measurement detection schemes). In particular this prospect highlights the demand for further theoretical tools enable to explore its feasible metrological power. In this sense, the present treatment could render a valuable theoretical support to envisage a new series of experiments in the realm of quantum phase estimation.

\acknowledgments
This material is based upon work supported by the Air Force Office of Scientific Research under award number FA2386-18-1-4019. The authors also acknowledge support from Spanish project PGC2018-094792-B-100 (MCIU/AEI/FEDER,EU).

\appendix

\begin{widetext}

\section{Basics of phase estimation}\label{app1}

In this section we briefly sketch the derivation of the general expressions \eqref{CFGEX} and \eqref{QFIGGS} by using results from matrix analysis theory \cite{bernstein20051,petersen20121} and the matrix identities \eqref{SPCVII} and \eqref{SPST} just relying on the interested set of probe iso-thermal Gaussian states. We start from the formal definition of the FI, which reads \cite{toth20141,paris20091,sidhu20191}
\begin{eqnarray}
F(\varphi)=\int d^{2m}\vect \lambda\frac{1}{p(\vect \lambda|\varphi)}\left( \frac{\partial p(\vect \lambda|\varphi)}{\partial \varphi}\right)^2. 
\label{CFI}
\end{eqnarray}
Thanks to the probability distribution characterizing the Gaussian phase-estimation scheme is a Gaussian function, the result of the integral involved in\ \eqref{CFI} is a Gaussian function as well. This can be seen more clearly once computed the derivative of the probability distribution, i.e.,
\begin{eqnarray}
\frac{\partial p(\vect \lambda|\varphi)}{\partial \varphi}&=&\frac{1}{2}p(\vect \lambda|\varphi)\Bigg((\vect \lambda- \left\langle \vect \lambda\right\rangle)^{T}\vect \sigma^{-1}\partial_{\varphi} \vect \sigma\vect \sigma^{-1} ( \vect \lambda-  \left\langle \vect \lambda\right\rangle)+ 2( \vect \lambda- \left\langle \vect \lambda\right\rangle)^{T}\vect \sigma^{-1}\partial_{\varphi}\left\langle \vect \lambda\right\rangle) -\text{Tr}\big(\vect \sigma^{-1}\partial_{\varphi}\vect \sigma\big)
\Bigg), \label{DCFI}
\end{eqnarray}
where the last term of the right-hand side appears due to the dependence of the probability distribution normalization-constant with the desired phase shift \cite{petersen20121}. Hence, one may realize that the classical Fisher information\ \eqref{CFI} reduces to carry out the integral of a quadratic polynomial (in the variable $\vect \lambda$) weighted by $p(\vect \lambda|\varphi)$. After substituting Eq.~\eqref{DCFI} in \eqref{CFI}, it is convenient to swap from $p(\vect \lambda|\varphi)$ to a zero-mean Gaussian probability distribution $p(\tilde{\vect \lambda}|\varphi)$, with the CV $\vect \sigma$, by making the change of variables $\tilde{\vect \lambda}=\vect \sigma^{-1}(\vect \lambda- \left\langle \vect \lambda\right\rangle)$. Upon doing this, we obtain
\begin{eqnarray}
F(\varphi)&=&\frac{1}{4}\int d^{2m}\tilde{\vect \lambda} \ p(\tilde{\vect \lambda}|\varphi)\Bigg(\Big(\text{Tr}\big(\vect \sigma^{-1}\partial_{\varphi}\vect \sigma\big)\Big)^2-2\text{Tr}\big(\vect \sigma^{-1}\partial_{\varphi}\vect \sigma\big)\Big(\tilde{\vect \lambda}^{T} \partial_{\varphi}\vect \sigma\tilde{\vect \lambda}-2\partial_{\varphi}\left\langle \vect \lambda^{T}\right\rangle \tilde{\vect \lambda} \Big) \nonumber \\
&+&\Big((\tilde{\vect \lambda}^{T} \partial_{\varphi}\vect \sigma\tilde{\vect \lambda})(\tilde{\vect \lambda}^{T}\partial_{\varphi} \vect \sigma\tilde{\vect \lambda}) +4(\partial_{\varphi}\left\langle \vect \lambda\right\rangle \tilde{\vect \lambda} )(\partial_{\varphi}\left\langle \vect \lambda\right\rangle \tilde{\vect \lambda} ) -4(\partial_{\varphi}\left\langle \vect \lambda\right\rangle \tilde{\vect \lambda} )(\tilde{\vect \lambda}^{T} \partial_{\varphi}\vect \sigma\tilde{\vect \lambda})\Big)\Bigg). 
\label{CGFI}
\end{eqnarray}
Since $p(\tilde{\vect \lambda}|\varphi)$ is centered with respect to the origin $\tilde{\vect \lambda}=0$, the contribution coming from the linear and third-order terms in the right-hand side of \eqref{CGFI} must cancel. The rest of the contributions can be computed by using the following identity for multivariate Gaussian integrals \cite{petersen20121},
\begin{equation}\label{GMII}
\int d^{2m}\vect \lambda \ |\vect \lambda|^4 \text{exp}\Big(-\frac{1}{2}(\vect \lambda-\left\langle \vect \lambda\right\rangle)\vect \sigma^{-1} (\vect \lambda-\left\langle \vect \lambda\right\rangle)\Big)=|\!\left\langle \vect \lambda\right\rangle\!|^4+4\left\langle \vect \lambda^{T}\right\rangle\vect \sigma\left\langle \vect \lambda\right\rangle+\text{Tr}^2(\vect \sigma)+2\text{Tr}(\vect \sigma)|\!\left\langle \vect \lambda\right\rangle\!|^2+2\text{Tr}\big(\vect \sigma\vect \sigma^{T}\big),
\end{equation}
with $|\vect x|$ denoting the usual Euclidean norm of a $2m$-dimensional vector $\vect x$. After substituting the result (\ref{GMII}) in (\ref{CGFI}), we obtain an expression with several terms  involving the derivative of $\vect \sigma$, these terms can be further simplified by using the matrix identity $\partial_{\varphi}\vect \sigma^{-1}=-\vect \sigma^{-1}\partial_{\varphi}\vect \sigma \vect \sigma^{-1}$ \cite{petersen20121}. This finally leads us to Eq.~\eqref{CFGEX}.

Now we turn the attention to the formula \eqref{QFIGGS} of the QFI by virtue of the phase generator $\vect U_{N}(\varphi)=\vect U(\varphi)\oplus \vect I_{N-1}$ with
\begin{equation}
\vect U(\varphi)=\left( \begin{array}{cc}
\cos \varphi & \sin\varphi \\
-\sin\varphi  &\cos \varphi
\end{array}\right) .
\label{URR}
\end{equation}
This is worked out from the general expression of the QFI valid for any pure or mixed single-mode Gaussian state provided in \cite{pinel20131,pinel20121,jiang20141}. The latter takes the following form for the set $\mathcal{G}(1,\bar{n}_{t})$ of interesting states and the phase-shift generator \eqref{URR},   
\begin{equation}
\mathcal{F}= \left\langle \vect R_{1}'^{T}\right\rangle \partial_{\varphi}\vect U^{T}(\varphi)\vect U(\varphi)\vect V_{1}'^{-1}\vect U^{T}(\varphi)\partial_{\varphi}\vect U(\varphi)\left\langle \vect R_{1}'\right\rangle -\frac{1}{2(1+(2\bar{n}_{t}+1)^{-2})}       \text{Tr}\Big(\partial_{\varphi}(\vect U^{T}(\varphi)\vect V_{1}'^{-1}\vect U(\varphi))(\partial_{\varphi}(\vect U^{T}(\varphi)\vect V_{1}'\vect U(\varphi))\Big).                   
\end{equation}
By replacing the matrix identity \eqref{SPCVII}, we obtain upon some manipulation
\begin{eqnarray}
\mathcal{F}&=&\frac{1}{(2\bar{n}_{t}+1)^2}\left\langle \vect R_{1}'^{T}\right\rangle \partial_{\varphi}\vect U^{T}(\varphi)\vect U(\varphi)\vect V_{1}'\vect U^{T}(\varphi)\partial_{\varphi}\vect U(\varphi)\left\langle \vect R_{1}'\right\rangle  +\frac{(2\bar{n}_{t}+1)^{-2}}{2(1+(2\bar{n}_{t}+1)^{-2})}\text{Tr}\Big(\partial_{\varphi}\Big(\vect J\vect U^{T}(\varphi)\vect V_{1}'\vect U(\varphi)\Big)^2\Big) \nonumber \\
&=&\frac{1}{(2\bar{n}_{t}+1)^2}\Bigg(\left\langle \vect R_{1}'^{T}\right\rangle \partial_{\varphi}\vect U^{T}(\varphi)\vect U(\varphi)\vect V_{1}'\vect U^{T}(\varphi)\partial_{\varphi}\vect U(\varphi)\left\langle \vect R_{1}'\right\rangle  +\frac{1}{1+(2\bar{n}_{t}+1)^{-2}}\Big(\text{Tr}\Big(\Big(\vect J\partial_{\varphi}\vect U^{T}(\varphi)\vect V_{1}'\vect U(\varphi)\Big)^2\Big) \nonumber \\
&-&(2\bar{n}_{t}+1)^2\text{Tr}\Big(\partial_{\varphi}\vect U^{T}(\varphi)\partial_{\varphi}\vect U(\varphi)\Big)\Big)\Bigg),
\end{eqnarray}
which after substituting $\partial_{\varphi}\vect U(\varphi)=\vect J \vect U(\varphi)$ leads to the expression \eqref{QFIGGS}. Notice that in the pure case (i.e. $\bar{n}_{t}=0$) the expression \eqref{QFIGGS} identically coincides with the result independently obtained from the standard expression of the QFI $\mathcal{F}=4(\Delta\hat H)^2$.

\section{Gaussian phase estimation}\label{app2}

In this appendix we extensively illustrate the derivation of Eq.~\eqref{IFSST} appearing in Sect.\ \ref{SGLI}.  We firstly express the inverse of \eqref{GGMS2} in terms of the CV matrix $\vect \sigma_{S}$ by means of the Woodbury identity \cite{petersen20121,bernstein20051}, that is
\begin{eqnarray}
\vect \sigma^{-1}=\vect \sigma_{S}^{-1} - \vect \sigma_{S}^{-1}\vect S_{SA}\big(\vect V_{A}^{-1}+\vect S_{SA}^{T}\vect \sigma_{S}^{-1}\vect S_{SA}\big)^{-1}\vect S_{SA}^{T}\vect \sigma_{S}^{-1},
\label{WOODI}
\end{eqnarray}
which always holds as $\vect V_{A}^{-1}+\vect S_{SA}^{T}\vect \sigma_{S}^{-1}\vect S_{SA}$ is expected to be an invertible matrix. Notice that we have omitted the explicit dependence of the matrices with $\varphi$ for the sake of clarity. The above identity allows us to separate the FI contribution in\ \eqref{CFGEX} that is completely independent of the ancilla CV matrix. This yields the expression \eqref{CFI2}. In other words, we gather together in $F_{S}(\varphi)$ all dependence in $\sigma_S$.  Since $\vect \Sigma^{-1}_{S}+ \big(\vect S_{S}^{T}\big)^{-1}\vect V_{S}^{-1}\vect S_{S}^{-1}$ must be an invertible matrix as well (see remark 2.16.21 in \cite{bernstein20051}), we can employ again the  Woodbury identity in order to express the $\vect \sigma_S^{-1}$ in terms of $\vect V_{S}^{-1}$, which is
\begin{equation}
\vect \sigma_{S}^{-1}=\big(\vect S_{S}^{T}\big)^{-1}\vect V_{S}^{-1}\vect S_{S}^{-1}-\big(\vect S_{S}^{T}\big)^{-1}\vect V_{S}^{-1}\vect S_{S}^{-1}\Big(\vect \Sigma^{-1}_{S}+ \big(\vect S_{S}^{T}\big)^{-1}\vect V_{S}^{-1}\vect S_{S}^{-1}\Big)^{-1}\big(\vect S_{S}^{T}\big)^{-1}\vect V_{S}^{-1}\vect S_{S}^{-1}.
\end{equation}
Replacing this result in the obtained expression for $F_{S}(\varphi)$, we arrive to  
\begin{eqnarray}
F_{S}(\varphi)&=&\left\langle \vect R_{S}^{T}\right\rangle\partial_{\varphi}\vect S_{S}^{T}\big(\vect S_{S}^{T}\big)^{-1} \vect V_{S}^{-1}\vect S_{S}^{-1}\partial_{\varphi}\vect S_{S}\left\langle \vect R_{S}\right\rangle  -\frac{1}{2}\text{Tr}\Big(\partial_{\varphi}\Big(\big(\vect S_{S}^{T}\big)^{-1} \vect V_{S}^{-1}\vect S_{S}^{-1}\Big)\partial_{\varphi}\Big(\vect S_{S} \vect V_{S}\vect S_{S}^{T}\Big)\Big)-F_{\text{Meas}}(\varphi),
\label{CFIMSA} 
\end{eqnarray}
where we have identified $F_{\text{Meas}}(\varphi)$ as the residual contribution given by Eq.~\eqref{CFIMM}.  We can further simplify\ \eqref{CFIMSA} by substituting the following identities satisfied by the inverse of the sub-block matrices of $\vect S$ (see proposition 2.8.7 in \cite{bernstein20051}), 
\begin{eqnarray}
\vect S^{-1}_{S}&=&\vect S_{S}^{T}- \Delta \vect S_{S}, \label{SCS1} \\
\big(\vect S_{S}^{T}\big)^{-1}&=&\vect S_{S}- \Delta \vect S_{S}^{T}, 
\end{eqnarray}
with 
\begin{equation}
\Delta \vect S_{S}=\vect S^{-1}_{S}\vect S_{SA}\big(\vect S/\vect S_{S}\big)^{-1}\vect S_{AS} \vect S^{-1}_{S},
\end{equation}
where $\vect S/\vect S_{S}$ stands for the Schur complement of $\vect S_{S}$ in $\vect S$, i.e. $\vect S/\vect S_{S}=\vect S_{A}-\vect S_{AS}\vect S_{S}^{-1}\vect S_{SA}$, which is non-singular for a realistic transformation $\vect S$. By plugging the relations\ \eqref{SCS1} into\ \eqref{CFIMSA}, after some tedious manipulation we obtain 
\begin{eqnarray}
F_{S}(\varphi)&=&\left\langle \vect R_{S}^{T}\right\rangle\partial_{\varphi}\vect S_{S}^{T}\vect S_{S} \vect V_{S}^{-1}\vect S_{S}^{T}\partial_{\varphi}\vect S_{S}\left\langle \vect R_{S}\right\rangle -\frac{1}{2}\text{Tr}\Big(\partial_{\varphi}\Big(\vect S_{S} \vect V_{S}^{-1}\vect S_{S}^{T}\Big)\partial_{\varphi}\Big(\vect S_{S} \vect V_{S}\vect S_{S}^{T}\Big)\Big)+\tilde{F}_{\text{Int}}(\varphi)-F_{\text{Meas}}(\varphi),
\label{CFIMSA2} 
\end{eqnarray}
where
\begin{equation}
\tilde{F}_{\text{Int}}(\varphi)=F_{\text{Int}}(\varphi)-\text{Tr}\big(\partial_{\varphi}\vect S_{S}^{T}\partial_{\varphi}\vect S_{S} \vect V_{S}^{-1} \vect S_{SA}\vect S_{SA}^{T} \vect V_{S}\big),
\end{equation}
and $F_{\text{Int}}(\varphi)$ is given by 
\begin{eqnarray}
F_{\text{Int}}(\varphi)&=&\text{Tr}\big(\partial_{\varphi}\vect S_{S}^{T}\partial_{\varphi}\vect S_{S} \vect V_{S}^{-1} \vect S_{SA}\vect S_{SA}^{T} \vect V_{S}\big)-2\left\langle \vect R_{S}^{T}\right\rangle\partial_{\varphi}\vect S_{S}^{T}\vect S_{S} \vect V_{S}^{-1}\Delta\vect S_{S}\partial_{\varphi}\vect S_{S}\left\langle \vect R_{S}\right\rangle \label{IFINT2} \\
 &+&\left\langle \vect R_{S}^{T}\right\rangle\partial_{\varphi}\vect S_{S}^{T}\Delta\vect S_{S}^{T} \vect V_{S}^{-1}\Delta\vect S_{S}\partial_{\varphi}\vect S_{S}\left\langle \vect R_{S}\right\rangle +\frac{1}{2}\text{Tr}\Big(\partial_{\varphi}\Big(2\vect S_{S} \vect V_{S}^{-1}\Delta\vect S_{S}- \Delta\vect S_{S}^{T} \vect V_{S}^{-1}\Delta\vect S_{S}\Big) \partial_{\varphi}\Big(\vect S_{S} \vect V_{S}\vect S_{S}^{T} \Big)\Big). 
 \nonumber
\end{eqnarray}

Now we substitute the inverse matrix $\vect V_{S}^{-1}$ in \eqref{CFIMSA2} according to the symplectic-like identity\ \eqref{SPCVII}, and then, we use the relation\ \eqref{SPST} to cast Eq.~\eqref{CFIMSA2} in the following form
\begin{eqnarray}
F_{S}(\varphi)&=&\frac{1}{(2\bar{n}_{t}+1)^{2}}\Bigg(\left\langle \vect R_{S}^{T}\right\rangle\partial_{\varphi}(\vect J_{m}\vect S_{S})^{T}\vect S_{S} \vect V_{S}\vect S_{S}^{T}\partial_{\varphi}(\vect J_{m}\vect S_{S})\left\langle \vect R_{S}\right\rangle +\frac{1}{2}\text{Tr}\Bigg(\Bigg(\partial_{\varphi}\Big(\vect J_{m} \vect S_{S} \vect V_{S}\vect S_{S}^{T}\Big)\Bigg)^{2}\Bigg)\Bigg)\nonumber \\
&+&\tilde{F}_{\text{Int}}(\varphi)-F_{\text{Meas}}(\varphi).\label{FFFI}
\label{CFIMSA3} 
\end{eqnarray}
Let us now focus the attention on the trace term of the equation (\ref{FFFI}). This term can be simplified as follows,
\begin{eqnarray}
\text{Tr}\Bigg(\Bigg(\partial_{\varphi}\Big(\vect J_{m} \vect S_{S} \vect V_{S}\vect S_{S}^{T}\Big)\Bigg)^{2}\Bigg)&=&\text{Tr}\Bigg(\Big(\vect J_{m} \partial_{\varphi}\vect S_{S}\vect V_{S}\vect S_{S}^{T}+\vect J_{m} \vect S_{S} \vect V_{S}\partial_{\varphi}\vect S_{S}^{T}\Big)^{2} \Bigg)\nonumber \\
&=&2\text{Tr}\Bigg(\Big(\vect J_{m} \partial_{\varphi}\vect S_{S}\vect V_{S}\vect S_{S}^{T}\Big)^2\Bigg)+2\text{Tr}\Big( \partial_{\varphi}\vect S_{S}\vect J_{m}\vect V_{S}\vect J_{m}\vect S_{S}^{T} \vect S_{S} \vect V_{S}\partial_{\varphi}\vect S_{S}^{T}\Big) \nonumber \\
&=&2\text{Tr}\Bigg(\Big(\partial_{\varphi}(\vect J_{m}\vect S_{S})\vect V_{S} \vect S_{S}^{T}\Big)^2\Bigg)-2(2\bar{n}+1)^2\Big(\text{Tr}\big(\partial_{\varphi}\vect S_{S}^{T}\partial_{\varphi}\vect S_{S}\big)\nonumber \\
&-&\text{Tr}\big(\partial_{\varphi}\vect S_{S}^{T}\partial_{\varphi}\vect S_{S} \vect V_{S}^{-1} \vect S_{SA}\vect S_{SA}^{T} \vect V_{S}\big)\Big), 
\label{STFIA}
\end{eqnarray}
where once again we have made use of the linear properties of the trace, as well as the identities\ \eqref{SPCVII} and \eqref{SPST0}. Replacing the result\ \eqref{STFIA} in Eq.~\eqref{CFIMSA3} directly returns the expression 
\begin{eqnarray}
F_{S}(\varphi)&=&\Bigg(\left\langle \vect R_{S}^{T}\right\rangle\partial_{\varphi}(\vect J_{m}\vect S_{S})^{T}\vect S_{S} \vect V_{S}\vect S_{S}^{T}\partial_{\varphi}(\vect J_{m}\vect S_{S})\left\langle \vect R_{S}\right\rangle +\text{Tr}\Bigg(\Big(\partial_{\varphi}(\vect J_{m}\vect S_{S})\vect V_{S} \vect S_{S}^{T}\Big)^2\Bigg) \Bigg)\frac{1}{(2\bar{n}_{t}+1)^2}\nonumber \\
&-&\text{Tr}\big(\partial_{\varphi}\vect S_{S}^{T}\partial_{\varphi}\vect S_{S}\big)-F_{\text{Meas}}(\varphi)+F_{\text{Int}}(\varphi),
\label{CFIMS} 
\end{eqnarray}
after rearranging the contribution $\text{Tr}\big(\partial_{\varphi}\vect S_{S}^{T}\partial_{\varphi}\vect S_{S} \vect V_{S}^{-1} \vect S_{SA}\vect S_{SA}^{T} \vect V_{S} \big)$ into the definition of $F_{\text{Int}}(\varphi)$. One can proceed by noticing from\ \eqref{URR} that $\partial_{\varphi}\vect U(\varphi)=\vect U(\varphi)\vect J$. The latter combined with Eq.~\eqref{SMS1} directly yields 
\begin{eqnarray}
\partial_{\varphi}\vect S_{S}(\varphi)&=&\vect P_{\varphi}\vect J_{m}\vect L_{S}, \label{SMS1D} 
\end{eqnarray}
where $\vect P_{\varphi}=\vect U(\varphi) \oplus \vect 0_{m-1}$, which is a $2m\times 2m$ projection matrix (i.e., $\vect P_{\varphi}\vect P_{\varphi}^{T}=\vect P_{\varphi}^{T}\vect P_{\varphi}=\vect I_{1} \oplus \vect 0_{m-1}$ as well as $\vect J_{m}^{T}\vect P_{\varphi}\vect J_{m}=\vect P_{\varphi}$). Here $\vect 0_{m-1}$ stands for the $2(m-1)\times2(m-1)$ null matrix (i.e. all its entries are zero), so that the effect of $\vect P_{\varphi}$ through the subsequent computation is to drop the explicit dependence with the slice of matrix that is not supported by the phase space of the phase-shifted mode $(\hat q_{1},\hat p_{1})$: for instance, for value $\varphi=0$, $ \left\langle \vect R_{S}\right\rangle$ and $\vect V_{S}$ get projected into the displacement vector $\left\langle \vect R_{1}'\right\rangle$ and the CV matrix $\vect V_{1}'$ of the first probe mode immediately before undergoing the phase rotation, that is $\left\langle \vect R_{1}'\right\rangle=\vect P_{0}\vect L_{S}\left\langle \vect R_{S}\right\rangle$ and $\vect V_{1}'=\vect P_{0}\vect L_{S}\vect V_{S}\vect L_{S}^{T}\vect P_{0}^{T}$. By virtue of the latter, after some manipulation once replaced Eq.~\eqref{SMS1D} in \eqref{CFIMS}, one gets 
\begin{eqnarray}
F_{S}(\varphi)&=&\frac{1}{(2\bar{n}_{t}+1)^2}\Big(\left\langle \vect R_{1}'^{T}\right\rangle \vect V_{1}'\left\langle \vect R_{1}'\right\rangle  +\text{Tr}\big(\vect V_{1}'\vect V_{1}'\big)\Big) -F_{\text{Meas}}(\varphi)+F_{\text{Int}}(\varphi)-\text{Tr}\Big(\vect P_{0}\vect L_{S}\vect L_{S}^{T}\Big),
\label{IFSS1}
\end{eqnarray}
from which one can identify the QFI characteristic of the $m$-mode probe system upon close inspection. By conveniently manipulating\ \eqref{IFSS1} once plunged into \eqref{CFI2}, we arrive at the desired expression\ \eqref{IFSST} for the FI.

In the particular case of non-assisted phase-estimation schemes and pure input Gaussian states, the expression\ \eqref{IFSST} boils down to\ \eqref{IFSSTPW}. In this scenario, the aforementioned auxiliary matrix $\tilde{\vect \Sigma}$ further becomes $ \vect S\vect V^{-1}\vect S^{T}\big(\vect \Sigma^{-1}+\vect S\vect V^{-1}\vect S^{T} \big)^{-1}\vect S\vect V^{-1}\vect S^{T}$, so that the measurement contribution $F_{\text{Meas}}$, given by Eq.~\eqref{CFIMM}, can be substantially simplified as well. More specifically, by substituting this observation we obtain the first term in the right-hand side of \eqref{FICHS}, whereas the second terms may be further simplified by using the symplectic-like identities for $\vect \Sigma$, $\vect S(\varphi)$ and $\vect V$ as before (as well as $\partial_{\varphi}\vect A^{-1}=-\vect A^{-1}\partial_{\varphi}\vect A\vect A^{-1}$),
\begin{eqnarray}
\text{Tr}\Big(\partial_{\varphi}\tilde{\vect \Sigma}\partial_{\varphi}\Big(\vect S \vect V\vect S^{T} \Big)\Big)&=&\text{Tr}\Big(\partial_{\varphi}\big(\vect \Sigma^{-1}+\vect S\vect V^{-1}\vect S^{T} \big)^{-1}\vect S\vect V^{-1}\vect S^{T}\partial_{\varphi}\big(\vect S \vect V\vect S^{T}\big)\vect S\vect V^{-1}\vect S^{T}\Big) \label{FMEAS} \\
&+&2\text{Tr}\Big(\partial_{\varphi}\big(\vect S \vect V^{-1}\vect S^{T}\big)\big(\vect \Sigma^{-1}+\vect S\vect V^{-1}\vect S^{T} \big)^{-1}\big(\vect S \vect V^{-1}\vect S^{T}\big)\partial_{\varphi}\big(\vect S \vect V\vect S^{T}\big)\Big) \nonumber \\
&=&- \text{Tr}\Big(\vect J_{N}^{T}\partial_{\varphi}\big(\vect \Sigma+\vect S\vect V\vect S^{T} \big)^{-1}\vect J_{N}\partial_{\varphi}\big(\vect S \vect V^{-1}\vect S^{T}\big)\Big)  \nonumber\\
&+&2\text{Tr}\Big(\big(\vect \Sigma^{-1}+\vect S\vect V^{-1}\vect S^{T} \big)^{-1}\vect J_{N}\Big(\vect S \vect V\vect S^{T}\Big)\vect J_{N}^{T}\partial_{\varphi}\big(\vect S \vect V\vect S^{T} \big)\vect J_{N}\partial_{\varphi}\big(\vect S \vect V\vect S^{T}\big)\vect J_{N}^{T}\Big)     \nonumber \\
&=&\text{Tr}\Big(\big(\vect \Sigma+\vect S\vect V\vect S^{T} \big)^{-1}\partial_{\varphi}\big(\vect S \vect V\vect S^{T}\big)\big(\vect \Sigma+\vect S\vect V\vect S^{T} \big)^{-1}\partial_{\varphi}\big(\vect S \vect V\vect S^{T}\big)\Big)   \nonumber \\ 
&-&2\text{Tr}\Big(\big(\vect \Sigma+\vect S\vect V\vect S^{T} \big)^{-1}\big(\vect S \vect V\vect S^{T}\big)\partial_{\varphi}\big(\vect J_{N}\vect S \vect V\vect S^{T} \big)\partial_{\varphi}\big(\vect J_{N}\vect S \vect V\vect S^{T}\big)\Big), \nonumber
\end{eqnarray} 
where once again we have employed the linearity properties of the trace and $\vect J_{N}=-\vect J_{N}^{T}$. By substituting \eqref{FMEAS} in \eqref{CFIMM}, it is clear to see that we arrive at the desired expression \eqref{FICHS} for the measurement contribution.

\section{$N$-mode homodyne detection with input coherent states and without ancilla system}\label{app3}

\subsection{Explicit expressions from Sect.\ \ref{SQM}}
In this section we provide the explicit form corresponding to the particular QUMI transformation studied in Sect.\ \ref{SCAQUMI}, as well as the functions and matrices involved in the expressions from \eqref{CVMSMI} to \eqref{QCRBC} appearing in Sect.\ \ref{PSNDR}.

In the phase space notation, the transformation of the first probe mode due to any interferometric operation can be compactly expressed as follows,
\begin{equation}
     \vect R_{1}\rightarrow \sum_{i=1}^{N}V_{1i} \ \vect R_{i} \ \ \ \text{with} \ \ \ \vect R_{i}=( x_{i},p_{i}),
     \label{APNFM}
\end{equation}
where $V_{1i}$ is determined by the unitary evolution describing the interferometric operation. As stated in Sec. IV of Ref.\cite{olson20171}, the family of QUMI transformations is formally characterized by the constraint $|V_{1j}|=1/\sqrt{N}$ for $j=1,\dots,N$ (see, discussion below Eq.(5)). In particular, we will focus the attention on the subset of QUMI transformations which satisfies $V_{1j}=1/\sqrt{N}$. For this subset and for input coherent states with $\left\langle \vect R_{i}\right\rangle=(\sqrt{2\bar{n}_{c}},\sqrt{2\bar{n}_{c}})$ for $i\in\left[ 1,N\right] $, the average photon number of the first probe mode after transformation is equal to the average number of input photons, as mentioned in the discussion around Eq.(\ref{PSACQ1}). 

The interesting subset of QUMI schemes is provided by the following prescription \cite{olson20171}: The probe $N$-mode sequentially interferes with all the remains input modes, and the transmitivity between the probe $N$ and $j$ -mode is given by $\tau=1-1/j$ (notice that $\sqrt{1-\tau}$ retrieves the  transmitivity amplitude in the notation of Ref.\cite{olson20171}). Based on this prescription, we found out that the orthogonal matrix, say $\vect L_{\text{QUMI}}$, associated to a simple QUMI architecture takes the form
\begin{equation}
\vect L_{\text{QUMI}}=\left( \begin{array}{cccccccc}
\sqrt{\frac{1}{N}} & 0  & \sqrt{\frac{1}{N}} & 0 & \sqrt{\frac{1}{N}} & 0 &  \cdots & 0\\
0 &\sqrt{\frac{1}{N}} & 0  & \sqrt{\frac{1}{N}} & 0 & \sqrt{\frac{1}{N}} & \cdots  & \sqrt{\frac{1}{N}} \\
-\sqrt{\frac{N-1}{N}} & 0  & \sqrt{\frac{1}{N(N-1)}} & 0 & \sqrt{\frac{1}{N(N-1)}} & 0  & \cdots & 0 \\
0 &-\sqrt{\frac{N-1}{N}} & 0  & \sqrt{\frac{1}{N(N-1)}} & 0 & \sqrt{\frac{1}{N(N-1)}} & \cdots  & \sqrt{\frac{1}{N(N-1)}} \\
0 & 0 &-\sqrt{\frac{N-2}{N-1}} & 0  & \sqrt{\frac{1}{(N-1)(N-2)}} & 0 & \cdots & 0   \\
0 & 0 &0 &-\sqrt{\frac{N-2}{N-1}} & 0  & \sqrt{\frac{1}{(N-1)(N-2)}} & \cdots & \sqrt{\frac{1}{(N-1)(N-2)}}   \\
0 & 0 &0 & \cdots &   & \vdots &  & \vdots   \\
0  & 0 &0 & \cdots  & -\sqrt{\frac{1}{2}} & 0 & \sqrt{\frac{1}{2}} & 0  \\
0 & 0  & 0 &\cdots  &  0 & -\sqrt{\frac{1}{2}} & 0 & \sqrt{\frac{1}{2}}  \\
\end{array}\right) .
\label{DLQUMI}
\end{equation}
Note that $\vect L_{\text{QUMI}}$ satisfies the condition which defines the QUMI family of transformations: all matrix elements on the first row are identical to $1/\sqrt{N}$. Let us briefly sketch how to obtain the result (\ref{DLQUMI}) by following a mathematical induction procedure: we initially prove that $\vect L_{\text{QUMI}}$ with $N=2$ takes the form given by (\ref{DLQUMI}), and then, we show that (\ref{DLQUMI}) holds for the next value $N +1$. For $N=2$, $\vect L_{\text{QUMI}}$ just reduces to the sympletic transformation describing a 50/50 beam splitter \cite{wedbrook20111}, i.e.
\begin{equation}
   \vect L_{\text{QUMI}}= \sqrt{\frac{1}{2}}\left( \begin{array}{cc}
      \vect I_2  & \vect I_2 \\
      -\vect I_2  & \vect I_2
   \end{array}\right),
   \nonumber
\end{equation} 
with $I_{2}$ being the $N\times N$ identity matrix. For $N=3$, we must realize that both the probe third and second mode interfere with transmitivity $\tau_{2}=1/2$, and subsequently, the probe second mode interferes with the probe first mode with transmitivity $\tau_{3}=2/3$. Hence, $\vect L_{\text{QUMI}} $ must result from two subsequent beam splitter operations, i.e.
\begin{eqnarray}
   \vect L_{\text{QUMI}}&=&\left( \begin{array}{ccc}
      \sqrt{1/3}\vect I_2  & \sqrt{1-1/3}\vect I_2 & \vect 0_2 \\
      -\sqrt{1-1/3}\vect I_2  & \sqrt{1/3}\vect I_2 & \vect 0_2 \\
      \vect 0_2 & \vect 0_2 & \vect I_2
   \end{array}\right) \left( \begin{array}{ccc}
   \vect I_2 & \vect 0_2 & \vect 0_2 \\
     \vect 0_2 & \sqrt{1/2}\vect I_2  & \sqrt{1-1/2}\vect I_2 \\
     \vect 0_2 & -\sqrt{1-1/2}\vect I_2  & \sqrt{1/2}\vect I_2 
   \end{array}\right),
   \nonumber
\end{eqnarray} 
where  $\vect 0_{2}$ denoting the $N\times N$ zero matrix. After manipulation the above expression returns the result expected from (\ref{DLQUMI}) for $N=3$. We can repeat this procedure for $N=4$ in similar fashion, i.e.
\begin{eqnarray}
   \vect L_{\text{QUMI}}&=&\left( \begin{array}{cccc}
      \sqrt{1/4}\vect I_2  & \sqrt{1-1/4}\vect I_2 &  \vect 0_2  &\vect 0_2\\
    - \sqrt{1-1/4}\vect I_2  & \sqrt{1/4}\vect I_2 & \vect 0_2 &\vect 0_2 \\
     \vect 0_2 & \vect 0_2  & \vect I_2 & \vect 0_2 \\
     \vect 0_2 & \vect 0_2 & \vect 0_2 & \vect I_2 \\
   \end{array}\right)\left( \begin{array}{cccc}
   \vect I_2 & \vect 0_2 & \vect 0_2 & \vect 0_2 \\
     \vect 0_2  & \sqrt{1/3}\vect I_2  & \sqrt{1-1/3}\vect I_2 & \vect 0_2\\
     \vect 0_2 & -\sqrt{1-1/3}\vect I_2  & \sqrt{1/3}\vect I_2 & \vect 0_2 \\
     \vect 0_2 & \vect 0_2  & \vect 0_2 & \vect I_2
   \end{array}\right) \nonumber \\
   &\times& \left( \begin{array}{cccc}
   \vect I_2 & \vect 0_2 & \vect 0_2 & \vect 0_2 \\
   \vect 0_2 & \vect I_2  & \vect 0_2 & \vect 0_2\\
     \vect 0_2 & \vect 0_2 & \sqrt{1/2}\vect I_2  & \sqrt{1-1/2}\vect I_2 \\
     \vect 0_2 & \vect 0_2 & -\sqrt{1-1/2}\vect I_2  & \sqrt{1/2}\vect I_2 
   \end{array}\right).
   \nonumber
\end{eqnarray} 
and probe that $\vect L_{\text{QUMI}}$ for $N=4$ coincides again with the orthogonal matrix retrieved by (\ref{DLQUMI}). By repeating this procedure one may see that $\vect L_{\text{QUMI}}$ for arbitrary size $N$ takes the form provided by the expression (\ref{DLQUMI}), as we wanted to show.

By computing Eq.~\eqref{CVMSMI} once replaced\ \eqref{DLQUMI} for different small values of $N$, an induction procedure for greater $N$ reveals that 
\begin{equation}
\vect \Omega_{N}(\varphi,s_1,s_2)=\left( \begin{array}{cccc}
d_{1} & c_{1} & c_{2} & \frac{c_{3}}{s_{1}s_{2}} \\
c_{1} & d_{2} & c_{3} & -\frac{c_{2}}{s_{1}s_{2}} \\
c_{2} & c_{3} & d_{3} & 0 \\
\frac{c_{3}}{s_{1}s_{2}} & -\frac{c_{2}}{s_{1}s_{2}} & 0 & d_{4} \\
\end{array}\right) ,
\label{CVMSMII}
\end{equation}
whose diagonal entries are determined by
\begin{eqnarray}
d_{1}&=&\frac{1}{s_{1}s_{2}N}\big(s_{1}s_{2}a_{N}(s_{2},s_{1})\cos^2\varphi +a_{N}(s_{1},s_{2})\sin^2\varphi\big),   \\
d_{2}&=&\frac{1}{s_{1}s_{2}N}\big(a_{N}(s_{1},s_{2})\cos^2\varphi +s_{1}s_{2}a_{N}(s_{2},s_{1})\sin^2\varphi\big),   \\
d_{3}&=& \frac{a_{N}(s_{1},s_{2})}{N},    \\
d_{4}&=& \frac{a_{N}(s_{2},s_{1})}{s_{1}s_{2}N},
\end{eqnarray}
whereas the non-diagonal elements are given by,
\begin{eqnarray}
c_{1}&=& \frac{a_{N}(s_{1},s_{2})-s_{1}s_{2}a_{N}(s_{2},s_{1})}{2s_{1}s_{2}N}\sin(2\varphi),  \\
c_{2}&=& \frac{(s_{2}-s_{1})\sqrt{N-1}}{N}\cos\varphi,   \\
c_{3}&=& \frac{(s_{1}-s_{2})\sqrt{N-1}}{N}\sin\varphi.
\end{eqnarray}

On the other side, by replacing the generic form\ \eqref{CVMSMII} in \eqref{SXSQCS} and using results borrowed from matrix analysis to compute the Moore-Penrose pseudoinverse \cite{petersen20121}, one obtains the auxiliary matrix \eqref{SXSQCS} with
\begin{equation}
\vect A^{(x)}=\left( \begin{array}{cccc}
\frac{d_{3}}{d_{1}d_{3}-c_{2}^2} & 0 & -\frac{c_{2}}{d_{1}d_{3}-c_{2}^2} & 0 \\
0 & 0 & 0 & 0 \\
-\frac{c_{2}}{d_{1}d_{3}-c_{2}^2} & 0 & \frac{d_{1}}{d_{1}d_{3}-c_{2}^2} & 0 \\
0 & 0 & 0 & 0 \\
\end{array}\right) ,
\label{MSAX}
\end{equation}
whereas for the momentum quadrature measurement
\begin{equation}
\vect A^{(p)}=\left( \begin{array}{cccc}
0 & 0 & 0 & 0 \\
0 & \frac{d_{4}s_{1}^2s_{2}^2}{d_{2}d_{4}s_{1}^2s_{2}^2-c_{2}^2} & 0 & \frac{c_{2}s_{1}s_{2}}{d_{2}d_{4}s_{1}^2s_{2}^2-c_{2}^2} \\
0 & 0 & 0 & 0 \\
0 & \frac{c_{2}s_{1}s_{2}}{d_{2}d_{4}s_{1}^2s_{2}^2-c_{2}^2} & 0 & \frac{d_{2}s_{1}^2s_{2}^2}{d_{2}d_{4}s_{1}^2s_{2}^2-c_{2}^2} \\
\end{array}\right)  ,
\label{MSAP}
\end{equation}
which reduces to the expected results $\vect A^{(x)}=\text{diag}(1,0,1,0)$ or $\vect A^{(p)}=\text{diag}(0,1,0,1)$ when the initial squeezing vanishes \cite{wedbrook20111} (i.e. $s_{1}=s_{2}=1$).

Moreover, by substituting\ \eqref{CVMSMII} in \eqref{IFSSTPW}, one obtains the expression\ \eqref{FICHSXS} for the FI for position quadrature measurement after a long tedious calculation where we have introduced 
\begin{eqnarray}
\big(\vect W^{(x)}_{N}\big)_{11}&=&-\frac{i N s_{1} s_{2}}{2 N s_{1} s_{2} \cot (\varphi )+2 i ((N-1) s_{1}+s_{2})}+\frac{i N s_{1} s_{2} \sin (\varphi )}{2 N s_{1}
   s_{2} \cos (\varphi )-2 i \sin (\varphi ) ((N-1) s_{1}+s_{2})}+\frac{s_{1}-s_{2}}{N}+s_{2},\label{MWCS1}   \\ 
\big(\vect W^{(x)}_{N}\big)_{22}&=&\frac{((N-1) s_{1}+s_{2})^3}{N^3 s_{1}^3 s_{2}^3 \cot ^2(\varphi )+N s_{1} s_{2} ((N-1) s_{1}+s_{2})^2}  \\
\big(\vect W^{(x)}_{N}\big)_{12}&=&\big(\vect W^{(x)}_{N}\big)_{21}=\frac{N s_{1} s_{2} \sin (2 \varphi ) ((N-1) s_{1}+s_{2})}{2 N^2 s_{1}^2 s_{2}^2 \cos ^2(\varphi )+2 \sin ^2(\varphi ) ((N-1) s_{1}+s_{2})^2},\label{MWCS3}
\end{eqnarray}
or for the momentum quadrature measurement
\begin{eqnarray}
\big(\vect W^{(p)}_{N}\big)_{11}&=& \frac{((N-1) s_{2}+s_{1})^3}{N^3 \cot ^2(\varphi )+N ((N-1) s_{2}+s_{1})^2}  \\ 
\big(\vect W^{(p)}_{N}\big)_{22}&=& -\frac{N ((N-1) s_{2}+s_{1})}{N^2 \cot ^2(\varphi )+((N-1) s_{2}+s_{1})^2}+\frac{\frac{1}{s_{1}}-\frac{1}{s_{2}}}{N}+\frac{1}{s_{2}},    \\ 
\big(\vect W^{(p)}_{N}\big)_{12}&=&\big(\vect W^{(p)}_{N}\big)_{21}=-\frac{N \sin (2 \varphi ) ((N-1) s_{2}+s_{1})}{2 \left(N^2 \cos ^2(\varphi )+\sin ^2(\varphi ) ((N-1) s_{2}+s_{1})^2\right)},
\end{eqnarray}
as well as, the auxiliary functions determining the influence of the second-moment resources,
\begin{eqnarray}
f^{(x)}_{N}(\sin^2\varphi,s_{1},s_{2})&=& \Big((N-1) s_{1} s_{2}^5 \left(N^2 s_{1}^2+1\right)+2 s_{2}^4 \left(s_{1}^2 \left(N^2 ((N-1) N+1) s_{1}^2-N
   (N+3)+2\right)+1\right) \label{FFXCS} \\
   &-&2 (N-1)^2 s_{1}^2 s_{2}^2 \left(\left(2 N^2+N-2\right) s_{1}^2-6\right)+2 (N-1)^4 s_{1}^4 \nonumber \\
   &+&(N-1) s_{1} s_{2}^3 \left(s_{1}^2 \left(N \left(N \left(s_{1}^2-7\right)-6\right)+6\right)+8\right) \nonumber \\
   &+&\cos (2 \varphi ) (N s_{1} (s_{2}-1)+s_{1}-s_{2}) \big(2 s_{2}^2 \left(((N-1) N+1) s_{1}^2-1\right)+(N-1) s_{1} \left(s_{1}^2-4\right) s_{2}\nonumber \\
   &-&2 (N-1)^2 s_{1}^2+(N-1) s_{1} s_{2}^3\big)
   (s_{1} (N s_{2}+N-1)+s_{2})+(N-1)^3 s_{1}^3 \left(s_{1}^2+8\right) s_{2}\Big)\frac{\sin^{2}\varphi  }{2y_{x}^2(\varphi)},
\nonumber
\end{eqnarray}
or
\begin{eqnarray}
f^{(p)}_{N}(\sin^2\varphi,s_{1},s_{2})&=& \Big(2 N^4 s_{1} s_{2} \left(s_{2}^2-1\right)^2+N^3 (s_{1}-s_{2}) \left(8 s_{1} s_{2}^4-7 s_{1} s_{2}^2+s_{1}-s_{2}^3-s_{2}\right)
\nonumber \\
&+&\cos (2 \varphi )(N (s_{2}-1)+s_{1}-s_{2}) ((N-1) s_{2}+N+s_{1}) \big(2 s_{1} s_{2} \left(-N^2+N+s_{1}^2-1\right)\nonumber \\
&+&(N-1) \left(4 s_{1}^2-1\right) s_{2}^2+(1-N) s_{1}^2+2 (N-1)^2 s_{1} s_{2}^3\big)+N^2 (s_{1}-s_{2})^2 \left(12 s_{1} s_{2}^3-2 s_{1} s_{2}-3 s_{2}^2-1\right) \nonumber \\
&+&N (s_{1}-s_{2})^3 \left(8 s_{1} s_{2}^2+s_{1}-3 s_{2}\right)+(s_{1}-s_{2})^4 (2 s_{1} s_{2}-1)\Big)\frac{\sin^{2}\varphi  }{2s_{1}s_{2}y_{p}^2(\varphi)},
\label{FFPCS}
\end{eqnarray}
with $y_{x}(\varphi)=(N s_{1}s_{2})^2\cos^2 \varphi+a_{N}^2(s_{1},s_{2})\sin^{2}\varphi$ and $y_{p}(\varphi)=(N^2\cos^2 \varphi+a_{N}^2(s_{2},s_{1})\sin^{2}\varphi$.

When addressing the optimal working point, as stated in the main text (by demanding the second and third terms in the right-hand side of Eq.~\eqref{FICHSXS} cancel) we find out the second-order polynomial\ \eqref{QCRBC} with real coefficients given by,
\begin{eqnarray}
\alpha_{N}^{(x)}(s_{1},s_{2}) &=&-2 N^2 s_{1}^2 s_{2}^2 ((N-1) s_{1}+s_{2})^2 \left(2 N^2-\left(\frac{N-1}{s_{2}}+\frac{1}{s_{1}}\right)^2-((N-1)
   s_{2}+s_{1})^2\right) \nonumber \\
   &-&2 N^2 (N s_{1} (s_{2}-1)+s_{1}-s_{2}) \left(2 s_{2}^2 \left(((N-1) N+1) s_{1}^2-1\right)+(N-1) s_{1}\big(s_{1}^2-4\right) s_{2}-2 (N-1)^2 s_{1}^2 \nonumber \\
   &+&(N-1) s_{1} s_{2}^3\big) (s_{1} (N s_{2}+N-1)+s_{2})+((N-1) s_{1}+s_{2})^4 \left(2 N^2-\left(\frac{N-1}{s_{2}}+\frac{1}{s_{1}}\right)^2-((N-1) s_{2}+s_{1})^2\right) \nonumber \\
   &+&N^4 s_{1}^4 s_{2}^4 \left(2
   N^2-\left(\frac{N-1}{s_{2}}+\frac{1}{s_{1}}\right)^2-((N-1) s_{2}+s_{1})^2\right), \label{PSCCS1}  \\
\beta_{N}^{(x)} (s_{1},s_{2}) &=&2 N^2 \big((N-1)^2 s_{1}^2 s_{2}^6 \left(N^2 s_{1}^2-1\right)-(N-1) s_{1} s_{2}^3 \left(2 N^2 s_{1}^2+(N-2)^2
   s_{1}^4-4\right) \nonumber \\
   &-&(N-1)^2 s_{1}^2 s_{2}^2 \left(N^2 s_{1}^2+s_{1}^4-6\right)+s_{2}^4 \big(N^2 s_{1}^6-N^2 s_{1}^2-(N (N ((N-2) N+8)-12)+6) s_{1}^4+1\big)\nonumber \\
   &+&(N-1)^4 s_{1}^4+4 (N-1)^3 s_{1}^3 s_{2}+(N-1) s_{1}^3 s_{2}^5 \left(N \left(N \left(2 s_{1}^2-1\right)+4\right)-4\right)\big),  \\
\delta_{N}^{(x)} (s_{1},s_{2})&=&N^4 s_{1}^4 s_{2}^4 \left(2 N^2-\left(\frac{N-1}{s_{2}}+\frac{1}{s_{1}}\right)^2-((N-1) s_{2}+s_{1})^2\right).
\label{PSCCS3}
\end{eqnarray}

\subsection{Explicit expressions from Sec.\ref{PSNDR}}

Now we turn the attention to the polychromatic scenario described in Sect.\ \ref{PSNDR}. As stated in the discussion about the Fisher information, after some manipulation one can show that for the choices $\epsilon=\pm1$ the expression \eqref{FICHSPpol} boils down to 
\begin{eqnarray}
F_{\text{pol}}^{(x)}(\varphi,\epsilon)&=&\mathcal{F}_{\text{pol}}(\epsilon)-\frac{2\big(\text{sign}(\epsilon)\sinh^2(2s')+\cos(4\varphi)\sinh(4s')\big)^2}{\big(\cosh(2s')+\text{sign}(\epsilon)\cos(4\varphi)\sinh(2s')\big)^2},
\label{FPOLNX}
\end{eqnarray}
for a measurement quadrature in position, or 
\begin{eqnarray}
F_{\text{pol}}^{(p)}(\varphi,\epsilon)&=&\mathcal{F}_{\text{pol}}(\epsilon)-\frac{2\big(\text{sign}(\epsilon)\sinh^2(2s')-\cos(4\varphi)\sinh(4s')\big)^2}{\big(\cosh(2s')-\text{sign}(\epsilon)\cos(4\varphi)\sinh(2s')\big)^2},
\label{FPOLNP}
\end{eqnarray}
for a measurement quadrature in momentum. By paying attention to Eqs.\ \eqref{FPOLNX} and \eqref{FPOLNP}, it is clear that the optimal operating point is obtained by demanding the numerator of the second term in the right-hand side cancels. Upon doing this, one arrives to the relation determining the optimal angle for the choice $\epsilon=\pm 1$. For the most general case of modulation frequency (i.e. $\epsilon\neq \pm1$), one obtains the following subsidiary condition from a perturbative analysis 
\begin{eqnarray}
(3 + \cosh(4 s') &-& 2 \cos(4 \phi) \sinh^2(2 s'))f_{0}^{(x)}\big(\cos(4\phi),s'\big) +\epsilon f_{1}^{(x)}(\phi,s')\approx 0,
\label{QCRBTSI}
\end{eqnarray}
where we have introduced the auxiliary functions
\begin{eqnarray}
f_{0}^{(x)}\big(\cos(4\varphi),s'\big)&=& \sinh ^2(2 r) \left(-2 \sinh ^2(2 r) \cos (4 \varphi )+\cosh (4 r)+3\right) \big(2 \sinh ^2(4 r) \left(2 \cos ^2(4 \varphi )-1\right) \nonumber \\
&-&4 (\cosh (8 r)-9) \cos (4 \varphi )+3 \cosh (8 r)+29\big), \label{FEPS0} \\
f_{1}^{(x)}(\varphi,s')&=&16 \sinh ^3(2 r) \Big(2 \varphi  \sinh (2 r) \sinh (4 r) \sin (10 \varphi )+\cosh (6 r) (-14 \varphi  \sin (2 \varphi )+3 \varphi  \sin (6 \varphi )-3 \cos (6 \varphi )) \nonumber \\
&+&\cosh (2 r) \big(4 (\cosh (4 r)+3) \cos (2 \varphi )+16\sinh ^2(r) \cosh ^2(r) \cos (10 \varphi )-2 \varphi  \sin (2 \varphi )+45 \varphi  \sin (6 \varphi ) \nonumber \\
&-&13 \cos (6 \varphi )\big)\Big). \label{FEPS1}
\end{eqnarray}
Clearly, from Eq.~\eqref{QCRBTSI} follows that in the particular case $\epsilon=0$ the optimal operating point $\varphi_{\text{opt}}^{(x)}$ is figure out from solving the second-order polynomial $f_{0}^{(x)}\big(y,s'\big)=0$ with argument understood as $y=\cos(4\varphi)$. Doing this, one directly obtains
\begin{equation}
y= \frac{1}{2} \left(\cosh (8 r)\pm4 \sqrt{6-2 \cosh (8 r)}-9\right) \text{csch}^2(4 r),
\label{FPOLROOTS}
\end{equation}
which is greater than the unit except for the choice of coherent resources $s'=0$ when one of the roots becomes $x\rightarrow -1$, retrieving in turn the same result $\varphi_{\text{opt}}^{(x)}=-\pi/4$ as previously obtained in Sect.\ \ref{SQM}, as expected.  

\subsection{Explicit expressions from Sec.\ref{SDEFED}}\label{app4}

In this appendix, we briefly illustrate the derivation of Eqs.\ \eqref{GGMSPLID1}, \eqref{GGMSPLID2}  and \eqref{TFIDISS2} appearing in Sect.\ \ref{SDEFED}. Firts, the nonunit efficiency of a single-mode homodyne measurement mainly resides in the use of photon-detectors suffering from a limited resolution $\eta_{\text{eff}}\in \left[0,1 \right]$, which results in a vacuum noise contribution proportional to $\sqrt{1-\eta_{\text{eff}}}$ in the measurement outcomes, i.e.
\begin{equation}
\left[ \vect \lambda\right] =\frac{1}{2}\Bigg( \sqrt{\frac{1+r}{2}}q^{f},\sqrt{\frac{1+r}{2r}}p^{f}\Bigg)+\sqrt{1-\eta_{\text{eff}}}(q^{\text{vac}},p^{\text{vac}}).
\end{equation}
Without loss of generality, this source of noise may be well approximated by the combination of an ideal Gaussian detector (described by the CV matrix $\vect \Sigma$) preceding by a beam splitter with transmission coefficient identical to the photon-detector resolution factor, where the probe mode would fictitiously interfere with an input vacuum beam representing $(\hat q^{\text{vac}},\hat p^{\text{vac}})$. In our framework, this corresponds to take the CV matrix determining the non-ideal Gaussian measurement scheme as
\begin{equation}
\vect \Sigma=\eta_{\text{\text{eff}}}\vect \Sigma+(1-\eta_{\text{\text{eff}}})\vect I_{N},
\label{CVMMEI}
\end{equation}
which returns the lossless homodyne detection scenario for $\eta_{\text{eff}}=1$. On the other side, decoherence effects of the probe $N$-mode system taking place during the light field propagation through the interferometer can be formulated in terms of the interaction with an environment modelled by a continuum  of oscillators \cite{serafini20051}. When the system-environment interaction is essentially linear, the time evolution of our probe $N$-mode system is governed by the Fokker-Plank (or diffusion) equation expressed in the interaction picture \cite{valido20141},
\begin{equation}
\frac{\partial W(\vect R,t )}{\partial t}=\left(\left(\frac{\partial}{\partial \vect R} \right)^{T}\vect \Gamma \vect R +\left(\frac{\partial}{\partial \vect R} \right)^{T}\vect D\left(\frac{\partial}{\partial \vect R} \right) \right) W(\vect R,t ),
\label{F-PE}
\end{equation}
with $\left(\frac{\partial}{\partial \vect R} \right)^{T}=\bigoplus_{i=1}^{N}\left(\frac{\partial}{\partial q_{i}},\frac{\partial}{\partial p_{i}}\right) $; $\vect \Gamma $ and $\vect D$ are $2N\times 2N$ real, symmetric matrices that essentially encrypt the photon-losses and thermal noise effects, respectively. In the interesting dissipative scenario the above matrices take the following simple form 
\begin{eqnarray}
\vect \Gamma&=& \frac{\gamma}{2} \bigoplus_{i=1}^{N}\vect I_{1},  \\
\vect D & =&\frac{\gamma (1 + 2n_{\text{th}} )}{4}\bigoplus_{i=1}^{N}\vect I_{1},
\end{eqnarray}
where $\gamma$ is the usual dissipative coefficient. Equation\ \eqref{F-PE} is a linear Fokker-Plank equation that can be solved by using the Green function method \cite{valido20141}. Furthermore, thanks to the diagonal form of the above dissipative and noise matrices, the decoherence evolution commutes with the phase shift rotation \cite{jarzyna20171,oh20181,oh20171}, and we obtain the results \eqref{GGMSPLID1} and \eqref{GGMSPLID2}. Substituting these in Eq.~\eqref{IFSSTPW}, and following a similar procedure as to compute the expression \eqref{CFI2}, we obtain
\begin{eqnarray}
F_{\text{deco}}(\varphi,\eta_{\text{loss}},\eta_{\text{eff}},n_{\text{th}})&=&F(\varphi,\eta_{\text{loss}},\eta_{\text{eff}})-\eta_{\text{loss}}\left\langle \vect R^{T}\right\rangle \vect L^{T}\vect P_{\varphi}^{T}\vect \Sigma_{\text{deco}}^{-1}\vect P_{\varphi}\vect L\left\langle \vect R\right\rangle 
+\frac{1}{2}\text{Tr}\big(\partial_{\varphi}\vect \Sigma_{\text{deco}}^{-1}\partial_{\varphi}\vect \sigma_{\text{deco}}\big),\label{TFIDISS}
\end{eqnarray}
with 
\begin{equation}
\vect \Sigma_{\text{deco}}(\eta_{\text{loss}},\eta_{\text{eff}},n_{\text{th}})=\vect \sigma_{\text{deco}} \Big(\big(1-\eta_{\text{eff}}+(1-\eta_{\text{loss}})(1+n_{\text{th}})\big)^{-1}\vect I_{N} +\vect \sigma_{\text{deco}}^{-1}\Big)\vect \sigma_{\text{deco}},
\label{SDISS}
\end{equation}
where $F(\varphi,\eta_{\text{loss}},\eta_{\text{eff}})$ comprises the Fisher information obtained from the noiseless expression\ \eqref{IFSSTPW} after substituting $\vect S(\varphi)\rightarrow \sqrt{\eta_{\text{loss}}}\vect S(\varphi)$, and $\vect \Sigma\rightarrow \eta_{\text{eff}}\vect \Sigma$. It is worthwhile to realize that the second contribution in the right-hand side of \eqref{TFIDISS} will be always negative for any $\varphi\in \mathbb{R}$ and $\left\langle \vect R\right\rangle\in \mathbb{R}^{2N}$, since the CV matrix $\vect \Sigma_{\text{deco}}^{-1}$ is positive-semidefinite by construction, and further, it asymptotically cancels in the limit of an ideal phase-estimation scenario (i.e. $F_{\text{deco}}\rightarrow F$ when $\eta_{\text{eff}}, \eta_{\text{loss}}\rightarrow 1$), as expected. Notice that the corresponding QFI is formally obtained from Eq.\ \eqref{QFIGGS} (with $n_{t}=0$) by replacing $\left\langle \vect R_{1}'\right\rangle\rightarrow\sqrt{\eta_{\text{loss}}}\vect P_{0}\vect L\left\langle \vect R\right\rangle$, and $\vect V_{1}'\rightarrow \vect P_{0}\big( \eta_{\text{loss}}\vect L\vect V\vect L^{T}+(1- \eta_{\text{loss}})(1+n_{\text{th}})\vect I_{N}\big)\vect P_{0}^{T}$.

In particular, in the dissipative scenario $\eta_{\text{eff}}=\eta_{\text{loss}}=\eta$, we find the FI for the previously-studied coherent resources and homdyne detection, i.e.
\begin{equation}
F_{\text{deco}}^{(x/p)}(\varphi,\eta,n_{th})=2\bar{n}_{c}N(1\mp\sin(2\varphi))\Bigg(\eta^2+\frac{(2+n_{\text{th}})(1-\eta)}{\eta+(1-\eta)n_{\text{th}}-2}\Bigg)=2\tilde{\eta}^2\bar{n}_{c}N(1\mp\sin(2\varphi))\label{FIDISS}.
\end{equation}
By comparing with the ideal result (\ref{PSACQ}) for $\mathcal{F}$, it is clear from the above equation\ \eqref{FIDISS} that the "optimal" working point defined in Sect.\ \ref{SDEFED} is obtained from demanding $\sin(2\varphi^{(x/p)})=\mp(2(\eta/\tilde{\eta})^2-1)$, which returns a result that substantially differs from the ideal case (i.e. $\varphi=\mp\pi/4$).

\end{widetext}
\bibliographystyle{apsrev4-1}
\bibliography{references}

\begin{thebibliography}{89}%
\makeatletter
\providecommand \@ifxundefined [1]{%
 \@ifx{#1\undefined}
}%
\providecommand \@ifnum [1]{%
 \ifnum #1\expandafter \@firstoftwo
 \else \expandafter \@secondoftwo
 \fi
}%
\providecommand \@ifx [1]{%
 \ifx #1\expandafter \@firstoftwo
 \else \expandafter \@secondoftwo
 \fi
}%
\providecommand \natexlab [1]{#1}%
\providecommand \enquote  [1]{``#1''}%
\providecommand \bibnamefont  [1]{#1}%
\providecommand \bibfnamefont [1]{#1}%
\providecommand \citenamefont [1]{#1}%
\providecommand \href@noop [0]{\@secondoftwo}%
\providecommand \href [0]{\begingroup \@sanitize@url \@href}%
\providecommand \@href[1]{\@@startlink{#1}\@@href}%
\providecommand \@@href[1]{\endgroup#1\@@endlink}%
\providecommand \@sanitize@url [0]{\catcode `\\12\catcode `\$12\catcode
  `\&12\catcode `\#12\catcode `\^12\catcode `\_12\catcode `\%12\relax}%
\providecommand \@@startlink[1]{}%
\providecommand \@@endlink[0]{}%
\providecommand \url  [0]{\begingroup\@sanitize@url \@url }%
\providecommand \@url [1]{\endgroup\@href {#1}{\urlprefix }}%
\providecommand \urlprefix  [0]{URL }%
\providecommand \Eprint [0]{\href }%
\providecommand \doibase [0]{http://dx.doi.org/}%
\providecommand \selectlanguage [0]{\@gobble}%
\providecommand \bibinfo  [0]{\@secondoftwo}%
\providecommand \bibfield  [0]{\@secondoftwo}%
\providecommand \translation [1]{[#1]}%
\providecommand \BibitemOpen [0]{}%
\providecommand \bibitemStop [0]{}%
\providecommand \bibitemNoStop [0]{.\EOS\space}%
\providecommand \EOS [0]{\spacefactor3000\relax}%
\providecommand \BibitemShut  [1]{\csname bibitem#1\endcsname}%
\let\auto@bib@innerbib\@empty
\bibitem [{\citenamefont {Serafini}(2017)}]{serafini20171}%
  \BibitemOpen
  \bibfield  {author} {\bibinfo {author} {\bibfnamefont {A.}~\bibnamefont
  {Serafini}},\ }\href@noop {} {\emph {\bibinfo {title} {Quantum Continuous
  Variables: A primer of theoretical methods}}}\ (\bibinfo  {publisher} {CRC
  Press},\ \bibinfo {year} {2017})\BibitemShut {NoStop}%
\bibitem [{\citenamefont {Jiang}(2014)}]{jiang20141}%
  \BibitemOpen
  \bibfield  {author} {\bibinfo {author} {\bibfnamefont {Z.}~\bibnamefont
  {Jiang}},\ }\href {\doibase 10.1103/PhysRevA.89.032128} {\bibfield  {journal}
  {\bibinfo  {journal} {Phys. Rev. A}\ }\textbf {\bibinfo {volume} {89}},\
  \bibinfo {pages} {032128} (\bibinfo {year} {2014})}\BibitemShut {NoStop}%
\bibitem [{\citenamefont {{\v{S}}afr{\'{a}}nek}\ \emph
  {et~al.}(2015)\citenamefont {{\v{S}}afr{\'{a}}nek}, \citenamefont {Lee},\
  and\ \citenamefont {Fuentes}}]{safranek20151}%
  \BibitemOpen
  \bibfield  {author} {\bibinfo {author} {\bibfnamefont {D.}~\bibnamefont
  {{\v{S}}afr{\'{a}}nek}}, \bibinfo {author} {\bibfnamefont {A.~R.}\
  \bibnamefont {Lee}}, \ and\ \bibinfo {author} {\bibfnamefont
  {I.}~\bibnamefont {Fuentes}},\ }\href {\doibase
  10.1088/1367-2630/17/7/073016} {\bibfield  {journal} {\bibinfo  {journal}
  {New J. Phys.}\ }\textbf {\bibinfo {volume} {17}},\ \bibinfo {pages} {073016}
  (\bibinfo {year} {2015})}\BibitemShut {NoStop}%
\bibitem [{\citenamefont {Gao}\ and\ \citenamefont {Lee}(2014)}]{gao20141}%
  \BibitemOpen
  \bibfield  {author} {\bibinfo {author} {\bibfnamefont {Y.}~\bibnamefont
  {Gao}}\ and\ \bibinfo {author} {\bibfnamefont {H.}~\bibnamefont {Lee}},\
  }\href {\doibase 10.1140/epjd/e2014-50560-1} {\bibfield  {journal} {\bibinfo
  {journal} {Eur. Phys. J. D}\ }\textbf {\bibinfo {volume} {68}},\ \bibinfo
  {pages} {347} (\bibinfo {year} {2014})}\BibitemShut {NoStop}%
\bibitem [{\citenamefont {Monras}()}]{monras20131}%
  \BibitemOpen
  \bibfield  {author} {\bibinfo {author} {\bibfnamefont {A.}~\bibnamefont
  {Monras}},\ }\href@noop {} {\bibinfo  {journal} {arXiv:1303.3682v1}\
  }\BibitemShut {NoStop}%
\bibitem [{\citenamefont {Giovannetti}\ \emph {et~al.}(2004)\citenamefont
  {Giovannetti}, \citenamefont {Lloyd},\ and\ \citenamefont
  {Maccone}}]{giovannetti20041}%
  \BibitemOpen
\bibfield  {journal} {  }\bibfield  {author} {\bibinfo {author} {\bibfnamefont
  {V.}~\bibnamefont {Giovannetti}}, \bibinfo {author} {\bibfnamefont
  {S.}~\bibnamefont {Lloyd}}, \ and\ \bibinfo {author} {\bibfnamefont
  {L.}~\bibnamefont {Maccone}},\ }\href {\doibase 10.1126/science.1104149}
  {\bibfield  {journal} {\bibinfo  {journal} {Science}\ }\textbf {\bibinfo
  {volume} {306}},\ \bibinfo {pages} {1330} (\bibinfo {year}
  {2004})}\BibitemShut {NoStop}%
\bibitem [{\citenamefont {Giovannetti}\ \emph {et~al.}(2011)\citenamefont
  {Giovannetti}, \citenamefont {Lloyd},\ and\ \citenamefont
  {Maccone}}]{giovannetti20111}%
  \BibitemOpen
  \bibfield  {author} {\bibinfo {author} {\bibfnamefont {V.}~\bibnamefont
  {Giovannetti}}, \bibinfo {author} {\bibfnamefont {S.}~\bibnamefont {Lloyd}},
  \ and\ \bibinfo {author} {\bibfnamefont {L.}~\bibnamefont {Maccone}},\ }\href
  {\doibase 10.1038/nphoton.2011.35} {\bibfield  {journal} {\bibinfo  {journal}
  {Nature Photonics}\ }\textbf {\bibinfo {volume} {5}},\ \bibinfo {pages} {222}
  (\bibinfo {year} {2011})}\BibitemShut {NoStop}%
\bibitem [{\citenamefont {Paris}(2009)}]{paris20091}%
  \BibitemOpen
  \bibfield  {author} {\bibinfo {author} {\bibfnamefont {M.~G.~A.}\
  \bibnamefont {Paris}},\ }\href {\doibase 10.1142/S0219749909004839}
  {\bibfield  {journal} {\bibinfo  {journal} {Int. J. Quant. Inf.}\ }\textbf
  {\bibinfo {volume} {7}},\ \bibinfo {pages} {125} (\bibinfo {year}
  {2009})}\BibitemShut {NoStop}%
\bibitem [{\citenamefont {T{\'{o}}th}\ and\ \citenamefont
  {Apellaniz}(2014)}]{toth20141}%
  \BibitemOpen
  \bibfield  {author} {\bibinfo {author} {\bibfnamefont {G.}~\bibnamefont
  {T{\'{o}}th}}\ and\ \bibinfo {author} {\bibfnamefont {I.}~\bibnamefont
  {Apellaniz}},\ }\href {\doibase 10.1088/1751-8113/47/42/424006} {\bibfield
  {journal} {\bibinfo  {journal} {J. Phys. A: Math Theor.}\ }\textbf {\bibinfo
  {volume} {47}},\ \bibinfo {pages} {424006} (\bibinfo {year}
  {2014})}\BibitemShut {NoStop}%
\bibitem [{\citenamefont {Dowling}(2008)}]{dowling20081}%
  \BibitemOpen
  \bibfield  {author} {\bibinfo {author} {\bibfnamefont {J.~P.}\ \bibnamefont
  {Dowling}},\ }\href {\doibase 10.1080/00107510802091298} {\bibfield
  {journal} {\bibinfo  {journal} {Comtemporary Physics}\ }\textbf {\bibinfo
  {volume} {49}},\ \bibinfo {pages} {125} (\bibinfo {year} {2008})}\BibitemShut
  {NoStop}%
\bibitem [{\citenamefont {Braun}\ \emph {et~al.}(2018)\citenamefont {Braun},
  \citenamefont {Adesso}, \citenamefont {Benatti}, \citenamefont {Floreanini},
  \citenamefont {Marzolino}, \citenamefont {Mitchell},\ and\ \citenamefont
  {Pirandola}}]{braun20181}%
  \BibitemOpen
  \bibfield  {author} {\bibinfo {author} {\bibfnamefont {D.}~\bibnamefont
  {Braun}}, \bibinfo {author} {\bibfnamefont {G.}~\bibnamefont {Adesso}},
  \bibinfo {author} {\bibfnamefont {F.}~\bibnamefont {Benatti}}, \bibinfo
  {author} {\bibfnamefont {R.}~\bibnamefont {Floreanini}}, \bibinfo {author}
  {\bibfnamefont {U.}~\bibnamefont {Marzolino}}, \bibinfo {author}
  {\bibfnamefont {M.~W.}\ \bibnamefont {Mitchell}}, \ and\ \bibinfo {author}
  {\bibfnamefont {S.}~\bibnamefont {Pirandola}},\ }\href {\doibase
  10.1103/RevModPhys.90.035006} {\bibfield  {journal} {\bibinfo  {journal}
  {Rev. Mod. Phys.}\ }\textbf {\bibinfo {volume} {10}},\ \bibinfo {pages}
  {035006} (\bibinfo {year} {2018})}\BibitemShut {NoStop}%
\bibitem [{\citenamefont {Pirandola}\ \emph {et~al.}(2018)\citenamefont
  {Pirandola}, \citenamefont {Bardhan}, \citenamefont {Gehring}, \citenamefont
  {Weedbrook},\ and\ \citenamefont {Lloyd}}]{pirandola20181}%
  \BibitemOpen
  \bibfield  {author} {\bibinfo {author} {\bibfnamefont {S.}~\bibnamefont
  {Pirandola}}, \bibinfo {author} {\bibfnamefont {B.~R.}\ \bibnamefont
  {Bardhan}}, \bibinfo {author} {\bibfnamefont {T.}~\bibnamefont {Gehring}},
  \bibinfo {author} {\bibfnamefont {C.}~\bibnamefont {Weedbrook}}, \ and\
  \bibinfo {author} {\bibfnamefont {S.}~\bibnamefont {Lloyd}},\ }\href
  {\doibase 10.1038/s41566-018-0301-6} {\bibfield  {journal} {\bibinfo
  {journal} {Nature Photonics}\ }\textbf {\bibinfo {volume} {12}},\ \bibinfo
  {pages} {724} (\bibinfo {year} {2018})}\BibitemShut {NoStop}%
\bibitem [{\citenamefont {Lee}\ \emph {et~al.}(2002)\citenamefont {Lee},
  \citenamefont {Kok},\ and\ \citenamefont {Dowling}}]{lee20021}%
  \BibitemOpen
  \bibfield  {author} {\bibinfo {author} {\bibfnamefont {H.}~\bibnamefont
  {Lee}}, \bibinfo {author} {\bibfnamefont {P.}~\bibnamefont {Kok}}, \ and\
  \bibinfo {author} {\bibfnamefont {J.~P.}\ \bibnamefont {Dowling}},\ }\href
  {\doibase 10.1080/0950034021000011536} {\bibfield  {journal} {\bibinfo
  {journal} {J. Mod. Opt.}\ }\textbf {\bibinfo {volume} {49}},\ \bibinfo
  {pages} {2325} (\bibinfo {year} {2002})}\BibitemShut {NoStop}%
\bibitem [{\citenamefont {Demkowicz-Dobrza{\'{n}}ski}\ \emph
  {et~al.}(2015)\citenamefont {Demkowicz-Dobrza{\'{n}}ski}, \citenamefont
  {Jarzyna},\ and\ \citenamefont {Ko{\l}ody{\'{n}}ski}}]{demkowicz20151}%
  \BibitemOpen
  \bibfield  {author} {\bibinfo {author} {\bibfnamefont {R.}~\bibnamefont
  {Demkowicz-Dobrza{\'{n}}ski}}, \bibinfo {author} {\bibfnamefont
  {M.}~\bibnamefont {Jarzyna}}, \ and\ \bibinfo {author} {\bibfnamefont
  {J.}~\bibnamefont {Ko{\l}ody{\'{n}}ski}},\ }\href {\doibase
  10.1016/bs.po.2015.02.003} {\bibfield  {journal} {\bibinfo  {journal}
  {Progress in Optics}\ }\textbf {\bibinfo {volume} {60}},\ \bibinfo {pages}
  {345} (\bibinfo {year} {2015})}\BibitemShut {NoStop}%
\bibitem [{\citenamefont {Sidhu}\ and\ \citenamefont {Kok}(2014)}]{sidhu20191}%
  \BibitemOpen
  \bibfield  {author} {\bibinfo {author} {\bibfnamefont {J.~S.}\ \bibnamefont
  {Sidhu}}\ and\ \bibinfo {author} {\bibfnamefont {P.}~\bibnamefont {Kok}},\
  }\href {\doibase 10.1116/1.5119961} {\bibfield  {journal} {\bibinfo
  {journal} {AVS Quantum Sci.}\ }\textbf {\bibinfo {volume} {2}},\ \bibinfo
  {pages} {014701} (\bibinfo {year} {2014})}\BibitemShut {NoStop}%
\bibitem [{\citenamefont {Ataman}(2020{\natexlab{a}})}]{ataman20191}%
  \BibitemOpen
  \bibfield  {author} {\bibinfo {author} {\bibfnamefont {S.}~\bibnamefont
  {Ataman}},\ }\href {\doibase 10.1103/PhysRevA.102.013704} {\bibfield
  {journal} {\bibinfo  {journal} {Phys. Rev. A}\ }\textbf {\bibinfo {volume}
  {102}},\ \bibinfo {pages} {013704} (\bibinfo {year}
  {2020}{\natexlab{a}})}\BibitemShut {NoStop}%
\bibitem [{\citenamefont {Ataman}(2020{\natexlab{b}})}]{ataman20201}%
  \BibitemOpen
  \bibfield  {author} {\bibinfo {author} {\bibfnamefont {S.}~\bibnamefont
  {Ataman}},\ }\href {\doibase 10.1103/PhysRevA.102.013704} {\bibfield
  {journal} {\bibinfo  {journal} {Phys. Rev. A}\ }\textbf {\bibinfo {volume}
  {102}},\ \bibinfo {pages} {013704} (\bibinfo {year}
  {2020}{\natexlab{b}})}\BibitemShut {NoStop}%
\bibitem [{\citenamefont {Gessner}\ \emph {et~al.}(2020)\citenamefont
  {Gessner}, \citenamefont {Smerzi},\ and\ \citenamefont
  {Pezz{\`{e}}}}]{gessner20201}%
  \BibitemOpen
  \bibfield  {author} {\bibinfo {author} {\bibfnamefont {M.}~\bibnamefont
  {Gessner}}, \bibinfo {author} {\bibfnamefont {A.}~\bibnamefont {Smerzi}}, \
  and\ \bibinfo {author} {\bibfnamefont {L.}~\bibnamefont {Pezz{\`{e}}}},\
  }\href {\doibase 10.1038/s41467-020-17471-3} {\bibfield  {journal} {\bibinfo
  {journal} {Nat. Commun.}\ }\textbf {\bibinfo {volume} {11}},\ \bibinfo
  {pages} {3817} (\bibinfo {year} {2020})}\BibitemShut {NoStop}%
\bibitem [{\citenamefont {Braunstein}\ and\ \citenamefont
  {Caves}(1994)}]{braunstein19941}%
  \BibitemOpen
  \bibfield  {author} {\bibinfo {author} {\bibfnamefont {S.~L.}\ \bibnamefont
  {Braunstein}}\ and\ \bibinfo {author} {\bibfnamefont {C.~M.}\ \bibnamefont
  {Caves}},\ }\href {\doibase 10.1103/PhysRevLett.72.3439} {\bibfield
  {journal} {\bibinfo  {journal} {Phys. Rev. Lett.}\ }\textbf {\bibinfo
  {volume} {72}},\ \bibinfo {pages} {3439} (\bibinfo {year}
  {1994})}\BibitemShut {NoStop}%
\bibitem [{\citenamefont {{\v{S}}afr{\'{a}}nek}(2018)}]{safranek20181}%
  \BibitemOpen
  \bibfield  {author} {\bibinfo {author} {\bibfnamefont {D.}~\bibnamefont
  {{\v{S}}afr{\'{a}}nek}},\ }\href {\doibase 10.1088/1751-8121/aaf068
  Manuscript} {\bibfield  {journal} {\bibinfo  {journal} {J. Phys. A: Math.
  Theor.}\ }\textbf {\bibinfo {volume} {52}},\ \bibinfo {pages} {035304}
  (\bibinfo {year} {2018})}\BibitemShut {NoStop}%
\bibitem [{\citenamefont {Gard}\ \emph {et~al.}(2017)\citenamefont {Gard},
  \citenamefont {You}, \citenamefont {Mishra}, \citenamefont {Singh},
  \citenamefont {Lee}, \citenamefont {Corbitt},\ and\ \citenamefont
  {Dowling}}]{gard20171}%
  \BibitemOpen
  \bibfield  {author} {\bibinfo {author} {\bibfnamefont {B.~T.}\ \bibnamefont
  {Gard}}, \bibinfo {author} {\bibfnamefont {C.}~\bibnamefont {You}}, \bibinfo
  {author} {\bibfnamefont {D.~K.}\ \bibnamefont {Mishra}}, \bibinfo {author}
  {\bibfnamefont {R.}~\bibnamefont {Singh}}, \bibinfo {author} {\bibfnamefont
  {H.}~\bibnamefont {Lee}}, \bibinfo {author} {\bibfnamefont {T.~R.}\
  \bibnamefont {Corbitt}}, \ and\ \bibinfo {author} {\bibfnamefont {J.~P.}\
  \bibnamefont {Dowling}},\ }\href {\doibase 10.1140/epjqt/s40507-017-0058-8}
  {\bibfield  {journal} {\bibinfo  {journal} {EPJ Quantum Technol.}\ }\textbf
  {\bibinfo {volume} {4}},\ \bibinfo {pages} {4} (\bibinfo {year}
  {2017})}\BibitemShut {NoStop}%
\bibitem [{\citenamefont {Oh}\ \emph {et~al.}(2017)\citenamefont {Oh},
  \citenamefont {Lee}, \citenamefont {Nha},\ and\ \citenamefont
  {Jeong}}]{oh20171}%
  \BibitemOpen
  \bibfield  {author} {\bibinfo {author} {\bibfnamefont {C.}~\bibnamefont
  {Oh}}, \bibinfo {author} {\bibfnamefont {S.-y.}\ \bibnamefont {Lee}},
  \bibinfo {author} {\bibfnamefont {H.}~\bibnamefont {Nha}}, \ and\ \bibinfo
  {author} {\bibfnamefont {H.}~\bibnamefont {Jeong}},\ }\href {\doibase
  10.1103/PhysRevA.96.062304} {\bibfield  {journal} {\bibinfo  {journal} {Phys.
  Rev. A}\ }\textbf {\bibinfo {volume} {96}},\ \bibinfo {pages} {062304}
  (\bibinfo {year} {2017})}\BibitemShut {NoStop}%
\bibitem [{\citenamefont {Steuernagel}\ and\ \citenamefont
  {Scheel}(2004)}]{steuernagel20041}%
  \BibitemOpen
  \bibfield  {author} {\bibinfo {author} {\bibfnamefont {O.}~\bibnamefont
  {Steuernagel}}\ and\ \bibinfo {author} {\bibfnamefont {S.}~\bibnamefont
  {Scheel}},\ }\href {\doibase 10.1088/1464-4266/6/3/011} {\bibfield  {journal}
  {\bibinfo  {journal} {J. Opt B: Quantum and Semiclass. Opt.}\ }\textbf
  {\bibinfo {volume} {6}},\ \bibinfo {pages} {566} (\bibinfo {year}
  {2004})}\BibitemShut {NoStop}%
\bibitem [{\citenamefont {Bondurant}\ and\ \citenamefont
  {Shapiro}(1984)}]{bondurant19841}%
  \BibitemOpen
  \bibfield  {author} {\bibinfo {author} {\bibfnamefont {R.~S.}\ \bibnamefont
  {Bondurant}}\ and\ \bibinfo {author} {\bibfnamefont {J.~H.}\ \bibnamefont
  {Shapiro}},\ }\href {\doibase 10.1103/PhysRevD.30.2548} {\bibfield  {journal}
  {\bibinfo  {journal} {Phys. Rev. D}\ }\textbf {\bibinfo {volume} {30}},\
  \bibinfo {pages} {2548} (\bibinfo {year} {1984})}\BibitemShut {NoStop}%
\bibitem [{\citenamefont {Li}\ \emph {et~al.}(2014)\citenamefont {Li},
  \citenamefont {Yuan}, \citenamefont {Ou},\ and\ \citenamefont
  {Zhang}}]{li20141}%
  \BibitemOpen
  \bibfield  {author} {\bibinfo {author} {\bibfnamefont {D.}~\bibnamefont
  {Li}}, \bibinfo {author} {\bibfnamefont {C.~H.}\ \bibnamefont {Yuan}},
  \bibinfo {author} {\bibfnamefont {Z.~Y.}\ \bibnamefont {Ou}}, \ and\ \bibinfo
  {author} {\bibfnamefont {W.}~\bibnamefont {Zhang}},\ }\href {\doibase
  10.1088/1367-2630/16/7/073020} {\bibfield  {journal} {\bibinfo  {journal}
  {New J. Phys.}\ }\textbf {\bibinfo {volume} {16}},\ \bibinfo {pages} {073020}
  (\bibinfo {year} {2014})}\BibitemShut {NoStop}%
\bibitem [{\citenamefont {Pinel}\ \emph {et~al.}(2012)\citenamefont {Pinel},
  \citenamefont {Fade}, \citenamefont {Braun}, \citenamefont {Jian},
  \citenamefont {Treps},\ and\ \citenamefont {Fabre}}]{pinel20121}%
  \BibitemOpen
  \bibfield  {author} {\bibinfo {author} {\bibfnamefont {O.}~\bibnamefont
  {Pinel}}, \bibinfo {author} {\bibfnamefont {J.}~\bibnamefont {Fade}},
  \bibinfo {author} {\bibfnamefont {D.}~\bibnamefont {Braun}}, \bibinfo
  {author} {\bibfnamefont {P.}~\bibnamefont {Jian}}, \bibinfo {author}
  {\bibfnamefont {N.}~\bibnamefont {Treps}}, \ and\ \bibinfo {author}
  {\bibfnamefont {C.}~\bibnamefont {Fabre}},\ }\href {\doibase
  10.1103/PhysRevA.85.010101} {\bibfield  {journal} {\bibinfo  {journal} {Phys.
  Rev. A}\ }\textbf {\bibinfo {volume} {85}},\ \bibinfo {pages} {010101(R)}
  (\bibinfo {year} {2012})}\BibitemShut {NoStop}%
\bibitem [{\citenamefont {Gagatsos}\ \emph {et~al.}(2016)\citenamefont
  {Gagatsos}, \citenamefont {Branford},\ and\ \citenamefont
  {Datta}}]{gagatsos20161}%
  \BibitemOpen
  \bibfield  {author} {\bibinfo {author} {\bibfnamefont {C.~N.}\ \bibnamefont
  {Gagatsos}}, \bibinfo {author} {\bibfnamefont {D.}~\bibnamefont {Branford}},
  \ and\ \bibinfo {author} {\bibfnamefont {A.}~\bibnamefont {Datta}},\ }\href
  {\doibase 10.1103/PhysRevA.94.042342} {\bibfield  {journal} {\bibinfo
  {journal} {Phys. Rev. A}\ }\textbf {\bibinfo {volume} {94}},\ \bibinfo
  {pages} {042342} (\bibinfo {year} {2016})}\BibitemShut {NoStop}%
\bibitem [{\citenamefont {Nichols}\ \emph {et~al.}(2018)\citenamefont
  {Nichols}, \citenamefont {Liuzzo-Scorpo}, \citenamefont {Knott},\ and\
  \citenamefont {Adesso}}]{nichols20181}%
  \BibitemOpen
  \bibfield  {author} {\bibinfo {author} {\bibfnamefont {R.}~\bibnamefont
  {Nichols}}, \bibinfo {author} {\bibfnamefont {P.}~\bibnamefont
  {Liuzzo-Scorpo}}, \bibinfo {author} {\bibfnamefont {P.~A.}\ \bibnamefont
  {Knott}}, \ and\ \bibinfo {author} {\bibfnamefont {G.}~\bibnamefont
  {Adesso}},\ }\href {\doibase 10.1103/PhysRevA.98.012114} {\bibfield
  {journal} {\bibinfo  {journal} {Phys. Rev. A}\ }\textbf {\bibinfo {volume}
  {98}},\ \bibinfo {pages} {012114} (\bibinfo {year} {2018})}\BibitemShut
  {NoStop}%
\bibitem [{\citenamefont {Matsubara}\ \emph {et~al.}(2019)\citenamefont
  {Matsubara}, \citenamefont {Facchi}, \citenamefont {Giovannetti},\ and\
  \citenamefont {Yuasa}}]{matsubara20191}%
  \BibitemOpen
  \bibfield  {author} {\bibinfo {author} {\bibfnamefont {T.}~\bibnamefont
  {Matsubara}}, \bibinfo {author} {\bibfnamefont {P.}~\bibnamefont {Facchi}},
  \bibinfo {author} {\bibfnamefont {V.}~\bibnamefont {Giovannetti}}, \ and\
  \bibinfo {author} {\bibfnamefont {K.}~\bibnamefont {Yuasa}},\ }\href
  {\doibase 10.1088/1367-2630/ab0604} {\bibfield  {journal} {\bibinfo
  {journal} {New J. Phys.}\ }\textbf {\bibinfo {volume} {21}},\ \bibinfo
  {pages} {033014} (\bibinfo {year} {2019})}\BibitemShut {NoStop}%
\bibitem [{\citenamefont {Oh}\ \emph {et~al.}(2019)\citenamefont {Oh},
  \citenamefont {Lee}, \citenamefont {Banchi}, \citenamefont {Lee},
  \citenamefont {Rockstuhl},\ and\ \citenamefont {Jeong}}]{oh20191}%
  \BibitemOpen
  \bibfield  {author} {\bibinfo {author} {\bibfnamefont {C.}~\bibnamefont
  {Oh}}, \bibinfo {author} {\bibfnamefont {C.}~\bibnamefont {Lee}}, \bibinfo
  {author} {\bibfnamefont {L.}~\bibnamefont {Banchi}}, \bibinfo {author}
  {\bibfnamefont {S.-y.}\ \bibnamefont {Lee}}, \bibinfo {author} {\bibfnamefont
  {C.}~\bibnamefont {Rockstuhl}}, \ and\ \bibinfo {author} {\bibfnamefont
  {H.}~\bibnamefont {Jeong}},\ }\href {\doibase 10.1103/PhysRevA.100.012323}
  {\bibfield  {journal} {\bibinfo  {journal} {Phys . Rev. A}\ }\textbf
  {\bibinfo {volume} {100}},\ \bibinfo {pages} {012323} (\bibinfo {year}
  {2019})}\BibitemShut {NoStop}%
\bibitem [{\citenamefont {{\v{S}}afr{\'{a}}nek}\ and\ \citenamefont
  {Fuentes}(2016)}]{safranek20161}%
  \BibitemOpen
  \bibfield  {author} {\bibinfo {author} {\bibfnamefont {D.}~\bibnamefont
  {{\v{S}}afr{\'{a}}nek}}\ and\ \bibinfo {author} {\bibfnamefont
  {I.}~\bibnamefont {Fuentes}},\ }\href {\doibase 10.1103/PhysRevA.94.062313}
  {\bibfield  {journal} {\bibinfo  {journal} {Phys. Rev. A}\ }\textbf {\bibinfo
  {volume} {94}},\ \bibinfo {pages} {062313} (\bibinfo {year}
  {2016})}\BibitemShut {NoStop}%
\bibitem [{\citenamefont {Oh}\ \emph {et~al.}(2020)\citenamefont {Oh},
  \citenamefont {Lee}, \citenamefont {Lie},\ and\ \citenamefont
  {Jeong}}]{oh20201}%
  \BibitemOpen
  \bibfield  {author} {\bibinfo {author} {\bibfnamefont {C.}~\bibnamefont
  {Oh}}, \bibinfo {author} {\bibfnamefont {C.}~\bibnamefont {Lee}}, \bibinfo
  {author} {\bibfnamefont {S.~H.}\ \bibnamefont {Lie}}, \ and\ \bibinfo
  {author} {\bibfnamefont {H.}~\bibnamefont {Jeong}},\ }\href {\doibase
  10.1103/physrevresearch.2.023030} {\bibfield  {journal} {\bibinfo  {journal}
  {Physi. Rev. Research}\ }\textbf {\bibinfo {volume} {2}},\ \bibinfo {pages}
  {023030} (\bibinfo {year} {2020})}\BibitemShut {NoStop}%
\bibitem [{\citenamefont {Yonezawa}\ \emph {et~al.}(2012)\citenamefont
  {Yonezawa}, \citenamefont {Nakane}, \citenamefont {Wheatley}, \citenamefont
  {Iwasawa}, \citenamefont {Takeda}, \citenamefont {Arao}, \citenamefont
  {Ohki}, \citenamefont {Tsumura}, \citenamefont {Berry}, \citenamefont
  {Ralph}, \citenamefont {Wiseman}, \citenamefont {Huntington},\ and\
  \citenamefont {Furusawa}}]{yonezawa20121}%
  \BibitemOpen
  \bibfield  {author} {\bibinfo {author} {\bibfnamefont {H.}~\bibnamefont
  {Yonezawa}}, \bibinfo {author} {\bibfnamefont {D.}~\bibnamefont {Nakane}},
  \bibinfo {author} {\bibfnamefont {T.~A.}\ \bibnamefont {Wheatley}}, \bibinfo
  {author} {\bibfnamefont {K.}~\bibnamefont {Iwasawa}}, \bibinfo {author}
  {\bibfnamefont {S.}~\bibnamefont {Takeda}}, \bibinfo {author} {\bibfnamefont
  {H.}~\bibnamefont {Arao}}, \bibinfo {author} {\bibfnamefont {K.}~\bibnamefont
  {Ohki}}, \bibinfo {author} {\bibfnamefont {K.}~\bibnamefont {Tsumura}},
  \bibinfo {author} {\bibfnamefont {D.~W.}\ \bibnamefont {Berry}}, \bibinfo
  {author} {\bibfnamefont {T.~C.}\ \bibnamefont {Ralph}}, \bibinfo {author}
  {\bibfnamefont {H.~M.}\ \bibnamefont {Wiseman}}, \bibinfo {author}
  {\bibfnamefont {E.~H.}\ \bibnamefont {Huntington}}, \ and\ \bibinfo {author}
  {\bibfnamefont {A.}~\bibnamefont {Furusawa}},\ }\href {\doibase
  10.1126/science.1225258} {\bibfield  {journal} {\bibinfo  {journal}
  {Science}\ }\textbf {\bibinfo {volume} {337}},\ \bibinfo {pages} {1514}
  (\bibinfo {year} {2012})}\BibitemShut {NoStop}%
\bibitem [{\citenamefont {Slussarenko}(2017)}]{slussarenko20171}%
  \BibitemOpen
  \bibfield  {author} {\bibinfo {author} {\bibfnamefont {S.}~\bibnamefont
  {Slussarenko}},\ }\href {\doibase 10.1038/s41566-017-0011-5 Unconditional}
  {\bibfield  {journal} {\bibinfo  {journal} {Nature Photonics}\ }\textbf
  {\bibinfo {volume} {11}},\ \bibinfo {pages} {700} (\bibinfo {year}
  {2017})}\BibitemShut {NoStop}%
\bibitem [{\citenamefont {et~al.}(2013)}]{aasi20131}%
  \BibitemOpen
  \bibfield  {author} {\bibinfo {author} {\bibfnamefont {J.~A.}\ \bibnamefont
  {et~al.}},\ }\href {\doibase 10.1038/nphoton.2013.177} {\bibfield  {journal}
  {\bibinfo  {journal} {Nature Photonics}\ }\textbf {\bibinfo {volume} {7}},\
  \bibinfo {pages} {613} (\bibinfo {year} {2013})}\BibitemShut {NoStop}%
\bibitem [{\citenamefont {Takeoka}\ \emph {et~al.}(2017)\citenamefont
  {Takeoka}, \citenamefont {Seshadreesan}, \citenamefont {You}, \citenamefont
  {Izumi},\ and\ \citenamefont {Dowling}}]{takeoka20171}%
  \BibitemOpen
  \bibfield  {author} {\bibinfo {author} {\bibfnamefont {M.}~\bibnamefont
  {Takeoka}}, \bibinfo {author} {\bibfnamefont {K.~P.}\ \bibnamefont
  {Seshadreesan}}, \bibinfo {author} {\bibfnamefont {C.}~\bibnamefont {You}},
  \bibinfo {author} {\bibfnamefont {S.}~\bibnamefont {Izumi}}, \ and\ \bibinfo
  {author} {\bibfnamefont {J.~P.}\ \bibnamefont {Dowling}},\ }\href {\doibase
  10.1103/PhysRevA.96.052118} {\bibfield  {journal} {\bibinfo  {journal} {Phys.
  Rev. A}\ }\textbf {\bibinfo {volume} {96}},\ \bibinfo {pages} {052118}
  (\bibinfo {year} {2017})}\BibitemShut {NoStop}%
\bibitem [{\citenamefont {Jarzyna}\ and\ \citenamefont
  {Demkowicz-Dobrza{\'{n}}ski}(2012)}]{jarzyna20121}%
  \BibitemOpen
  \bibfield  {author} {\bibinfo {author} {\bibfnamefont {M.}~\bibnamefont
  {Jarzyna}}\ and\ \bibinfo {author} {\bibfnamefont {R.}~\bibnamefont
  {Demkowicz-Dobrza{\'{n}}ski}},\ }\href {\doibase 10.1103/PhysRevA.85.011801}
  {\bibfield  {journal} {\bibinfo  {journal} {Phys. Rev. A}\ }\textbf {\bibinfo
  {volume} {85}},\ \bibinfo {pages} {011801 (R)} (\bibinfo {year}
  {2012})}\BibitemShut {NoStop}%
\bibitem [{\citenamefont {Oh}\ \emph {et~al.}(2018)\citenamefont {Oh},
  \citenamefont {Lee}, \citenamefont {Rockstuhl}, \citenamefont {Jeong},
  \citenamefont {Kim}, \citenamefont {Nha},\ and\ \citenamefont
  {Lee}}]{oh20181}%
  \BibitemOpen
  \bibfield  {author} {\bibinfo {author} {\bibfnamefont {C.}~\bibnamefont
  {Oh}}, \bibinfo {author} {\bibfnamefont {C.}~\bibnamefont {Lee}}, \bibinfo
  {author} {\bibfnamefont {C.}~\bibnamefont {Rockstuhl}}, \bibinfo {author}
  {\bibfnamefont {H.}~\bibnamefont {Jeong}}, \bibinfo {author} {\bibfnamefont
  {J.}~\bibnamefont {Kim}}, \bibinfo {author} {\bibfnamefont {H.}~\bibnamefont
  {Nha}}, \ and\ \bibinfo {author} {\bibfnamefont {S.-Y.}\ \bibnamefont
  {Lee}},\ }\href {\doibase 10.1038/s41534-019-0124-4} {\bibfield  {journal}
  {\bibinfo  {journal} {npj Quantum Information}\ }\textbf {\bibinfo {volume}
  {5}},\ \bibinfo {pages} {10} (\bibinfo {year} {2018})}\BibitemShut {NoStop}%
\bibitem [{\citenamefont {Chaboyer}\ \emph {et~al.}(2015)\citenamefont
  {Chaboyer}, \citenamefont {Meany}, \citenamefont {Helt}, \citenamefont
  {Withford},\ and\ \citenamefont {Steel}}]{chaboyer20151}%
  \BibitemOpen
  \bibfield  {author} {\bibinfo {author} {\bibfnamefont {Z.}~\bibnamefont
  {Chaboyer}}, \bibinfo {author} {\bibfnamefont {T.}~\bibnamefont {Meany}},
  \bibinfo {author} {\bibfnamefont {L.~G.}\ \bibnamefont {Helt}}, \bibinfo
  {author} {\bibfnamefont {M.~J.}\ \bibnamefont {Withford}}, \ and\ \bibinfo
  {author} {\bibfnamefont {M.~J.}\ \bibnamefont {Steel}},\ }\href {\doibase
  10.1038/srep09601} {\bibfield  {journal} {\bibinfo  {journal} {Scientific
  Reports}\ }\textbf {\bibinfo {volume} {5}},\ \bibinfo {pages} {10} (\bibinfo
  {year} {2015})}\BibitemShut {NoStop}%
\bibitem [{\citenamefont {Gramegna}\ \emph {et~al.}()\citenamefont {Gramegna},
  \citenamefont {Triggiani}, \citenamefont {Facchi}, \citenamefont {Narducci},\
  and\ \citenamefont {Tamma}}]{gramegna20201}%
  \BibitemOpen
  \bibfield  {author} {\bibinfo {author} {\bibfnamefont {G.}~\bibnamefont
  {Gramegna}}, \bibinfo {author} {\bibfnamefont {D.}~\bibnamefont {Triggiani}},
  \bibinfo {author} {\bibfnamefont {P.}~\bibnamefont {Facchi}}, \bibinfo
  {author} {\bibfnamefont {F.~A.}\ \bibnamefont {Narducci}}, \ and\ \bibinfo
  {author} {\bibfnamefont {V.}~\bibnamefont {Tamma}},\ }\href
  {http://arxiv.org/abs/2003.12550} {\ }\Eprint
  {http://arxiv.org/abs/arXiv:2003.12550} {arXiv:2003.12550} \BibitemShut
  {NoStop}%
\bibitem [{\citenamefont {Gramegna}\ \emph {et~al.}(2021)\citenamefont
  {Gramegna}, \citenamefont {Triggiani}, \citenamefont {Facchi}, \citenamefont
  {Narducci},\ and\ \citenamefont {Tamma}}]{gramegna20211}%
  \BibitemOpen
  \bibfield  {author} {\bibinfo {author} {\bibfnamefont {G.}~\bibnamefont
  {Gramegna}}, \bibinfo {author} {\bibfnamefont {D.}~\bibnamefont {Triggiani}},
  \bibinfo {author} {\bibfnamefont {P.}~\bibnamefont {Facchi}}, \bibinfo
  {author} {\bibfnamefont {F.~A.}\ \bibnamefont {Narducci}}, \ and\ \bibinfo
  {author} {\bibfnamefont {V.}~\bibnamefont {Tamma}},\ }\href {\doibase
  10.1103/PhysRevResearch.3.013152} {\bibfield  {journal} {\bibinfo  {journal}
  {Phys. Rev. Research}\ }\textbf {\bibinfo {volume} {3}},\ \bibinfo {pages}
  {013152} (\bibinfo {year} {2021})},\ \Eprint
  {http://arxiv.org/abs/2003.12551} {arXiv:2003.12551} \BibitemShut {NoStop}%
\bibitem [{\citenamefont {Polino}\ \emph {et~al.}(2019)\citenamefont {Polino},
  \citenamefont {Riva}, \citenamefont {Valeri}, \citenamefont {Silvestri},
  \citenamefont {Corrielli}, \citenamefont {Crespi}, \citenamefont {Spagnolo},
  \citenamefont {Osellame},\ and\ \citenamefont {Sciarrino}}]{polino20191}%
  \BibitemOpen
  \bibfield  {author} {\bibinfo {author} {\bibfnamefont {E.}~\bibnamefont
  {Polino}}, \bibinfo {author} {\bibfnamefont {M.}~\bibnamefont {Riva}},
  \bibinfo {author} {\bibfnamefont {M.}~\bibnamefont {Valeri}}, \bibinfo
  {author} {\bibfnamefont {R.}~\bibnamefont {Silvestri}}, \bibinfo {author}
  {\bibfnamefont {G.}~\bibnamefont {Corrielli}}, \bibinfo {author}
  {\bibfnamefont {A.}~\bibnamefont {Crespi}}, \bibinfo {author} {\bibfnamefont
  {N.}~\bibnamefont {Spagnolo}}, \bibinfo {author} {\bibfnamefont
  {R.}~\bibnamefont {Osellame}}, \ and\ \bibinfo {author} {\bibfnamefont
  {F.}~\bibnamefont {Sciarrino}},\ }\href {\doibase 10.1364/OPTICA.6.000288}
  {\bibfield  {journal} {\bibinfo  {journal} {Optica}\ }\textbf {\bibinfo
  {volume} {6}},\ \bibinfo {pages} {288} (\bibinfo {year} {2019})}\BibitemShut
  {NoStop}%
\bibitem [{\citenamefont {Paesani}\ \emph {et~al.}(2017)\citenamefont
  {Paesani}, \citenamefont {Gentile}, \citenamefont {Santagati}, \citenamefont
  {Wang}, \citenamefont {Wiebe}, \citenamefont {Tew}, \citenamefont {O'Brien},\
  and\ \citenamefont {Thompson}}]{paesani20171}%
  \BibitemOpen
  \bibfield  {author} {\bibinfo {author} {\bibfnamefont {S.}~\bibnamefont
  {Paesani}}, \bibinfo {author} {\bibfnamefont {A.~A.}\ \bibnamefont
  {Gentile}}, \bibinfo {author} {\bibfnamefont {R.}~\bibnamefont {Santagati}},
  \bibinfo {author} {\bibfnamefont {J.}~\bibnamefont {Wang}}, \bibinfo {author}
  {\bibfnamefont {N.}~\bibnamefont {Wiebe}}, \bibinfo {author} {\bibfnamefont
  {D.~P.}\ \bibnamefont {Tew}}, \bibinfo {author} {\bibfnamefont {J.~L.}\
  \bibnamefont {O'Brien}}, \ and\ \bibinfo {author} {\bibfnamefont {M.~G.}\
  \bibnamefont {Thompson}},\ }\href {\doibase 10.1103/PhysRevLett.118.100503}
  {\bibfield  {journal} {\bibinfo  {journal} {Phys. Rev. Lett.}\ }\textbf
  {\bibinfo {volume} {118}},\ \bibinfo {pages} {100503} (\bibinfo {year}
  {2017})}\BibitemShut {NoStop}%
\bibitem [{\citenamefont {Pinel}\ \emph {et~al.}(2013)\citenamefont {Pinel},
  \citenamefont {Jian}, \citenamefont {Treps}, \citenamefont {Fabre},\ and\
  \citenamefont {Braun}}]{pinel20131}%
  \BibitemOpen
  \bibfield  {author} {\bibinfo {author} {\bibfnamefont {O.}~\bibnamefont
  {Pinel}}, \bibinfo {author} {\bibfnamefont {P.}~\bibnamefont {Jian}},
  \bibinfo {author} {\bibfnamefont {N.}~\bibnamefont {Treps}}, \bibinfo
  {author} {\bibfnamefont {C.}~\bibnamefont {Fabre}}, \ and\ \bibinfo {author}
  {\bibfnamefont {D.}~\bibnamefont {Braun}},\ }\href {\doibase
  10.1103/PhysRevA.88.040102} {\bibfield  {journal} {\bibinfo  {journal} {Phys.
  Rev. A}\ }\textbf {\bibinfo {volume} {88}},\ \bibinfo {pages} {040102 (R)}
  (\bibinfo {year} {2013})}\BibitemShut {NoStop}%
\bibitem [{\citenamefont {You}\ \emph {et~al.}(2017)\citenamefont {You},
  \citenamefont {Adhikari}, \citenamefont {Chi}, \citenamefont {Laborde},
  \citenamefont {Matyas}, \citenamefont {Zhang}, \citenamefont {Su},
  \citenamefont {Byrnes}, \citenamefont {Lu}, \citenamefont {Dowling},\ and\
  \citenamefont {Olson}}]{you20171}%
  \BibitemOpen
  \bibfield  {author} {\bibinfo {author} {\bibfnamefont {C.}~\bibnamefont
  {You}}, \bibinfo {author} {\bibfnamefont {S.}~\bibnamefont {Adhikari}},
  \bibinfo {author} {\bibfnamefont {Y.}~\bibnamefont {Chi}}, \bibinfo {author}
  {\bibfnamefont {M.~L.}\ \bibnamefont {Laborde}}, \bibinfo {author}
  {\bibfnamefont {C.~T.}\ \bibnamefont {Matyas}}, \bibinfo {author}
  {\bibfnamefont {C.}~\bibnamefont {Zhang}}, \bibinfo {author} {\bibfnamefont
  {Z.}~\bibnamefont {Su}}, \bibinfo {author} {\bibfnamefont {T.}~\bibnamefont
  {Byrnes}}, \bibinfo {author} {\bibfnamefont {C.}~\bibnamefont {Lu}}, \bibinfo
  {author} {\bibfnamefont {J.~P.}\ \bibnamefont {Dowling}}, \ and\ \bibinfo
  {author} {\bibfnamefont {J.~P.}\ \bibnamefont {Olson}},\ }\href {\doibase
  10.1088/2040-8986/aa9133} {\bibfield  {journal} {\bibinfo  {journal} {J.
  Opt.}\ }\textbf {\bibinfo {volume} {19}},\ \bibinfo {pages} {124002}
  (\bibinfo {year} {2017})}\BibitemShut {NoStop}%
\bibitem [{\citenamefont {Olson}\ \emph {et~al.}(2017)\citenamefont {Olson},
  \citenamefont {Motes}, \citenamefont {Birchall}, \citenamefont {Studer},
  \citenamefont {LaBorde}, \citenamefont {Moulder}, \citenamefont {Rohde},\
  and\ \citenamefont {Dowling}}]{olson20171}%
  \BibitemOpen
  \bibfield  {author} {\bibinfo {author} {\bibfnamefont {J.~P.}\ \bibnamefont
  {Olson}}, \bibinfo {author} {\bibfnamefont {K.~R.}\ \bibnamefont {Motes}},
  \bibinfo {author} {\bibfnamefont {P.~M.}\ \bibnamefont {Birchall}}, \bibinfo
  {author} {\bibfnamefont {N.~M.}\ \bibnamefont {Studer}}, \bibinfo {author}
  {\bibfnamefont {M.}~\bibnamefont {LaBorde}}, \bibinfo {author} {\bibfnamefont
  {T.}~\bibnamefont {Moulder}}, \bibinfo {author} {\bibfnamefont {P.~P.}\
  \bibnamefont {Rohde}}, \ and\ \bibinfo {author} {\bibfnamefont {J.~P.}\
  \bibnamefont {Dowling}},\ }\href {\doibase 10.1103/PhysRevA.96.013810}
  {\bibfield  {journal} {\bibinfo  {journal} {Phys. Rev. A}\ }\textbf {\bibinfo
  {volume} {96}},\ \bibinfo {pages} {013810} (\bibinfo {year}
  {2017})}\BibitemShut {NoStop}%
\bibitem [{\citenamefont {Motes}\ \emph {et~al.}(2015)\citenamefont {Motes},
  \citenamefont {Olson}, \citenamefont {Rabeaux}, \citenamefont {Dowling},
  \citenamefont {Olson},\ and\ \citenamefont {Rohde}}]{motes20151}%
  \BibitemOpen
  \bibfield  {author} {\bibinfo {author} {\bibfnamefont {K.~R.}\ \bibnamefont
  {Motes}}, \bibinfo {author} {\bibfnamefont {J.~P.}\ \bibnamefont {Olson}},
  \bibinfo {author} {\bibfnamefont {E.~J.}\ \bibnamefont {Rabeaux}}, \bibinfo
  {author} {\bibfnamefont {J.~P.}\ \bibnamefont {Dowling}}, \bibinfo {author}
  {\bibfnamefont {S.~J.}\ \bibnamefont {Olson}}, \ and\ \bibinfo {author}
  {\bibfnamefont {P.~P.}\ \bibnamefont {Rohde}},\ }\href {\doibase
  10.1103/PhysRevLett.114.170802} {\bibfield  {journal} {\bibinfo  {journal}
  {Phys. Rev. Lett.}\ }\textbf {\bibinfo {volume} {114}},\ \bibinfo {pages}
  {170802} (\bibinfo {year} {2015})}\BibitemShut {NoStop}%
\bibitem [{\citenamefont {Su}\ \emph {et~al.}(2017)\citenamefont {Su},
  \citenamefont {Li}, \citenamefont {Rohde}, \citenamefont {Huang},
  \citenamefont {Wang}, \citenamefont {Li}, \citenamefont {Liu}, \citenamefont
  {Dowling}, \citenamefont {Lu},\ and\ \citenamefont {Pan}}]{su20171}%
  \BibitemOpen
  \bibfield  {author} {\bibinfo {author} {\bibfnamefont {Z.-e.}\ \bibnamefont
  {Su}}, \bibinfo {author} {\bibfnamefont {Y.}~\bibnamefont {Li}}, \bibinfo
  {author} {\bibfnamefont {P.~P.}\ \bibnamefont {Rohde}}, \bibinfo {author}
  {\bibfnamefont {H.-l.}\ \bibnamefont {Huang}}, \bibinfo {author}
  {\bibfnamefont {X.-l.}\ \bibnamefont {Wang}}, \bibinfo {author}
  {\bibfnamefont {L.}~\bibnamefont {Li}}, \bibinfo {author} {\bibfnamefont
  {N.-l.}\ \bibnamefont {Liu}}, \bibinfo {author} {\bibfnamefont {J.~P.}\
  \bibnamefont {Dowling}}, \bibinfo {author} {\bibfnamefont {C.-y.}\
  \bibnamefont {Lu}}, \ and\ \bibinfo {author} {\bibfnamefont {J.-w.}\
  \bibnamefont {Pan}},\ }\href {\doibase 10.1103/PhysRevLett.119.080502}
  {\bibfield  {journal} {\bibinfo  {journal} {Phys. Rev. Lett.}\ }\textbf
  {\bibinfo {volume} {119}},\ \bibinfo {pages} {080502} (\bibinfo {year}
  {2017})}\BibitemShut {NoStop}%
\bibitem [{\citenamefont {Ferraro}\ \emph {et~al.}(2005)\citenamefont
  {Ferraro}, \citenamefont {Olivares},\ and\ \citenamefont
  {Paris}}]{ferraro20051}%
  \BibitemOpen
  \bibfield  {author} {\bibinfo {author} {\bibfnamefont {A.}~\bibnamefont
  {Ferraro}}, \bibinfo {author} {\bibfnamefont {S.}~\bibnamefont {Olivares}}, \
  and\ \bibinfo {author} {\bibfnamefont {M.}~\bibnamefont {Paris}},\
  }\href@noop {} {\emph {\bibinfo {title} {Gaussian states in continuous
  variable quantum information}}}\ (\bibinfo  {publisher} {Bibliopolis,
  Napoli},\ \bibinfo {year} {2005})\BibitemShut {NoStop}%
\bibitem [{\citenamefont {Braunstein}\ and\ \citenamefont {van
  Loock}(2005)}]{braunstein20071}%
  \BibitemOpen
  \bibfield  {author} {\bibinfo {author} {\bibfnamefont {S.~L.}\ \bibnamefont
  {Braunstein}}\ and\ \bibinfo {author} {\bibfnamefont {P.}~\bibnamefont {van
  Loock}},\ }\href {\doibase 10.1103/RevModPhys.77.513} {\bibfield  {journal}
  {\bibinfo  {journal} {Rev. Mod. Phys.}\ }\textbf {\bibinfo {volume} {77}},\
  \bibinfo {pages} {513} (\bibinfo {year} {2005})}\BibitemShut {NoStop}%
\bibitem [{\citenamefont {Weedbrook}\ \emph {et~al.}(2012)\citenamefont
  {Weedbrook}, \citenamefont {Pirandola}, \citenamefont {Cerf}, \citenamefont
  {Ralph}, \citenamefont {Shapiro},\ and\ \citenamefont
  {Lloyd}}]{wedbrook20111}%
  \BibitemOpen
  \bibfield  {author} {\bibinfo {author} {\bibfnamefont {C.}~\bibnamefont
  {Weedbrook}}, \bibinfo {author} {\bibfnamefont {S.}~\bibnamefont
  {Pirandola}}, \bibinfo {author} {\bibfnamefont {N.~J.}\ \bibnamefont {Cerf}},
  \bibinfo {author} {\bibfnamefont {T.~C.}\ \bibnamefont {Ralph}}, \bibinfo
  {author} {\bibfnamefont {J.~H.}\ \bibnamefont {Shapiro}}, \ and\ \bibinfo
  {author} {\bibfnamefont {S.}~\bibnamefont {Lloyd}},\ }\href {\doibase
  10.1103/RevModPhys.84.621} {\bibfield  {journal} {\bibinfo  {journal} {Rev.
  Mod. Phys.}\ }\textbf {\bibinfo {volume} {84}},\ \bibinfo {pages} {621}
  (\bibinfo {year} {2012})},\ \Eprint {http://arxiv.org/abs/arXiv:1110.3234v1}
  {arXiv:1110.3234v1} \BibitemShut {NoStop}%
\bibitem [{\citenamefont {Boixo}\ \emph {et~al.}(2007)\citenamefont {Boixo},
  \citenamefont {Flammia}, \citenamefont {Caves},\ and\ \citenamefont
  {Geremia}}]{boixo20071}%
  \BibitemOpen
  \bibfield  {author} {\bibinfo {author} {\bibfnamefont {S.}~\bibnamefont
  {Boixo}}, \bibinfo {author} {\bibfnamefont {S.~T.}\ \bibnamefont {Flammia}},
  \bibinfo {author} {\bibfnamefont {C.~M.}\ \bibnamefont {Caves}}, \ and\
  \bibinfo {author} {\bibfnamefont {J.~M.}\ \bibnamefont {Geremia}},\ }\href
  {\doibase 10.1103/PhysRevLett.98.090401} {\bibfield  {journal} {\bibinfo
  {journal} {Phys. Rev. Lett.}\ }\textbf {\bibinfo {volume} {98}},\ \bibinfo
  {pages} {090401} (\bibinfo {year} {2007})}\BibitemShut {NoStop}%
\bibitem [{\citenamefont {Mathieu Elias~Fra{\"{i}}sse}\ and\ \citenamefont
  {Braun}(2017)}]{fraisse20171}%
  \BibitemOpen
  \bibfield  {author} {\bibinfo {author} {\bibfnamefont {J.}~\bibnamefont
  {Mathieu Elias~Fra{\"{i}}sse}}\ and\ \bibinfo {author} {\bibfnamefont
  {D.}~\bibnamefont {Braun}},\ }\href {\doibase 10.1103/PhysRevA.95.062342}
  {\bibfield  {journal} {\bibinfo  {journal} {Phys. Rev. A}\ }\textbf {\bibinfo
  {volume} {95}},\ \bibinfo {pages} {062342} (\bibinfo {year}
  {2017})}\BibitemShut {NoStop}%
\bibitem [{\citenamefont {Monras}(2006)}]{monras20061}%
  \BibitemOpen
  \bibfield  {author} {\bibinfo {author} {\bibfnamefont {A.}~\bibnamefont
  {Monras}},\ }\href {\doibase 10.1103/PhysRevA.73.033821} {\bibfield
  {journal} {\bibinfo  {journal} {Phys. Revi. A}\ }\textbf {\bibinfo {volume}
  {73}},\ \bibinfo {pages} {033821} (\bibinfo {year} {2006})}\BibitemShut
  {NoStop}%
\bibitem [{\citenamefont {Aspachs}\ \emph {et~al.}(2009)\citenamefont
  {Aspachs}, \citenamefont {Calsamiglia}, \citenamefont {Mu{\~{n}}oz-Tapia},\
  and\ \citenamefont {Bagan}}]{aspachs20091}%
  \BibitemOpen
  \bibfield  {author} {\bibinfo {author} {\bibfnamefont {M.}~\bibnamefont
  {Aspachs}}, \bibinfo {author} {\bibfnamefont {J.}~\bibnamefont
  {Calsamiglia}}, \bibinfo {author} {\bibfnamefont {R.}~\bibnamefont
  {Mu{\~{n}}oz-Tapia}}, \ and\ \bibinfo {author} {\bibfnamefont
  {E.}~\bibnamefont {Bagan}},\ }\href {\doibase 10.1103/PhysRevA.79.033834}
  {\bibfield  {journal} {\bibinfo  {journal} {Phys. Rev. A}\ }\textbf {\bibinfo
  {volume} {79}},\ \bibinfo {pages} {033834} (\bibinfo {year}
  {2009})}\BibitemShut {NoStop}%
\bibitem [{\citenamefont {Sparaciari}\ \emph {et~al.}(2016)\citenamefont
  {Sparaciari}, \citenamefont {Olivares},\ and\ \citenamefont
  {Paris}}]{sparaciari20161}%
  \BibitemOpen
  \bibfield  {author} {\bibinfo {author} {\bibfnamefont {C.}~\bibnamefont
  {Sparaciari}}, \bibinfo {author} {\bibfnamefont {S.}~\bibnamefont
  {Olivares}}, \ and\ \bibinfo {author} {\bibfnamefont {M.~G.~A.}\ \bibnamefont
  {Paris}},\ }\href {\doibase 10.1103/PhysRevA.93.023810} {\bibfield  {journal}
  {\bibinfo  {journal} {Phys. Rev. A}\ }\textbf {\bibinfo {volume} {93}},\
  \bibinfo {pages} {023810} (\bibinfo {year} {2016})}\BibitemShut {NoStop}%
\bibitem [{\citenamefont {Olivares}(2012)}]{olivares20121}%
  \BibitemOpen
  \bibfield  {author} {\bibinfo {author} {\bibfnamefont {S.}~\bibnamefont
  {Olivares}},\ }\href {\doibase 10.1140/epjst/e2012-01532-4} {\bibfield
  {journal} {\bibinfo  {journal} {Eur. Phys. J. Special Topics}\ }\textbf
  {\bibinfo {volume} {203}},\ \bibinfo {pages} {3} (\bibinfo {year}
  {2012})}\BibitemShut {NoStop}%
\bibitem [{\citenamefont {Giedke}\ and\ \citenamefont {{Ignacio
  Cirac}}(2002)}]{giedke20021}%
  \BibitemOpen
  \bibfield  {author} {\bibinfo {author} {\bibfnamefont {G.}~\bibnamefont
  {Giedke}}\ and\ \bibinfo {author} {\bibfnamefont {J.}~\bibnamefont {{Ignacio
  Cirac}}},\ }\href {\doibase 10.1103/PhysRevA.66.032316} {\bibfield  {journal}
  {\bibinfo  {journal} {Phys. Rev. A}\ }\textbf {\bibinfo {volume} {66}},\
  \bibinfo {pages} {032316} (\bibinfo {year} {2002})}\BibitemShut {NoStop}%
\bibitem [{\citenamefont {Genoni}\ \emph {et~al.}(2014)\citenamefont {Genoni},
  \citenamefont {Mancini},\ and\ \citenamefont {Serafini}}]{genoni20141}%
  \BibitemOpen
  \bibfield  {author} {\bibinfo {author} {\bibfnamefont {M.~G.}\ \bibnamefont
  {Genoni}}, \bibinfo {author} {\bibfnamefont {S.}~\bibnamefont {Mancini}}, \
  and\ \bibinfo {author} {\bibfnamefont {A.}~\bibnamefont {Serafini}},\ }\href
  {\doibase 10.1134/S1061920814030054} {\bibfield  {journal} {\bibinfo
  {journal} {Russ. J. Math. Phys.}\ }\textbf {\bibinfo {volume} {21}},\
  \bibinfo {pages} {329} (\bibinfo {year} {2014})}\BibitemShut {NoStop}%
\bibitem [{\citenamefont {Kim}\ and\ \citenamefont {Sanders}(1996)}]{kim19961}%
  \BibitemOpen
  \bibfield  {author} {\bibinfo {author} {\bibfnamefont {M.~S.}\ \bibnamefont
  {Kim}}\ and\ \bibinfo {author} {\bibfnamefont {B.~C.}\ \bibnamefont
  {Sanders}},\ }\href {\doibase 10.1103/PhysRevA.53.3694} {\bibfield  {journal}
  {\bibinfo  {journal} {Phys. Rev. A}\ }\textbf {\bibinfo {volume} {53}},\
  \bibinfo {pages} {3694} (\bibinfo {year} {1996})}\BibitemShut {NoStop}%
\bibitem [{\citenamefont {Pezz{\'{e}}}\ and\ \citenamefont
  {Smerzi}(2008)}]{pezze20081}%
  \BibitemOpen
  \bibfield  {author} {\bibinfo {author} {\bibfnamefont {L.}~\bibnamefont
  {Pezz{\'{e}}}}\ and\ \bibinfo {author} {\bibfnamefont {A.}~\bibnamefont
  {Smerzi}},\ }\href {\doibase 10.1103/PhysRevLett.100.073601} {\bibfield
  {journal} {\bibinfo  {journal} {Phys. Rev. Lett.}\ }\textbf {\bibinfo
  {volume} {100}},\ \bibinfo {pages} {073601} (\bibinfo {year}
  {2008})}\BibitemShut {NoStop}%
\bibitem [{\citenamefont {Braunstein}\ \emph {et~al.}(1996)\citenamefont
  {Braunstein}, \citenamefont {Caves},\ and\ \citenamefont
  {Milburn}}]{braunstein19961}%
  \BibitemOpen
  \bibfield  {author} {\bibinfo {author} {\bibfnamefont {S.~L.}\ \bibnamefont
  {Braunstein}}, \bibinfo {author} {\bibfnamefont {C.~M.}\ \bibnamefont
  {Caves}}, \ and\ \bibinfo {author} {\bibfnamefont {G.~J.}\ \bibnamefont
  {Milburn}},\ }\href {\doibase 10.1006/aphy.1996.0040} {\bibfield  {journal}
  {\bibinfo  {journal} {Ann. Phys.}\ }\textbf {\bibinfo {volume} {173}},\
  \bibinfo {pages} {135} (\bibinfo {year} {1996})}\BibitemShut {NoStop}%
\bibitem [{\citenamefont {Pezze}\ \emph {et~al.}(2017)\citenamefont {Pezze},
  \citenamefont {Ciampini}, \citenamefont {Humphreys}, \citenamefont {Datta},
  \citenamefont {Walmsley}, \citenamefont {Barbieri}, \citenamefont
  {Sciarrino},\ and\ \citenamefont {Smerzi}}]{pezze20171}%
  \BibitemOpen
  \bibfield  {author} {\bibinfo {author} {\bibfnamefont {L.}~\bibnamefont
  {Pezze}}, \bibinfo {author} {\bibfnamefont {M.~A.}\ \bibnamefont {Ciampini}},
  \bibinfo {author} {\bibfnamefont {P.~C.}\ \bibnamefont {Humphreys}}, \bibinfo
  {author} {\bibfnamefont {A.}~\bibnamefont {Datta}}, \bibinfo {author}
  {\bibfnamefont {I.~A.}\ \bibnamefont {Walmsley}}, \bibinfo {author}
  {\bibfnamefont {M.}~\bibnamefont {Barbieri}}, \bibinfo {author}
  {\bibfnamefont {F.}~\bibnamefont {Sciarrino}}, \ and\ \bibinfo {author}
  {\bibfnamefont {A.}~\bibnamefont {Smerzi}},\ }\href {\doibase
  10.1103/PhysRevLett.119.130504} {\bibfield  {journal} {\bibinfo  {journal}
  {Phys. Rev. Lett.}\ }\textbf {\bibinfo {volume} {119}},\ \bibinfo {pages}
  {130504} (\bibinfo {year} {2017})}\BibitemShut {NoStop}%
\bibitem [{\citenamefont {Petersen}\ and\ \citenamefont
  {Pedersen}(2012)}]{petersen20121}%
  \BibitemOpen
  \bibfield  {author} {\bibinfo {author} {\bibfnamefont {K.}~\bibnamefont
  {Petersen}}\ and\ \bibinfo {author} {\bibfnamefont {M.}~\bibnamefont
  {Pedersen}},\ }\href@noop {} {\emph {\bibinfo {title} {The Matrix
  Cookbook}}}\ (\bibinfo  {publisher} {Technical University of Denmark},\
  \bibinfo {year} {2012})\BibitemShut {NoStop}%
\bibitem [{\citenamefont {Bernstein}(2002)}]{bernstein20051}%
  \BibitemOpen
  \bibfield  {author} {\bibinfo {author} {\bibfnamefont {D.}~\bibnamefont
  {Bernstein}},\ }\href@noop {} {\emph {\bibinfo {title} {Matrix mathematics:
  theory, facts, and formulas with application to linear systems theory}}}\
  (\bibinfo  {publisher} {Princeton University Press},\ \bibinfo {year}
  {2002})\BibitemShut {NoStop}%
\bibitem [{\citenamefont {Eisert}\ \emph {et~al.}(2002)\citenamefont {Eisert},
  \citenamefont {Scheel},\ and\ \citenamefont {Plenio}}]{eisert20021}%
  \BibitemOpen
  \bibfield  {author} {\bibinfo {author} {\bibfnamefont {J.}~\bibnamefont
  {Eisert}}, \bibinfo {author} {\bibfnamefont {S.}~\bibnamefont {Scheel}}, \
  and\ \bibinfo {author} {\bibfnamefont {M.~B.}\ \bibnamefont {Plenio}},\
  }\href {\doibase 10.1103/PhysRevLett.89.137903} {\bibfield  {journal}
  {\bibinfo  {journal} {Phys. Rev. Lett.}\ }\textbf {\bibinfo {volume} {89}},\
  \bibinfo {pages} {137903} (\bibinfo {year} {2002})}\BibitemShut {NoStop}%
\bibitem [{Note1()}]{Note1}%
  \BibitemOpen
  \bibinfo {note} {Physically, the tensorial product of identical Gaussian
  states (up to an arbitrary displacement) with diagonal CV matrix remains
  invariant under a beam splitter transformation (which induces no phase shift
  between transmitted and reflected modes) for any transmission coefficient
  \cite {springer20091,kim20021}.}\BibitemShut {Stop}%
\bibitem [{\citenamefont {Olivares}\ and\ \citenamefont
  {Paris}(2009)}]{olivares20091}%
  \BibitemOpen
  \bibfield  {author} {\bibinfo {author} {\bibfnamefont {S.}~\bibnamefont
  {Olivares}}\ and\ \bibinfo {author} {\bibfnamefont {M.~G.}\ \bibnamefont
  {Paris}},\ }\href {\doibase 10.1088/0953-4075/42/5/055506} {\bibfield
  {journal} {\bibinfo  {journal} {J. Phys. B: At. Mol. Opt. Phys.}\ }\textbf
  {\bibinfo {volume} {42}},\ \bibinfo {pages} {055506} (\bibinfo {year}
  {2009})}\BibitemShut {NoStop}%
\bibitem [{\citenamefont {DePasquale}\ \emph {et~al.}(2015)\citenamefont
  {DePasquale}, \citenamefont {Facchi}, \citenamefont {Florio}, \citenamefont
  {Giovannetti}, \citenamefont {Matsuoka},\ and\ \citenamefont
  {Yuasa}}]{pasquale20151}%
  \BibitemOpen
  \bibfield  {author} {\bibinfo {author} {\bibfnamefont {A.}~\bibnamefont
  {DePasquale}}, \bibinfo {author} {\bibfnamefont {P.}~\bibnamefont {Facchi}},
  \bibinfo {author} {\bibfnamefont {G.}~\bibnamefont {Florio}}, \bibinfo
  {author} {\bibfnamefont {V.}~\bibnamefont {Giovannetti}}, \bibinfo {author}
  {\bibfnamefont {K.}~\bibnamefont {Matsuoka}}, \ and\ \bibinfo {author}
  {\bibfnamefont {K.}~\bibnamefont {Yuasa}},\ }\href {\doibase
  10.1103/PhysRevA.92.042115} {\bibfield  {journal} {\bibinfo  {journal} {Phys.
  Rev. A}\ }\textbf {\bibinfo {volume} {92}},\ \bibinfo {pages} {042115}
  (\bibinfo {year} {2015})}\BibitemShut {NoStop}%
\bibitem [{\citenamefont {Gaiba}\ and\ \citenamefont
  {Paris}(2009)}]{gaiba20091}%
  \BibitemOpen
  \bibfield  {author} {\bibinfo {author} {\bibfnamefont {R.}~\bibnamefont
  {Gaiba}}\ and\ \bibinfo {author} {\bibfnamefont {M.~G.}\ \bibnamefont
  {Paris}},\ }\href {\doibase 10.1016/j.physleta.2009.01.026} {\bibfield
  {journal} {\bibinfo  {journal} {Phys. Lett. A}\ }\textbf {\bibinfo {volume}
  {373}},\ \bibinfo {pages} {934} (\bibinfo {year} {2009})}\BibitemShut
  {NoStop}%
\bibitem [{\citenamefont {Lang}\ and\ \citenamefont {Caves}(2014)}]{lang20141}%
  \BibitemOpen
  \bibfield  {author} {\bibinfo {author} {\bibfnamefont {M.~D.}\ \bibnamefont
  {Lang}}\ and\ \bibinfo {author} {\bibfnamefont {C.~M.}\ \bibnamefont
  {Caves}},\ }\href {\doibase 10.1103/PhysRevA.90.025802} {\bibfield  {journal}
  {\bibinfo  {journal} {Phys. Rev. A}\ }\textbf {\bibinfo {volume} {90}},\
  \bibinfo {pages} {025802} (\bibinfo {year} {2014})}\BibitemShut {NoStop}%
\bibitem [{\citenamefont {Sparaciari}\ \emph {et~al.}(2015)\citenamefont
  {Sparaciari}, \citenamefont {Olivares},\ and\ \citenamefont
  {Paris}}]{sparaciari20151}%
  \BibitemOpen
  \bibfield  {author} {\bibinfo {author} {\bibfnamefont {C.}~\bibnamefont
  {Sparaciari}}, \bibinfo {author} {\bibfnamefont {S.}~\bibnamefont
  {Olivares}}, \ and\ \bibinfo {author} {\bibfnamefont {M.~G.~A.}\ \bibnamefont
  {Paris}},\ }\href {\doibase 10.1364/josab.32.001354} {\bibfield  {journal}
  {\bibinfo  {journal} {JOSAB}\ }\textbf {\bibinfo {volume} {32}},\ \bibinfo
  {pages} {001354} (\bibinfo {year} {2015})}\BibitemShut {NoStop}%
\bibitem [{\citenamefont {Caves}(1981)}]{caves19811}%
  \BibitemOpen
  \bibfield  {author} {\bibinfo {author} {\bibfnamefont {C.~M.}\ \bibnamefont
  {Caves}},\ }\href {\doibase 10.1103/PhysRevD.23.1693} {\bibfield  {journal}
  {\bibinfo  {journal} {Phys. Rev. D}\ }\textbf {\bibinfo {volume} {23}},\
  \bibinfo {pages} {1693} (\bibinfo {year} {1981})}\BibitemShut {NoStop}%
\bibitem [{\citenamefont {Zhang}\ \emph {et~al.}(2018)\citenamefont {Zhang},
  \citenamefont {Jin}, \citenamefont {Zhang}, \citenamefont {Cen},
  \citenamefont {Hu},\ and\ \citenamefont {Zhao}}]{hang20181}%
  \BibitemOpen
  \bibfield  {author} {\bibinfo {author} {\bibfnamefont {J.-D.}\ \bibnamefont
  {Zhang}}, \bibinfo {author} {\bibfnamefont {C.-F.}\ \bibnamefont {Jin}},
  \bibinfo {author} {\bibfnamefont {Z.-J.}\ \bibnamefont {Zhang}}, \bibinfo
  {author} {\bibfnamefont {L.-Z.}\ \bibnamefont {Cen}}, \bibinfo {author}
  {\bibfnamefont {J.-Y.}\ \bibnamefont {Hu}}, \ and\ \bibinfo {author}
  {\bibfnamefont {Y.}~\bibnamefont {Zhao}},\ }\href {\doibase
  10.1364/OE.26.033080} {\bibfield  {journal} {\bibinfo  {journal} {Opt.
  Express}\ }\textbf {\bibinfo {volume} {26}},\ \bibinfo {pages} {33080}
  (\bibinfo {year} {2018})}\BibitemShut {NoStop}%
\bibitem [{\citenamefont {Yurke}\ \emph {et~al.}(1986)\citenamefont {Yurke},
  \citenamefont {Mccall},\ and\ \citenamefont {Klauder}}]{yurke19861}%
  \BibitemOpen
  \bibfield  {author} {\bibinfo {author} {\bibfnamefont {B.}~\bibnamefont
  {Yurke}}, \bibinfo {author} {\bibfnamefont {S.~L.}\ \bibnamefont {Mccall}}, \
  and\ \bibinfo {author} {\bibfnamefont {J.~R.}\ \bibnamefont {Klauder}},\
  }\href {\doibase 10.1103/PhysRevA.33.4033} {\bibfield  {journal} {\bibinfo
  {journal} {Phys. Rev. A}\ }\textbf {\bibinfo {volume} {33}},\ \bibinfo
  {pages} {4033} (\bibinfo {year} {1986})}\BibitemShut {NoStop}%
\bibitem [{\citenamefont {Anisimov}\ \emph {et~al.}(2010)\citenamefont
  {Anisimov}, \citenamefont {Raterman}, \citenamefont {Chiruvelli},
  \citenamefont {Plick}, \citenamefont {Huver}, \citenamefont {Lee},\ and\
  \citenamefont {Dowling}}]{anisimov20101}%
  \BibitemOpen
  \bibfield  {author} {\bibinfo {author} {\bibfnamefont {P.~M.}\ \bibnamefont
  {Anisimov}}, \bibinfo {author} {\bibfnamefont {G.~M.}\ \bibnamefont
  {Raterman}}, \bibinfo {author} {\bibfnamefont {A.}~\bibnamefont
  {Chiruvelli}}, \bibinfo {author} {\bibfnamefont {W.~N.}\ \bibnamefont
  {Plick}}, \bibinfo {author} {\bibfnamefont {S.~D.}\ \bibnamefont {Huver}},
  \bibinfo {author} {\bibfnamefont {H.}~\bibnamefont {Lee}}, \ and\ \bibinfo
  {author} {\bibfnamefont {J.~P.}\ \bibnamefont {Dowling}},\ }\href {\doibase
  10.1103/PhysRevLett.104.103602} {\bibfield  {journal} {\bibinfo  {journal}
  {Phys. Rev. Lett.}\ }\textbf {\bibinfo {volume} {104}},\ \bibinfo {pages}
  {103602} (\bibinfo {year} {2010})}\BibitemShut {NoStop}%
\bibitem [{\citenamefont {Plick}\ \emph {et~al.}(2010)\citenamefont {Plick},
  \citenamefont {Anisimov}, \citenamefont {Dowling}, \citenamefont {Lee},\ and\
  \citenamefont {Agarwal}}]{plick20101}%
  \BibitemOpen
  \bibfield  {author} {\bibinfo {author} {\bibfnamefont {W.~N.}\ \bibnamefont
  {Plick}}, \bibinfo {author} {\bibfnamefont {P.~M.}\ \bibnamefont {Anisimov}},
  \bibinfo {author} {\bibfnamefont {J.~P.}\ \bibnamefont {Dowling}}, \bibinfo
  {author} {\bibfnamefont {H.}~\bibnamefont {Lee}}, \ and\ \bibinfo {author}
  {\bibfnamefont {G.~S.}\ \bibnamefont {Agarwal}},\ }\href {\doibase
  10.1088/1367-2630/12/11/113025} {\bibfield  {journal} {\bibinfo  {journal}
  {New J. Phys.}\ }\textbf {\bibinfo {volume} {12}},\ \bibinfo {pages} {113025}
  (\bibinfo {year} {2010})}\BibitemShut {NoStop}%
\bibitem [{\citenamefont {Birrittella}\ \emph {et~al.}(2015)\citenamefont
  {Birrittella}, \citenamefont {Gura},\ and\ \citenamefont
  {Gerry}}]{birrittella20151}%
  \BibitemOpen
  \bibfield  {author} {\bibinfo {author} {\bibfnamefont {R.}~\bibnamefont
  {Birrittella}}, \bibinfo {author} {\bibfnamefont {A.}~\bibnamefont {Gura}}, \
  and\ \bibinfo {author} {\bibfnamefont {C.~C.}\ \bibnamefont {Gerry}},\ }\href
  {\doibase 10.1103/PhysRevA.91.053801} {\bibfield  {journal} {\bibinfo
  {journal} {Physi. Rev. A}\ }\textbf {\bibinfo {volume} {91}},\ \bibinfo
  {pages} {053801} (\bibinfo {year} {2015})}\BibitemShut {NoStop}%
\bibitem [{\citenamefont {Kim}\ \emph {et~al.}(2002)\citenamefont {Kim},
  \citenamefont {Son}, \citenamefont {Bu{\v z}ek},\ and\ \citenamefont
  {Knight}}]{kim20021}%
  \BibitemOpen
  \bibfield  {author} {\bibinfo {author} {\bibfnamefont {M.~S.}\ \bibnamefont
  {Kim}}, \bibinfo {author} {\bibfnamefont {W.}~\bibnamefont {Son}}, \bibinfo
  {author} {\bibfnamefont {V.}~\bibnamefont {Bu{\v z}ek}}, \ and\ \bibinfo
  {author} {\bibfnamefont {P.~L.}\ \bibnamefont {Knight}},\ }\href {\doibase
  10.1103/PhysRevA.65.032323} {\bibfield  {journal} {\bibinfo  {journal} {Phys.
  Rev. A}\ }\textbf {\bibinfo {volume} {65}},\ \bibinfo {pages} {032323}
  (\bibinfo {year} {2002})}\BibitemShut {NoStop}%
\bibitem [{\citenamefont {You}\ \emph {et~al.}(2019)\citenamefont {You},
  \citenamefont {Adhikari}, \citenamefont {Ma}, \citenamefont {Sasaki},
  \citenamefont {Takeoka},\ and\ \citenamefont {Dowling}}]{you20191}%
  \BibitemOpen
  \bibfield  {author} {\bibinfo {author} {\bibfnamefont {C.}~\bibnamefont
  {You}}, \bibinfo {author} {\bibfnamefont {S.}~\bibnamefont {Adhikari}},
  \bibinfo {author} {\bibfnamefont {X.}~\bibnamefont {Ma}}, \bibinfo {author}
  {\bibfnamefont {M.}~\bibnamefont {Sasaki}}, \bibinfo {author} {\bibfnamefont
  {M.}~\bibnamefont {Takeoka}}, \ and\ \bibinfo {author} {\bibfnamefont
  {J.~P.}\ \bibnamefont {Dowling}},\ }\href {\doibase
  10.1103/PhysRevA.99.042122} {\bibfield  {journal} {\bibinfo  {journal}
  {Physical Review A}\ }\textbf {\bibinfo {volume} {99}},\ \bibinfo {pages}
  {042122} (\bibinfo {year} {2019})}\BibitemShut {NoStop}%
\bibitem [{\citenamefont {Jarzyna}\ and\ \citenamefont
  {Zwierz}(2017)}]{jarzyna20171}%
  \BibitemOpen
  \bibfield  {author} {\bibinfo {author} {\bibfnamefont {M.}~\bibnamefont
  {Jarzyna}}\ and\ \bibinfo {author} {\bibfnamefont {M.}~\bibnamefont
  {Zwierz}},\ }\href {\doibase 10.1103/PhysRevA.95.012109} {\bibfield
  {journal} {\bibinfo  {journal} {Phys. Rev. A}\ }\textbf {\bibinfo {volume}
  {95}},\ \bibinfo {pages} {012109} (\bibinfo {year} {2017})}\BibitemShut
  {NoStop}%
\bibitem [{\citenamefont {Gao}(2016)}]{gao20161}%
  \BibitemOpen
  \bibfield  {author} {\bibinfo {author} {\bibfnamefont {Y.}~\bibnamefont
  {Gao}},\ }\href {\doibase 10.1103/PhysRevA.94.023834} {\bibfield  {journal}
  {\bibinfo  {journal} {Phys. Rev. A}\ }\textbf {\bibinfo {volume} {94}},\
  \bibinfo {pages} {023834} (\bibinfo {year} {2016})}\BibitemShut {NoStop}%
\bibitem [{\citenamefont {Gagatsos}\ \emph {et~al.}(2017)\citenamefont
  {Gagatsos}, \citenamefont {Bash}, \citenamefont {Guha},\ and\ \citenamefont
  {Datta}}]{gagatsos20171}%
  \BibitemOpen
  \bibfield  {author} {\bibinfo {author} {\bibfnamefont {C.~N.}\ \bibnamefont
  {Gagatsos}}, \bibinfo {author} {\bibfnamefont {B.~A.}\ \bibnamefont {Bash}},
  \bibinfo {author} {\bibfnamefont {S.}~\bibnamefont {Guha}}, \ and\ \bibinfo
  {author} {\bibfnamefont {A.}~\bibnamefont {Datta}},\ }\href {\doibase
  10.1103/PhysRevA.96.062306} {\bibfield  {journal} {\bibinfo  {journal} {Phys.
  Rev. A}\ }\textbf {\bibinfo {volume} {96}},\ \bibinfo {pages} {062306}
  (\bibinfo {year} {2017})}\BibitemShut {NoStop}%
\bibitem [{\citenamefont {Escher}\ \emph {et~al.}(2011)\citenamefont {Escher},
  \citenamefont {{De Matos Filho}},\ and\ \citenamefont
  {Davidovich}}]{escher20111}%
  \BibitemOpen
  \bibfield  {author} {\bibinfo {author} {\bibfnamefont {B.~M.}\ \bibnamefont
  {Escher}}, \bibinfo {author} {\bibfnamefont {R.~L.}\ \bibnamefont {{De Matos
  Filho}}}, \ and\ \bibinfo {author} {\bibfnamefont {L.}~\bibnamefont
  {Davidovich}},\ }\href {\doibase 10.1038/nphys1958} {\bibfield  {journal}
  {\bibinfo  {journal} {Nature Phys.}\ }\textbf {\bibinfo {volume} {7}},\
  \bibinfo {pages} {406} (\bibinfo {year} {2011})}\BibitemShut {NoStop}%
\bibitem [{\citenamefont {Ko{\l}odynski}\ and\ \citenamefont
  {Demkowicz-Dobrzanski}(2013)}]{koodynski20131}%
  \BibitemOpen
  \bibfield  {author} {\bibinfo {author} {\bibfnamefont {J.}~\bibnamefont
  {Ko{\l}odynski}}\ and\ \bibinfo {author} {\bibfnamefont {R.}~\bibnamefont
  {Demkowicz-Dobrzanski}},\ }\href {\doibase 10.1088/1367-2630/15/7/073043}
  {\bibfield  {journal} {\bibinfo  {journal} {New J. Phys}\ }\textbf {\bibinfo
  {volume} {15}},\ \bibinfo {pages} {073043} (\bibinfo {year}
  {2013})}\BibitemShut {NoStop}%
\bibitem [{\citenamefont {Valido}\ \emph {et~al.}(2014)\citenamefont {Valido},
  \citenamefont {Levi},\ and\ \citenamefont {Mintert}}]{valido20141}%
  \BibitemOpen
  \bibfield  {author} {\bibinfo {author} {\bibfnamefont {A.~A.}\ \bibnamefont
  {Valido}}, \bibinfo {author} {\bibfnamefont {F.}~\bibnamefont {Levi}}, \ and\
  \bibinfo {author} {\bibfnamefont {F.}~\bibnamefont {Mintert}},\ }\href
  {\doibase 10.1103/PhysRevA.90.052321} {\bibfield  {journal} {\bibinfo
  {journal} {Phys. Rev. A}\ }\textbf {\bibinfo {volume} {90}},\ \bibinfo
  {pages} {052321} (\bibinfo {year} {2014})}\BibitemShut {NoStop}%
\bibitem [{\citenamefont {Serafini}\ \emph {et~al.}(2005)\citenamefont
  {Serafini}, \citenamefont {Paris}, \citenamefont {Illuminati},\ and\
  \citenamefont {{De Siena}}}]{serafini20051}%
  \BibitemOpen
  \bibfield  {author} {\bibinfo {author} {\bibfnamefont {A.}~\bibnamefont
  {Serafini}}, \bibinfo {author} {\bibfnamefont {M.~G.}\ \bibnamefont {Paris}},
  \bibinfo {author} {\bibfnamefont {F.}~\bibnamefont {Illuminati}}, \ and\
  \bibinfo {author} {\bibfnamefont {S.}~\bibnamefont {{De Siena}}},\ }\href
  {\doibase 10.1088/1464-4266/7/4/R01} {\bibfield  {journal} {\bibinfo
  {journal} {J. Opt. B: Quantum Semiclass. Opt.}\ }\textbf {\bibinfo {volume}
  {7}},\ \bibinfo {pages} {R19} (\bibinfo {year} {2005})}\BibitemShut {NoStop}%
\bibitem [{\citenamefont {Harris}\ \emph {et~al.}(2017)\citenamefont {Harris},
  \citenamefont {Steinbrecher}, \citenamefont {Mower}, \citenamefont {Lahini},
  \citenamefont {Prabhu}, \citenamefont {Baehr-Jones}, \citenamefont
  {Hochberg}, \citenamefont {Lloyd},\ and\ \citenamefont
  {Englund}}]{harris20171}%
  \BibitemOpen
  \bibfield  {author} {\bibinfo {author} {\bibfnamefont {N.~C.}\ \bibnamefont
  {Harris}}, \bibinfo {author} {\bibfnamefont {G.~R.}\ \bibnamefont
  {Steinbrecher}}, \bibinfo {author} {\bibfnamefont {J.}~\bibnamefont {Mower}},
  \bibinfo {author} {\bibfnamefont {Y.}~\bibnamefont {Lahini}}, \bibinfo
  {author} {\bibfnamefont {M.}~\bibnamefont {Prabhu}}, \bibinfo {author}
  {\bibfnamefont {T.}~\bibnamefont {Baehr-Jones}}, \bibinfo {author}
  {\bibfnamefont {M.}~\bibnamefont {Hochberg}}, \bibinfo {author}
  {\bibfnamefont {S.}~\bibnamefont {Lloyd}}, \ and\ \bibinfo {author}
  {\bibfnamefont {D.}~\bibnamefont {Englund}},\ }\href {\doibase
  10.1038/NPHOTON.2017.95} {\bibfield  {journal} {\bibinfo  {journal} {Nature
  Photonics}\ }\textbf {\bibinfo {volume} {11}},\ \bibinfo {pages} {447}
  (\bibinfo {year} {2017})}\BibitemShut {NoStop}%
\bibitem [{\citenamefont {Springer}\ \emph {et~al.}(2009)\citenamefont
  {Springer}, \citenamefont {Lee}, \citenamefont {Bellini},\ and\ \citenamefont
  {Kim}}]{springer20091}%
  \BibitemOpen
  \bibfield  {author} {\bibinfo {author} {\bibfnamefont {S.~C.}\ \bibnamefont
  {Springer}}, \bibinfo {author} {\bibfnamefont {J.}~\bibnamefont {Lee}},
  \bibinfo {author} {\bibfnamefont {M.}~\bibnamefont {Bellini}}, \ and\
  \bibinfo {author} {\bibfnamefont {M.~S.}\ \bibnamefont {Kim}},\ }\href
  {\doibase 10.1103/PhysRevA.79.062303} {\bibfield  {journal} {\bibinfo
  {journal} {Phys. Rev. A}\ }\textbf {\bibinfo {volume} {79}},\ \bibinfo
  {pages} {062303} (\bibinfo {year} {2009})}\BibitemShut {NoStop}%
\end{thebibliography}%

\end{document}